\documentclass[fleqn,usenatbib]{mnras}

\usepackage{newtxtext,newtxmath}

\usepackage[T1]{fontenc}
\usepackage[utf8]{inputenc}

\usepackage{graphicx}	
\usepackage{amsmath}	
\usepackage{amssymb}	
\usepackage{adjustbox}
\usepackage[version=3]{mhchem}
\usepackage{lscape}
\usepackage{longtable}
\usepackage{booktabs}
\usepackage{caption}
\usepackage[dvipsnames]{xcolor}

\title[Colour-magnitude diagrams III]{Colour-magnitude diagrams of transiting Exoplanets - III.  A public code, nine strange planets, and the role of Phosphine.}

\author[G. Dransfield et al.]{
Georgina Dransfield,$^{1}$\thanks{E-mail: gxg831@bham.ac.uk}
Amaury H.M.J. Triaud,$^{1}$
\\
$^{1}$School of Physics \& Astronomy, University of Birmingham, Edgbaston, Birmingham B15 2TT, United Kingdom\\
}

\date{Accepted XXX. Received YYY; in original form ZZZ}

\pubyear{2020}

\begin{document}
\label{firstpage}
\pagerange{\pageref{firstpage}--\pageref{lastpage}}
\maketitle

\begin{abstract}

Colour-Magnitude Diagrams provide a convenient way of comparing populations of similar objects. When well populated with precise measurements, they allow quick inferences to be made about the bulk properties of an astronomic object simply from its proximity on a diagram to other objects. We present here a Python toolkit which allows a user to produce colour-magnitude diagrams of transiting exoplanets, comparing planets to populations of ultra-cool dwarfs, of directly imaged exoplanets, to theoretical models of planetary atmospheres, and to other transiting exoplanets.
Using a selection of near- and mid-infrared colour-magnitude diagrams, we show how outliers can be identified for further investigation, and how emerging sub-populations can be identified. Additionally, we present evidence that observed differences in the {\it \textit{Spitzer}}'s 4.5$\mu$m flux, between irradiated Jupiters, and field brown dwarfs, might be attributed to phosphine, which is susceptible to photolysis. The presence of phosphine in low irradiation environments may negate the need for thermal inversions to explain eclipse measurements. 
We speculate that the anomalously low 4.5$\mu$m flux flux of the nightside of HD\,189733b and the daysides of GJ\,436b and GJ\,3470b might be caused by phosphine absorption. Finally, we use our toolkit to include {\it Hubble} WFC3 spectra, creating a new photometric band called the `Water band' (\textit{W$_{JH}$}-band) in the process. We show that the colour index [\textit{W$_{JH}$-H}] can be used to constrain the C/O ratio of exoplanets, showing that future observations with {\it JWST} and \textcolor{Black}{{\it Ariel}} will be able to distinguish these populations if they exist, and select members for future follow-up.

\end{abstract}

\begin{keywords}
planets and satellites: atmospheres -- binaries: eclipsing -- brown dwarfs -- Hertzsprung-Russell and colour-magnitude diagrams -- planetary systems
\end{keywords}



\section{Introduction}

We have come a long way since \cite{1995Natur.378..355M} discovered the first hot Jupiter orbiting a sun-like star. We now know of over 4000 exoplanets and there are over 2000 candidates waiting for confirmation of their planetary status\footnote{\url{https://exoplanetarchive.ipac.caltech.edu/docs/counts_detail.html}}. Our ambition has grown with the broadening scope of the field: we are now not content to simply know of a system's geometry, but are probing various layers of exoplanetary atmospheres through transit spectroscopy and multi-waveband photometry \citep[for a review:][]{2019arXiv190403190M}.
We are able to infer the presence of atomic \citep[e.g.][]{2002ApJ...568..377C,2008ApJ...673L..87R} and molecular species \citep[e.g.][]{2014ApJ...791...55M,2014ApJ...793L..27K,2017ApJ...850L..32S}. This in turn reveals some of the chemical and thermal transport processes taking place at different pressure levels within the atmosphere \citep{2014Sci...346..838S}. A clearer picture of the chemical composition of a planet's atmosphere allows us to compute useful parameters such as the carbon-to-oxygen (C/O) ratio \citep{2013ApJ...763...25M}. 
The C/O ratio is particularly useful  to probe the nebular gas in which the planet formed, and hence the location of its formation within a disc \citep{2011ApJ...743L..16O,2011ApJ...743..191M}. 

Inferences for the C/O ratio have regularly been performed on transmission spectra \citep[e.g.][]{2019MNRAS.482.1485P}. However transmission only probes a special location of the atmosphere of a tidally locked planet, its terminator, which might not be representative of the whole. In addition, transmission spectra can be strongly affected by opacity on the line of sight, with haze and clouds often masking important features \citep{2016AAS...22730603S}. Furthermore transmission can be affected by stellar contamination \citep[e.g.][]{2013ApJ...778..184J, 2017DPS....4941620R}.

A solution is to measure a planet's integrated dayside thermal emission, which is obtained during a secondary eclipse event (or occultation) when the planet passes behind its parent star \citep{2018haex.bookE.100K}. When this has been detected in several photometric bands a low resolution emission spectrum emerges \citep{2018haex.bookE..40A}, which we can use to retrieve atmospheric compositions.

The process of atmospheric retrieval relies on reliable atmospheric models, \textcolor{Black}{some of} which require accurate chemical networks and complete line lists for all the main opacity sources (see \cite{2018haex.bookE.104M} for a detailed review of atmospheric retrieval processes). At present we have access to several state-of-the-art modelling codes, but they are very much still in flux, with chemical networks and line lists being updated frequently \citep[e.g.][]{2017ApJ...850..150B, 2019arXiv191207246V, 2019MNRAS.487.2242H}. Retrieval codes being computationally intensive, there is interest in including as small a number of species as possible. However if a certain molecule is present in the atmosphere, but absent in the code, the retrieved abundances will be inaccurate \citep{2015ESS.....311920W, 2017ApJ...850L..15M}. Obtaining diagnostics about which atomic or molecular species are present in a spectrum is therefore important. This is where colour-magnitude diagrams can help.

The field of exoplanet physics is in its infancy when compared to the field of stellar physics, and this latter hit a turning point with the plotting of the first Hertzsprung-Russell diagram \citep{1911POPot..63.....H, 1914PA.....22..275R}. The H-R diagram was crucial as it allowed astronomers to statistically characterise a single object by placing it in the context of a well-studied population. In this way, sub-populations could be identified as well as their formation and evolution mechanisms \citep{1920SciMo..11..297E}, allowing future observational strategies to be shaped.

We are now approaching a turning point in the field of exoplanetary observations: the launch of the James Webb Space Telescope ({\it \textit{\textit{JWST}}}), which is scheduled for March of 2021, will allow more detailed characterisation of planetary atmospheres than ever before (See \cite{2019arXiv190403190M}, Figure 10). As such, it is vital that we use the data already available to select the very best targets for further investigation. Looking further into the future, \textcolor{Black}{{\it Ariel}} is to be launched in 2028, with the goal of observing approximately 1000 planets to compile the planetary equivalent of an H-R Diagram \citep{2018ExA....46..135T}. While a Bolometric Luminosity vs. Spectral Type H-R diagram is not yet achievable for planets, mid- and near-infrared colour-magnitude and colour-colour diagrams are already used in the field to help characterise atmospheres \citep{2015MNRAS.454.3002Z, 2015ApJ...810..118K, 2015MNRAS.450.2279T,2018haex.bookE..40A, 2019PASP..131a3001D}. Additionally, plots similar to colour-magnitude diagrams have been produced, for instance Figure 10 in \cite{2019ApJ...873...32M}, but rely on some model-independent parameters (such as temperature). Similar diagrams can also be made using transmission spectra (see Extended Data Figures 1 \& 2 in \cite{2016Natur.529...59S}).

Direct imaging of exoplanets yields a straightforward measurement of the planet's brightness; for planets observed in this way, the use of a colour-magnitude diagram to compare it with objects of similar brightness is intuitive \citep[e.g.][]{2007ApJ...657.1064M, 2008Sci...322.1348M, 2014ApJ...783..112B, 2016PASP..128j2001B}. The use of colour-magnitude diagrams for \textit{transiting} exoplanets was first produced in \cite{2014MNRAS.439L..61T}, and later expanded on by \cite{2014MNRAS.444..711T}.  Colour-magnitude diagrams serve a similar purpose to H-R diagrams in that they allow for planets to be compared to a larger population. In the most recent paper in this series, distance measurements were photometrically estimated due to the lack of availability of parallaxes for most systems. 

Colour-magnitude diagrams presented in \cite{2014MNRAS.444..711T} showed that in general the planets are compatible in magnitude with the M and L sequence of dwarfs, although there is more diversity in colour shown by the exoplanets, most of which were hot Jupiters. In some bands, planets appeared to be equally compatible with the blackbody sequence as they sat at the intersection of the two. \cite{2019AJ....157..101M} used a near-infrared colour-magnitude diagram to show that in \textit{J} and \textit{H} bands brown dwarfs are good spectral matches for hot Jupiters.

In the present work we expand significantly on the number of planets plotted. We also include new photometric bands and use alternative populations for comparison. Additionally, we have developed a publicly available Python toolkit which produces colour-magnitude diagrams, in a variety of formats and in any combination of photometric bands. We also show how a colour-magnitude diagram can be used to make an initial diagnostic, in order to select stand-out objects for rapid follow-up with upcoming missions.

The structure of our paper is as follows: in section \ref{sec:methods} we outline how we compiled our data sets, as well as how this data is processed in our Python tools to produce colour-magnitude diagrams. We then present a selection of new colour-magnitude diagrams plotted with our tools, and describe the results we infer from the positions of planets. Finally, we conclude and discuss the uses of colour-magnitude diagrams in the context of the next generation of telescopes.

\section{Methods}
\label{sec:methods}

In these sections, we describe the methods we have used to process spectra and secondary eclipse data found in the literature. We also outline the functionality of our Python modules, which we are releasing alongside this paper in order to facilitate similar data handling by other astronomers.

We first describe how we have assembled our data-set and the data contained therein. We then explain how we processed spectra to produce our comparison samples in Section \ref{sec:Comparisons} along with our motivations for each choice. Finally, in section \ref{sec:python} we outline the functionality of the three modules which make up our Python toolkit.

\subsection{Database of transiting exoplanet emission measurements}
\label{sec:data}

Our starting point was the data set compiled by \cite{2014MNRAS.444..711T}. Since 2014, a handful of these measurements have been updated; additionally, there have been many secondary eclipses measured for the first time. \cite{2018haex.bookE..40A} provided a helpful list of planets with secondary eclipse measurements, together with the bands in which the data are available. \cite{2019arXiv190107040G} published secondary eclipses for 36 planets in \textit{Spitzer}'s Channels 1 and 2, 27 of which had been measured for the first time. We also made use of the NASA Exoplanet Archive\footnote{\url{https://exoplanetarchive.ipac.caltech.edu/cgi-bin/TblView/nph-tblView?app=ExoTbls\&config=emissionspec}} which provides secondary eclipse data in all bands, and we continuously searched the ADS and Arxiv for new publications containing planetary emissions. All of these resources allowed us to assemble an up-to-date database of fluxes measured at occultation for a sample of 83 exoplanets. 

Once our planet sample was assembled, we searched the 2MASS catalogue \citep{2003yCat.2246....0C} for host star apparent magnitudes in \textit{J}, \textit{H} and \textit{K}-bands. As \textit{Spitzer}'s IRAC instrument reached the end of its cryogenic lifetime before 2014, there have been no new measurements in the 5.8$\mu$m or 8$\mu$m channels. In order to obtain host star apparent magnitudes in the 3.6$\mu$m and 4.5$\mu$m channels, we made use of the \textit{WISE} All-Sky catalogue \citep{2012yCat.2311....0C} as WISE's channels W1 and W2 are very similar to \textit{Spitzer}'s Channels 1 and 2. \citep{2014MNRAS.444..711T}. Where a host star's apparent magnitude was not available in a certain band, we derived synthetic photometry making use of standard spectra from the Pickles Atlas\footnote{\url{http://www.stsci.edu/hst/instrumentation/reference-data-for-calibration-and-tools/astronomical-catalogs/pickles-atlas}}. For a detailed description of this process, see Appendix \ref{sec:pickles-ap}.

To compute absolute magnitudes, we need distance with {\it Gaia}'s DR2 providing the most recent parallaxes \citep{2016AAA...595A...1G, 2018AAA...616A...1G}. The distances could not be determined simply by inverting the parallaxes published in DR2 due to the non-linearity of the process of parallax estimation by {\it Gaia}; instead, we used the distances calculated by \cite{2018AJ....156...58B}.

Finally the planetary radii were retrieved from \url{exoplanet.eu} \citep{2011epsc.conf....3S} where available. Our compilation of planetary secondary eclipse measurements can be found in Appendix \ref{sec:database}.

\subsection{Transforming planetary flux into magnitudes}
\label{sec:pl_data}

In order to add planets to a colour-magnitude diagram, we convert the fluxes measured at occultation to apparent magnitudes using the usual relation
\begin{equation}
\label{eq: mags}
\centering
    m_p = -2.5 \times \log \left(\frac{F_p}{F_{\star}}\right) + m_{\star}, 
\end{equation}
where \textit{m$_p$} is the apparent magnitude of the planet, \textit{m$_{\star}$} is the apparent magnitude of the parent star in the same band, and \textit{\(F_p/F_{\star}\)} is the planet-to-star flux ratio measured during the secondary eclipse event \citep{2010exop.book...55W}. These are then converted to absolute magnitudes using  astrometric distances.

\begin{figure*}
	\centering
	\includegraphics[width=0.8\textwidth]{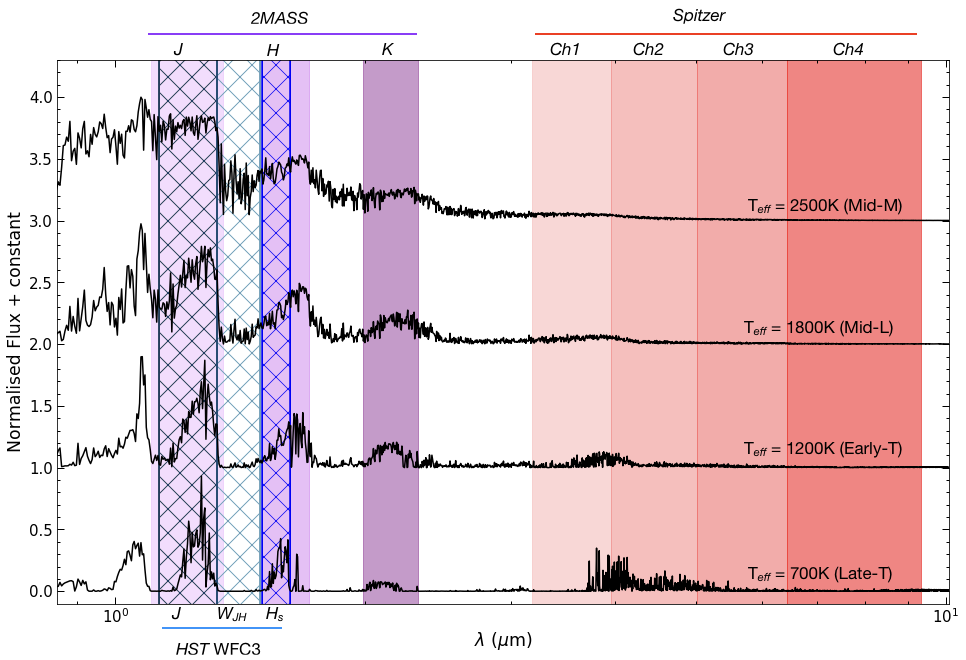}
    \caption{Model brown dwarf spectra, with photometric bands highlighted. In shades of red we have highlighted the position {\it Spitzer's} channels 1--4. The three bands in shades of lilac are 2MASS bands {\it J, H,} and {\it K}; the HST G141 grism we used are in blue hatching, \textcolor{Black}{with each band we made from it hatched in a different shade of blue.}}
    \label{fig:BDmodel}
\end{figure*}

\begin{figure*}
	\centering
	\includegraphics[width=0.8\textwidth]{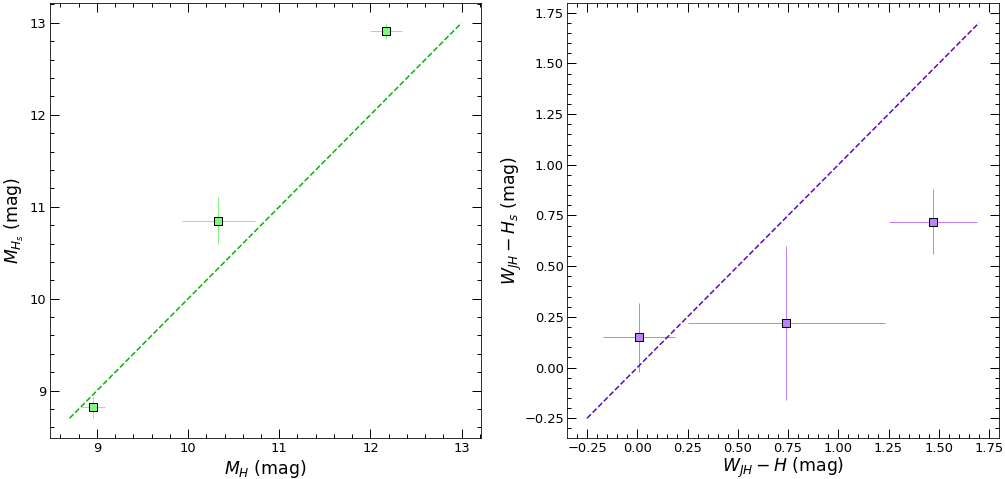}
    \caption{\textcolor{Black}{On the left panel we have show how {\it H}-short ({\it H$_s$}; obtained from WFC3 spectra) and {\it H} bands measurements cannot be assumed to be equivalent. We plot {\it H$_s$} absolute magnitudes vs. {\it H} absolute magnitudes for the three planets where these measurements exist (from left: WASP-12b, HAT-P-32b, WASP-43b). On the right panel we show ${\it W_{JH}}$ - {\it H$_s$} colours vs ${\it W_{JH}}$ - {\it H} colours for the same three planets. In both panels we show the position of the 1-1 line for ease of comparison.}}
    \label{fig:H_short}
\end{figure*}

In addition, we integrate low resolution emission spectra measured with the G141 grism on the Wide Field Camera 3 instrument (WFC3) on board the {\it Hubble} Space Telescope to produce additional planetary photometry. This particular instrument covers the wavelength range 1.1--1.7$\mu$m which overlaps with the majority of the \textit{J}-band. By cutting this grism between 1.130 and 1.325$\mu$m and integrating the planetary flux, we were able to compute \textit{J}-band photometry for eleven planets. The WFC3 grism extends into the {\it H}-band as well, but cuts short. We integrated the WFC3 spectra to create an {\it H}-short band ({\it H$_s$} thereafter) \textcolor{Black}{between 1.504 and 1.624$\mu$m}, as was done in \cite{2019AJ....157..101M} and \cite{2020arXiv200407431M}, \textcolor{Black}{and we include this photometry in our database. However, the difference in magnitudes between the {\it H} and {\it H$_s$} is significant, and contrary to \cite{2020arXiv200407431M}, we cannot assume {\it H} and {\it H$_s$} to be the same. The locations of the {\it Hubble} bands can be seen in Figure \ref{fig:BDmodel}. We illustrate the difference between the {\it H} and {\it H$_s$} bands in Figure \ref{fig:H_short}, where we have plotted {\it H$_s$} vs {\it H} magnitudes and colours for three planets for which we have {\it H}-band photometry and HST WFG3 low resolution spectra.} 

The WFC3 instrument is most often used to search for signs of water in emission or transmission spectra due to a key water absorption feature at 1.4$\mu$m \citep{2005ARAAA..43..195K}. In order to test whether we could diagnose the presence or absence of water using a colour-magnitude diagram, we create a photometric band centred on the water feature between the \textit{J} and \textit{H} bands (\textit{W$_{JH}$}-band hereafter) defined by integrating between 1.325 and 1.495$\mu$m. We therefore also add \textit{W$_{JH}$}-band photometry for eleven planets to our database. 

\subsection{Assembling a brown dwarf comparisons sample}
\label{sec:Comparisons}

The beauty of a colour magnitude diagram is to enable simple comparison between population samples in a given wavelength space. A handful of objects on a colour-magnitude diagram by themselves do not allow us to infer much about these objects. Therefore, it is crucial that we have a large sample of well-studied objects to compare with our planets. As was done in \cite{2014MNRAS.444..711T}, we make use of the detailed catalogue of near- and mid-infrared photometry of brown dwarfs produced by \cite{2012ApJS..201...19D} to populate the background of our diagrams. Brown dwarfs are an excellent comparison sample as they overlap with exoplanets in temperature and radius, which leads to comparable luminosities \citep{2014MNRAS.439L..61T}. For non standard bands, and for photometric bands that we defined, such as $W_{JH}$, there are no brown dwarfs catalogs we could use. We therefore synthetically create brown dwarf magnitudes and colours by integrating their spectra, which helps us populate the diagram and provide a comparison sample. 

The SpeX Prism Library\footnote{\url{http://svo2.cab.inta-csic.es/vocats/v2/spex/index.php}} provides normalised near-infrared spectra of brown dwarfs spanning the wavelength range 0.8--2.5$\mu$m. These data are collected from the ground but are corrected for telluric absorption caused by water in the atmosphere \citep{2003PASP..115..362R}.

We integrated the SpeX spectra to produce a catalogue of photometry of 119 brown dwarfs. These were cross-referenced with the parallaxes provided by \cite{2012ApJS..201...19D} in order to include distances in our catalogue. Our process of producing synthetic photometry is explained in Appendix \ref{sec:spex-ap}, along with how we validated our method. In Appendix \ref{sec:bd_photometry} we have included a table of the photometry we computed in several near- and mid-infrared bands. 




It is also important to verify whether exoplanet atmospheric models match observations. We produce synthetic photometry from atmospheric model spectra. 
In this paper we chose to use the publicly available model spectra produced by \cite{2015ApJ...813...47M} as they cover a wide parameter space, most importantly carbon-oxygen ratios of 0.35 to 1.40; effective temperatures of 1000 to 2500K in 250K increments; and five metallicity values, ranging from -0.5 to 2.0.  While we chose the Mollière models to demonstrate our code, it can adapted to use others as well. The data were processed using an adapted version of the code we used to produce magnitudes from Spex data.

\subsection{Description of our Python Toolkit}
\label{sec:python}

We have produced a selection of Python tools which automate all of the data analysis methods described above. Data handling is packaged into three modules: \textsc{Synth.py} to produce synthetic photometry of brown dwarfs from SpeX spectra; \textsc{Models.py} to produce synthetic photometry from model exoplanetary spectra; and \textsc{ExoCMD.py}: a plotting module which computes planetary magnitudes and plots colour-magnitude diagrams. Users can interacts with the modules via a Jupyter Notebook; the code and the notebook can be accessed at \url{https://github.com/gdransfield/ExoCMD}.

Below we give a brief outline of the modules; more detailed information can be found in Appendix \ref{sec:toolkit}, which also include a walkthrough tutorial.

\subsubsection{\textsc{Synth.py}}

This module provides a user with the flexibility to define bespoke photometric bands, in much the same way as we created our  $W_{JH}$-band. New bands should be designed to coincide with interesting absorption features, which can be selected by inspection of brown dwarf spectra, or from line lists \citep[e.g.][]{2018MNRAS.480.2597P}. Figure \ref{fig:BDmodel} shows model brown dwarf emission spectra \citep{2003AAA...402..701B, 2001ApJ...556..357A} on which we have highlighted the 2MASS \textit{J}, \textit{H}, and \textit{K} bands in lilac, along with \textit{Spitzer}'s mid-infrared bands in shades of red. The hatched area corresponds to the G141 grism we used, with the light blue hatched area indicating the position of our \textit{$W_{JH}$}-band. This plot shows how the 1.4$\mu$m water absorption widens and deepens with decreasing temperature.


It is important to bear in mind that a new band might not necessarily be useful for the full temperature range of brown dwarfs: the spectra of cooler objects is likely to be dominated by molecular species while hotter objects could have molecular, atomic or even ionised absorbers present. These changes are evident in the spectra shown in Figure \ref{fig:BDmodel}, as well as the changing width of absorption features. Cooler objects are likely to need wider photometric bands to detect molecular features, whereas narrower bands are increasingly useful for the narrow absorption features seen in objects with higher temperatures. 

Our \textsc{Synth.py} module contains seven built-in photometric bands, and all synthetic photometry will be produced in these bands along with a user-defined band. As well as the three {\it 2MASS} bands and our \textit{W$_{JH}$}-band, we have included {\it HAWK-I}'s two narrow bands ({\it NB1090} and {\it NB2190}) and {\it Sloan}'s {\it z'}-band. 


While this module has been written with SpeX spectra in mind specifically, it can easily be adapted to work with any other brown dwarf spectra. The function outputs either a text file or a spreadsheet with photometry in the desired bands, along with spectral types and astrometric distances.

\subsubsection{\textsc{Models.py}}

The \textsc{Models.py} module computes photometry from Molli\`ere's model spectra. Although the code has been written with this particular set of models in mind, it can be used for any model spectra that are produced in physical units. The functions make use of the map provided by \cite{2015ApJ...813...47M} to search for the spectrum which matches with the chosen parameters. The inputs required are constraints on metallicity, surface gravity, C/O ratio, effective temperature and host star spectral type. These constraints can be single values or lists of values; it is also possible to leave a parameter open which will result in all possible values being computed for that parameter. The two functions within the module output photometric magnitudes or colours respectively. There are eleven near- and mid-infrared bands built-in which can be called by name, and once again users can define bespoke bands if required.

\subsubsection{\textsc{ExoCMD.py}}

This module reads the planet database we have assembled and computes colours and magnitudes of exoplanets. There are five plotting functions which use these data to produce colour-magnitude diagrams. 

The first plotting function (\textsc{ExoCMD$\_$1}) produces diagrams in the style of those presented in \cite{2014MNRAS.444..711T}. We have added a keyword argument to this and all other plotting functions (\textsc{adjusted}) which when called will adjust the absolute magnitudes of the exoplanets to a size of 0.9R$_J$. This is to allow for a better comparison between planets, and to brown dwarfs. We chose this particular radius since typically brown dwarfs have a radius of $\approx$0.9R$_J$ \citep{2005ARAAA..43..195K} while hot Jupiters are more diverse in size. This only corrects the measurement with a simple translation up or down in absolute magnitude. \textcolor{Black}{See appendix \ref{sec:scaling} and Fig.~\ref{fig:Scaling} for an illustration of the effect of different adjustment factors. }

The second and third plotting functions (\textsc{ExoCMD$\_$2} and \textsc{ExoCMD$\_$3}) both show a polynomial to represent the mean trend of brown dwarfs in order to clarify and de-clutter the diagrams; this is especially valuable in colours where we now have many planets plotted. The polynomials are positioned using coefficients computed by \cite{2012ApJS..201...19D}. The key difference between the second and third plotting functions is a keyword argument (\textsc{highlight}) present in \textsc{ExoCMD$\_$3} which greys out all planets except those called by name. This allows objects of particular interest to be highlighted when needed.

The remaining two plotting functions make use of our new comparison samples. They call the functions from \textsc{Synth.py} and \textsc{Models.py} in order to compute the necessary photometry for the bands requested. When using the model plotting function (\textsc{ExoCMD$\_$model}), the model atmospheres can be coloured according to any of the five model parameters (C/O ratio, metallicity, surface gravity, effective temperature, or host star spectra type); in the synthetic brown dwarf function (\textsc{ExoCMD$\_$synth}) the ultra-cool dwarfs are coloured according to spectral type.

All plotting functions additionally include the ability to plot the position of a blackbody of comparable radius to the objects plotted on the diagram. 

\section{Results}
\label{sec:results}

In the sections that follow, we present a selection of our colour-magnitude diagrams along with some of the key inferences we have been able to make from them. These are designed to be illustrative of how powerful it can be to view results for individual planets in context.

We begin with a brief explanation of how to read a colour-magnitude diagram, along with some of the terminology to expect. In Section \ref{sec:outliers}, we show how a colour-magnitude diagram can allow us to select stand-out objects for rapid follow-up, and Section \ref{sec:molecules} outlines how inconsistency between colours of planets and brown dwarfs led us to investigate the absence of phosphine in irradiated objects. In Section \ref{sec:COratio} we demonstrate how a colour-magnitude diagram can be used to get a quick constraint on the C/O ratio.

\subsection{Notes on Terminology}

The x-axis of a colour-magnitude diagram is a colour index, calculated as the difference in magnitude between two photometric bands. It is conventional to subtract a longer wavelength magnitude from a shorter wavelength magnitude; this convention is observed throughout our paper. 

In a conventional colour-magnitude diagram objects can therefore be compared in terms of their x-position on the plot: an object on the left hand side would be described as `bluer' than one on the right hand side. This is due it having more flux, and therefore a higher magnitude, in the shorter, bluer wavelength than in the longer, redder wavelength. The converse is true of `redder' objects. When describing the spread of objects on our colour-magnitude diagrams, we will therefore use the terms `bluer' and `redder' to refer to placements on the left and right hand sides respectively.

\subsection{Outliers and Emerging Sub-populations}
\label{sec:outliers}

In this subsection, we go through a few examples on how colour-magnitude diagrams can be used to select target for additional observations, asking questions about patterns in the data, and diagnosing molecular signatures.

\subsubsection{Identifying oddball systems and measurements}

\begin{figure}
	\centering
	\includegraphics[width=\columnwidth]{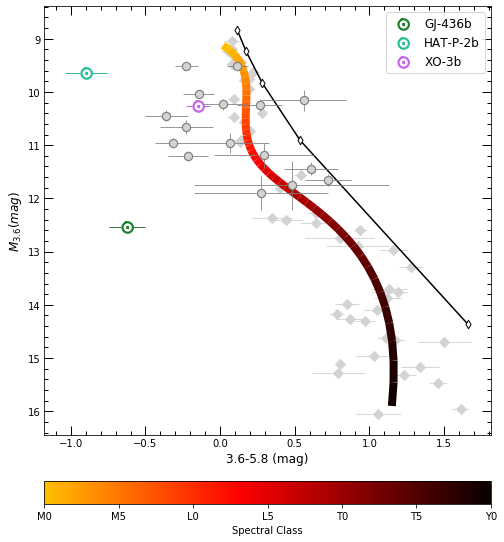}
    \caption{Colour-magnitude diagram in $M_{3.6}$ vs. [3.6 - 5.8$\mu$m] using our function \textsc{ExoCMD\_3}. \textcolor{Black}{Planets are plotted as circles in the foreground, while brown dwarfs are grey diamonds in the background. Three planets have been plotted in colours in order to highlight their positions when compared with those in grey. The polynomial shows the mean position of the brown dwarf sequence and is coloured according to their spectral type.} Planetary magnitudes have been scaled to a 0.9R$_J$ object for better comparison with the brown dwarfs. The black line shows the position of a 0.9R$_J$ blackbody with the white-filled diamonds showing the position of the blackbody at temperatures of 750K, 1750K, 2750K, 3750K and 4750K.  }
    \label{fig:out1}
\end{figure}

Figure \ref{fig:out1} is a colour magnitude diagram made with our plotting function \textsc{ExoCMD$\_$3}. As we are comparing planets and brown dwarfs we have used the key word argument \textsc{adjusted} to scale the magnitudes to the size of a typical brown dwarf. On this plot we have highlighted two objects which are clear outliers, and one which is not. 

On the top left we have HAT-P-2b, a highly eccentric planet \citep[e $\approx$ 0.5;][]{2013ApJ...766...95L}) that is speculated to have a dayside temperature inversion. In [3.6 - 4.5$\mu$m] it is consistent with both the L-Dwarf sequence and the mean position of other planets, which indicates that the 5.8$\mu$m flux is the one causing its very blue colour. The surprisingly shallow secondary eclipse in \textit{Spitzer}'s Channel 3 was noted at the time of measuring, as it yields a brightness temperature $\sim$700K lower than the secondary eclipses measured in Channels 1, 2 and 4. If this low flux in the 5.8$\mu$m band is caused by processes unique to eccentric hot Jupiters, then it is possible that further eccentric systems will be similarly blue in this colour.

Further down on the same plot, we have GJ 436b. This object is also eccentric \citep[e $\approx$ 0.14;][]{2014AcA....64..323M}) yet in this case the problematic flux is in the 3.6$\mu$m band. This is confirmed by its position on a M$_{5.8}$ vs [5.8 - 8.0$\mu$m] where it intersects exactly with the brown dwarf sequence. 

GJ 436b is a hot Neptune with an equilibrium temperature of $\approx$700K \citep{2016MNRAS.459..789T}, while HAT-P-2b is a hot Jupiter with an equilibrium temperature of 1540K \citep{2010MNRAS.401.2665P}. Following up on both of these objects will allow us to determine whether their blue colours in [3.6 - 5.8$\mu$m] are in any way caused by their eccentricity, and if so it could point to key population differences between Jupiter and Neptune-sized objects.

We have highlighted one other planet on Figure \ref{fig:out1}: XO-3b is the only other planet with a significant eccentricity (e  significant $> 0.1$) on this plot \citep[e $\approx$ 0.28;][]{2010ApJ...711..111M}). The 3.6$\mu$m and 5.8$\mu$m fluxes for XO-3b have not been updated since 2010, and the values come from single eclipse measurements. When the 4.5$\mu$m flux was remeasured in 2014 by \cite{2014ApJ...794..134W} they calculated a deeper eclipse of 0.158$\%$ which differs by 2.1$\sigma$ from the original. This new eclipse depth was derived from 12 consecutive secondary eclipse events, and the mean variation between them of just 5$\%$ indicates no consequential orbit-to-orbit variation. XO-3b's location within the population of circular planets throws a doubt on our initial hypothesis, an example on how a colour-magnitude diagram can be used.

Both of the HAT-P-2b fluxes were also calculated from single secondary eclipse events. For GJ 436b, the 3.6$\mu$m flux has been remeasured since the first observations of its thermal emission, but the 5.8$\mu$m has not; this latter value also came from a single eclipse event. \cite{2014MNRAS.444.3632H} has claimed that these flux ratios measured from single events have low reproducibility and underestimated errors as they do not adequately account for instrument systematics. This throws into question the significance of results inferred from single-eclipse photometry. 

A further stand-out system can be seen in Figure \ref{fig:out2}: on the far left with a colour of -0.9 we find WASP-65b. WASP-65b is one of the densest known hot Jupiters in its mass regime (R = 1.112R$_J$, M = 1.55M$_J$); it orbits in an area where inflated radii are the norm (a = 0.0334AU) yet it is denser than Jupiter itself \citep{2013AAA...559A..36G}. It has been suggested that its uninflated radius could be evidence of the advanced age of the system, and that if this is the case, the contraction of its atmosphere could lead to changes in its temperature-pressure profile giving rise to unexpected spectral features. The measurements for WASP-65b also result from observations of a single eclipse event \citep{2019arXiv190107040G}.

These four planets are clear candidates for follow-up. While at first look, the data so far is indicative that eccentricity and density might cause an important difference in atmospheric properties in two cases, it is however more likely that a lack of repeated measurement is the root cause. Regardless of what the answer turns out to be, using a colour-magnitude diagram simplifies the process of target selection.


\subsubsection{Planets near the T spectral class}

\begin{figure}
	\centering
	\includegraphics[width=\columnwidth]{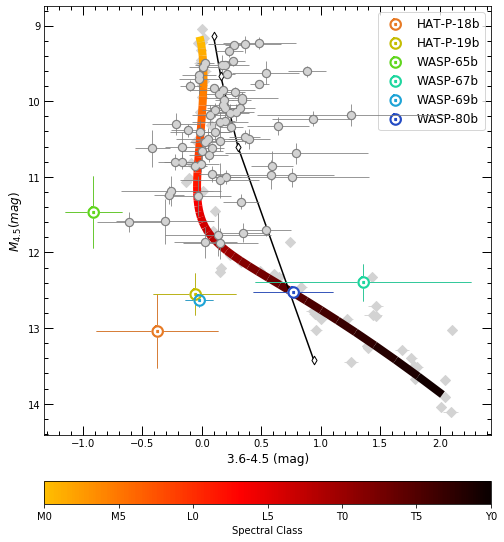}
    \caption{Colour-magnitude diagram in $M_{4.5}$ vs. [3.6 - 4.5$\mu$m] using our function \textsc{ExoCMD\_3}. \textcolor{Black}{Planets are plotted as circles in the foreground, while brown dwarfs are grey diamonds in the background. Six planets have been plotted in different colours in order to highlight their positions when compared with those in grey. As before, the polynomial shows the mean position of the brown dwarf sequence and is coloured according to their spectral type.} Planetary magnitudes have been scaled to a 0.9R$_J$ object for better comparison with the brown dwarfs. The black line shows the position of a 0.9R$_J$ blackbody with the white-filled diamonds showing the position of the blackbody at temperatures of 750K, 1750K, 2750K, 3750K and 4750K.}
    \label{fig:out2}
\end{figure}

In Figure \ref{fig:out2} we present a colour-magnitude diagram in M$_{3.6}$ vs. [3.6 - 4.5$\mu$m]. The five planets we have highlighted all have equilibrium temperatures of between 800-1000K, with HAT-P-18b and WASP-80b being the coolest of the set. \cite{2015MNRAS.450.2279T} pointed out that WASP-80b was the first planet whose measured dayside flux fell in a position consistent with the L-T transition experienced by ultra-cool dwarfs between 1100-1500K. This transition is characterised by the emerging spectral signature of methane, which has its fundamental band at 3.3$\mu$m and is therefore detectable by \textit{Spitzer}'s Channel 1. \cite{2015MNRAS.450.2279T} suggested that this could be indicative that planets undergo a similar transition but at a lower temperature. The fact that we now have fluxes measured for HAT-P-18b which is comparable in temperature and radius \citep{2019arXiv190800014W}, yet is significantly bluer, indicates that perhaps this is not true of all cool exoplanets. 

One way in which these two planets differ is in mass: WASP-80b is approximately three times more massive than HAT-P-18b (0.55M$_J$ vs. 0.183M$_J$). Additionally, we find that WASP-67b, whose colour is also consistent with that of an early T-dwarf, is more than twice as massive as HAT-P-18b \citep{2015ApJ...810..118K}. HAT-P-19b (0.292M$_J$) and WASP-69b (0.25M$_J$) fall between the others in both colour and mass. This is still too small a sample for a proper inference, however so far, there is an interesting indication in the transition from L to T class (CO to CH$_4$ chemistry) that increased mass might be correlated with a redder colour in [3.6 - 4.5$\mu$m]. 


This ties in well with the conclusions of \cite{2014ApJ...797...41Z}, who showed that the temperature of transition from CO-dominated to CH$_4$-dominated atmospheres scales with gravity. As the densest of the five, WASP-80b also has the highest surface gravity which would point to a higher temperature to undergo the planetary version of an L-T transition. 


An alternative interpretation for the range of colour that these planets cover might arise as differences in metallicity and C/O ratio. \cite{2015ApJ...810..118K} sought to find a link between mass, metallicity and C/O ratio for cool exoplanets, with HAT-P-19b and WASP-67b included in their sample. They found a tentative link between the masses of cool planets and the ratio of 3.6$\mu$m and 4.5$\mu$m magnitudes, which is consistent with the suggestion that less massive planets have higher metallicities \citep{2013ApJ...777...34M}. More recently, \cite{2019arXiv190800014W} found that for cool planets, extreme values of C/O ratio lead to big shifts in atmospheric chemistry, having large effects on the [3.6 - 4.5$\mu$m] colour. Our synthetic photometry of model atmospheres shows a very similar trend. 

\begin{figure}
	\includegraphics[width=\columnwidth]{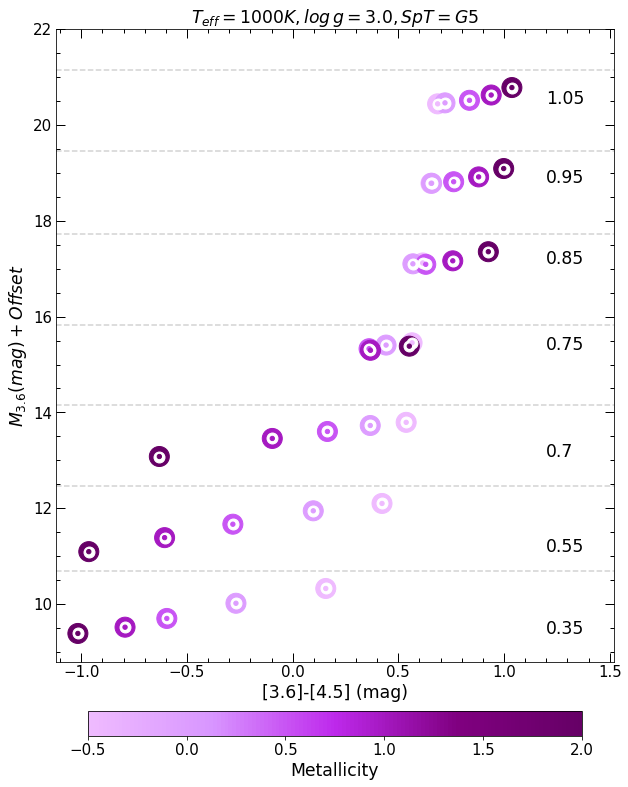}
    \caption{Colour-magnitude diagram of $M_{4.5}$ (plus an arbitrary offset) vs [3.6 - 4.5$\mu$m] using model spectra. The colours have been offset by -1 magnitudes (see Section \ref{sec:calibrations} for an explanation of the motivation). Points are coloured according to their assigned metallicity and each row of points represents model spectral with a different C/O ratio, as detailed on the right-hand-side of each row. }
    \label{fig:model36}
\end{figure}

Figure \ref{fig:model36} shows a colour-magnitude diagram created using \textsc{ExoCMD\_model()}. The Mollière spectra chosen correspond to a planet with T$_{\rm eff}$ = 1000K, log g = 3.0 and host star spectral type = G5. We found that this colour was not sensitive to host star spectral type, but showed some changes for values of log g $\geq$ 4. We can see that in this temperature regime, planets with a metallicity $>0$ (which we expect), experience a dramatic shift in colour with very small changes in oxygen abundance (between C/O = 0.7 and C/O = 0.85). 


This shows that in principle, we could diagnose limits on both C/O ratio and metallicity for exoplanets under 1000K simply by measuring thermal emission in these two bands, and without extensive retrieval methods. For example the only objects with colours of 0.25 or lower are oxygen-rich ones. We also see that the very bluest colours only occur with a combination of oxygen-richness and high metallicity. Colours close to 1 indicate both high metallicity and a C/O $\geq$ 0.85. 

At the time of writing this relationship is not yet calibrated. In Section \ref{sec:calibrations} we outline the problem of model spectra which are not fully calibrated to real data. To account for this, the colours in Figure \ref{fig:model36} have been offset by -1 magnitude. This offset is an approximation from inspection of the mean offsets in M$_{3.6}$ and M$_{4.5}$ as can be seen in Figure \ref{fig:offset}. \textcolor{Black}{We also note that the relationships outlined above are true for the Mollière model spectra used in this paper. An interesting next step would be to see if the same relationship holds for other model sets.}

\subsection{Identifying molecular Signatures}
\label{sec:molecules}

\begin{figure}
	\centering
	\includegraphics[width=\columnwidth]{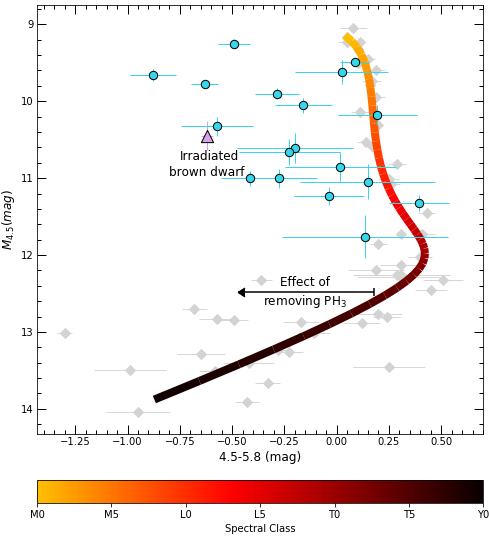}
    \caption{Colour-magnitude diagram showing the comparative blueness of planets with respect to brown dwarfs of similar brightness. \textcolor{Black}{Planets are plotted as blue circles in the foreground, while brown dwarfs are grey diamonds in the background. Once again, the polynomial shows the mean position of the brown dwarf sequence and is coloured according to their spectral type.} The planetary absolute magnitudes have been scaled to a size of 0.9R$_J$ for better comparison with brown dwarfs. \textcolor{Black}{Additionally, we have plotted the position of the irradiated brown dwarf WD0137-349B as a lilac triangle.} The black arrow indicates the effect on this colour of removing phosphine.}
    \label{fig:PH3}
\end{figure}

One interesting prospect for colour-magnitude diagram would be their ability to diagnose the presence of certain molecules, which would help setting up certain retrieval schemes.

A plot similar to Figure \ref{fig:PH3} appeared in \cite{2014MNRAS.444..711T}, who highlighted a large discrepancy between the colours of brown dwarfs and hot Jupiters. Even with our improved distance measurements and absolute magnitudes adjusted to 0.9R$_J$, we can clearly see that planets are systematically bluer than brown dwarfs. 
This is in contrast with most near- and mid-infrared colour-magnitude diagrams where the planets are largely consistent in both colour and magnitude with the L-dwarf sequence (see Appendix \ref{sec:updates} for up-to-date colour-magnitude diagrams).
In \cite{2014MNRAS.444..711T} we suggested that the discrepancy in [4.5 - 5.8$\mu$m] could perhaps be explained by an additional absorber within the 4.5$\mu$m band, present for the brown dwarfs but not for the exoplanets. We reached this conclusion because 5.8$\mu$m measurements of exoplanets appear consistent with brown dwarfs'.


An important difference between brown dwarfs and hot Jupiters is that while brown dwarfs are self-luminous, hot Jupiters are irradiated objects. We searched the literature to find out what irradiation could produce in relation to the 4.5 $\mu$m band. We identified phosphine, PH$_3$, as a molecule present within brown dwarf atmospheres, but most likely absent in hot Jupiters (due to photolysis),  as the cause of the discrepancy.


Phosphine has a strong absorption feature at approximately 4.3$\mu$m \citep{2007ApJS..168..140S} and is identified as the most likely Phosphorus-carrying gas in the \textcolor{Black}{observable} atmospheres of hot T-dwarfs and cool L-dwarfs, with temperatures in the range 1000K - 1400K \citep{2006ApJ...648.1181V}. \textcolor{Black}{\cite{2006ApJ...648.1181V} also notes that while in the atmospheres of warmer L-dwarfs PH$_3$ may be replaced by other phosphorus-bearing species, it may still be possible to detect the 4.3$\mu$m PH$_3$ feature {\it if} it can be distinguished from the 4.5$\mu$m CO feature.} However, PH$_3$ is highly susceptible to irradiation \citep{2019arXiv191005224S}, and \textcolor{Black}{ if present in the atmospheres of hot Jupiters, we expect it to be photodissociated and therefore not detectable. This would also then be true of other highly irradiated objects such as brown dwarf secondary to high-mass or high-temperature primaries.} 

We sought to verify our hypothesis by searching for an irradiated brown dwarf with an eclipse measurement in the bands that we considered. There is only one such object to our knowledge, WD0137-349B \citep{2015MNRAS.447.3218C}. This object is part of a white dwarf - brown dwarf binary and as such its dayside is subject to high levels of irradiation.
We plot WD0137-349B's irradiated side on Figure \ref{fig:PH3} \citep{2015MNRAS.447.3218C}.  Its position on the colour-magnitude diagram is more consistent with the most irradiated exoplanets rather than the ultra-cool dwarfs. We interpret this as indication that irradiation is likely the cause of a higher than usual flux in the 4.5 $\mu$m channel. Since phosphine does absorb in that particular band, and is expected to be within brown dwarfs' atmospheres, but not within hot Jupiter, we deduce that a lack of phosphine may provide a good explanation for the 4.5 $\mu$m measurements.


To further investigate whether phosphine can have the effect we thought, we used \textcolor{Black}{model spectra of GJ\,504b  produced with and without PH$_3$ present \citep{2017ApJ...850..150B}. We integrated these spectra and found that the removal of PH$_3$ from the atmosphere causes a blueward shift of 0.65 magnitudes in [4.5 - 5.8$\mu$m].} We have added an arrow of this size to Figure \ref{fig:PH3} to illustrate the impact of PH$_3$ in this colour, which has an amplitude consistent with the difference between brown dwarfs and hot Jupiters, and between the irradiated brown dwarf WD0137-349B and its field brethren.

One further interesting feature of Figure \ref{fig:PH3} is that the amplitude of the colour offset between brown dwarfs and hot Jupiters increases with decreasing absolute magnitude (i.e. increasing equilibrium temperature). Equilibrium temperature is obviously related with insolation. If the bluer colours of planets are caused by the photodissociation of PH$_3$, then higher levels of insolation would be expected to lead to higher PH$_3$ depletion.

Could it be something else? \cite{2011ApJ...729...41M} describe how \textit{Spitzer} fluxes, and therefore our colours, can be interpreted based on knowledge of the location of spectral features of the key absorbers present in an atmosphere. Most notably, they state that these interpretations are based on the assumption that H$_2$O, CH$_4$, CO and CO$_2$ are the four dominant molecules in all \textit{Spitzer} bands. Of these four, CO and CO$_2$ both have strong absorption features in the 4.5$\mu$m channel, so low fluxes in this band are usually attributed to one or both of these molecules. However, thermal equilibrium predicts that both brown dwarfs and hot Jupiters should have most of their atmospheric carbon locked into CO in this temperature regime. We therefore return to our `additional absorber' hypothesis.

An alternative explanation is to invoke thermal inversions in the vast majority of the hot Jupiters depicted in Figure \ref{fig:PH3}. With such an inversion, CO would be in emission and increase the flux in the 4.5$\mu$m band. However, as we discuss \ref{sec:upcoming}, \textcolor{Black}{planets with compelling evidence for a thermal inversion are scarce,} and their existence is doubted by several authors.

\textcolor{Black}{One other possibility worth considering is the effect of clouds. Clouds are believed to be the cause of the significant colour change seen in brown dwarfs close the L-T transition in near and mid-infrared colours \citep{2006ApJ...651..502P}, as the disappearance of silicate and iron condensates reveals atmospheric methane absorption \citep{2014MNRAS.444..711T}. It is unclear as yet whether similar colour changes in exoplanets could also be caused by clouds since very few planets corresponding to a T spectral class have been measured. In transmission spectroscopy, spectral features can present lower amplitudes than expected due to the presence clouds and haze opacities \citep{2015AAA...573A.122W}, which can lead to incorrect abundances being inferred. This is less the case when considering secondary eclipse data, as the daysides of planets as the spectrum probes at a higher altitude \citep{2018haex.bookE.104M}. Nevertheless, better understanding of the role of clouds and hazes in exoplanetary atmospheres is needed to rule this out as a contributing factor since difference in irradiation and gravity between brown dwarfs and hot Jupiters might produce differing cloud coverage, altitude, and dynamics. From visual inspection of Fig.~\ref{fig:nir}, and particularly of Fig.~\ref{fig:rising}, we note that irradiated hot Jupiters and field brown dwarfs do follow a similar behaviour in the near-IR, over the L-type range, where clouds are thought to dominate brown dwarf atmospheres.}

\subsection{Seeking to constrain the C/O ratio with colour-magnitude diagrams}
\label{sec:COratio}

\begin{figure*}
	\centering
	\includegraphics[width=\textwidth]{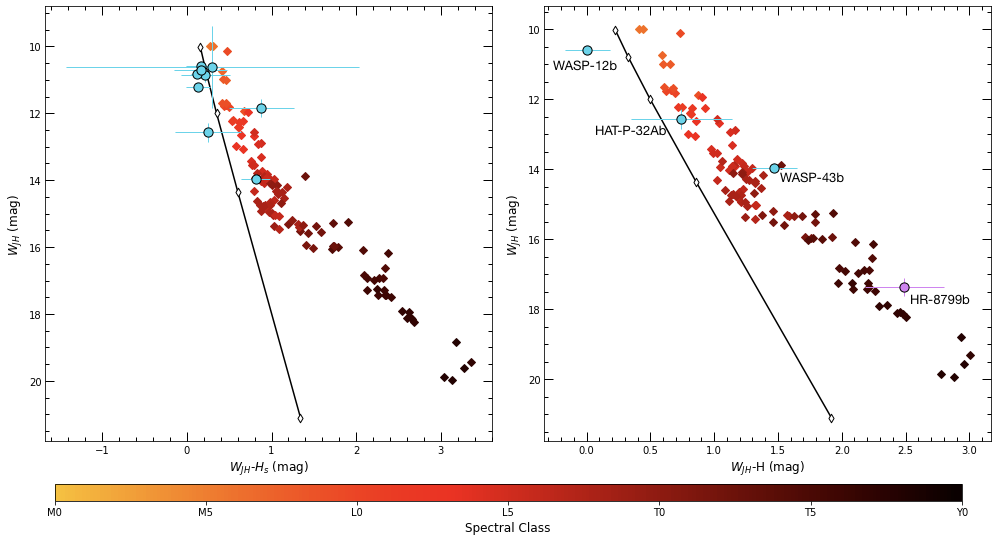}
    \caption{\textcolor{Black}{Colour-magnitude diagrams of $M_{W_{JH}}$ vs. [$W_{JH}-H$] made using our plotting function \textsc{ExoCMD\_synth}}. \textcolor{Black}{Planets are plotted as blue circles with their} magnitudes scaled to a 0.9R$_J$ sized object to allow better comparison with the brown dwarfs. \textcolor{Black}{Brown dwarfs are plotted in the background as diamonds coloured according to spectral type.} The black line shows the position of a 0.9R$_J$ blackbody, with the white-filled diamonds highlighting the position at temperatures of 1000-5000K in steps of 1000K. \textcolor{Black}{{\it Left panel:} $M_{W_{JH}}$ vs. [$W_{JH}$-$H_s$]. We note that in this colour planets are significantly less spread out in colour. {\it Right panel:} $M_{W_{JH}}$ vs. [$W_{JH}-H$]. HR8799b is highlighted in lilac as its photometry was taken with direct imaging rather than secondary eclipse observations \citep{2015ApJ...809L..33R, 2008Sci...322.1348M}.}}
    \label{fig:H2O}
\end{figure*}

\textcolor{Black}{In Figure \ref{fig:H2O} we present colour-magnitude diagrams featuring our new \textit{W$_{JH}$} band, made with our \textsc{ExoCMD$\_$synth()} function. On the left-hand panel we have plotted nine of the planets which have low resolution spectra measured with {\it HST}, combining {\it $W_{JH}$} photometry with {\it H$_s$)}. We include this as there are significantly more planets available than those with {\it H}-band photometry, but we see that their positions are shifted in colour with respect to the left panel. On the right-hand panel we have plotted three planets: from brightest they are WASP-12b, HAT-P-32Ab, WASP-43b. We computed the brown dwarf photometry for both panels with our \textsc{Synth.py} code, and we computed photometry for the three planets by integrating low resolution emission spectra measured with the Hubble Space Telescope's G141 grism. }

We have yet to identify a colour index where objects with confirmed water detections are easily distinguishable from those without, however we can see that three objects in \ref{fig:H2O} are widely spread in colour. Incidentally, all three of these planets have firm detections of water using these data. 

The \textit{H}-band is centered on 1.6$\mu$m and has a prominent CH$_4$ absorption feature, and a slightly weaker CO feature \citep{2007ApJS..168..140S}, while the \textit{W$_{JH}$}-band is dominated by water absorption. These four molecules are related by the following net equilibrium equation, as described in \cite{2012ApJ...758...36M}:
\begin{equation}
\label{eq: chem}
    \ce{CH4 + H2O <=>T[T$\ga$1000K][T$\la$1000K] CO + 3H2}.
\end{equation}

In objects cooler than 1000K the left hand side of the equation is favoured and methane is the dominant carbon-bearing molecule. For objects hotter than 1000K, carbon is found mainly in the form of carbon monoxide. However, if a hotter atmosphere is  also oxygen rich, we would expect the excess oxygen to react with the H$_2$ to form water. This indicates that C/O ratio should be the biggest indicator of both [\textit{W$_{JH}$ - H}] colour and water abundance: more excess oxygen will cause more water to be produced. This will deepen the absorption at 1.4$\mu$m leading to increased \textit{W$_{JH}$} magnitude, and therefore a redder colour. This is consistent with the retrieved water abundances for WASP-12b and WASP-43b by \cite{2014ApJ...783...70L}: WASP-43b's abundance was found to be greater than WASP-12b's by a factor of 10$^3$.

\begin{figure}
	\centering
	\includegraphics[width=\columnwidth]{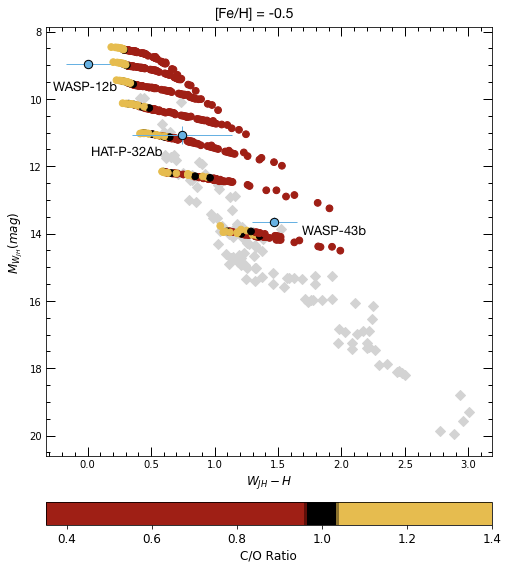}
    \caption{Colour-magnitude diagram of $M_{W_{JH}}$ vs. [$W_{JH}-H$], plotted using our function \textsc{ExoCMD\_model}. The following parameters were entered for the model: $T_{\rm eff}$ = 1000 -- 2500K, log g = 2.3, 3.0, 4.0 and 5.0, SpT = G5, and [Fe/H] = -0.5. Points have been coloured according to C/O ratio in order to highlight the differences between carbon- and oxygen-rich atmospheres. \textcolor{Black}{Brown dwarfs are plotted in the background as grey diamonds.}}
    \label{fig:COa}
\end{figure}

\begin{figure}
	\centering
	\includegraphics[width=\columnwidth]{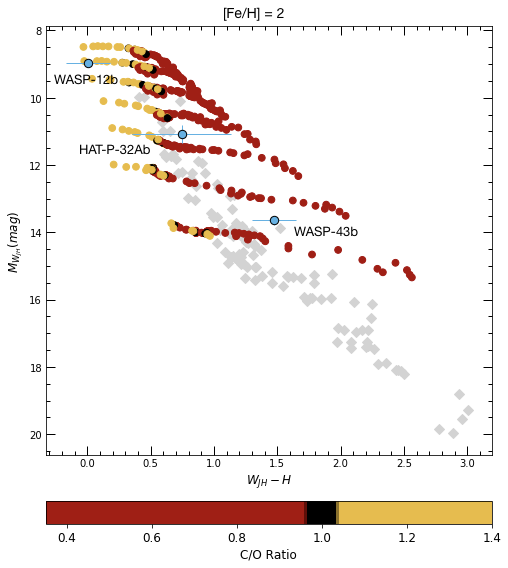}
    \caption{Colour-magnitude diagram of $M_{W_{JH}}$ vs. [$W_{JH}-H$], plotted using our function \textsc{ExoCMD\_model}. The following parameters were entered for the model: $T_{\rm eff}$ = 1000 -- 2500K, log g = 2.3, 3.0, 4.0 and 5.0, SpT = G5, and [Fe/H] = 2. As before, points have been coloured according to C/O ratio in order to highlight the differences between carbon- and oxygen-rich atmospheres. \textcolor{Black}{Brown dwarfs are plotted in the background as grey diamonds.}}
    \label{fig:COb}
\end{figure}

Figures \ref{fig:COa} and \ref{fig:COb} are colour-magnitude diagrams featuring Molli\`ere model atmospheres. Both have been plotted for four values of log g from 2.3 to 5.0, seven values of effective temperature from 1000--2500K, and a host star spectral type of G5. Figure \ref{fig:COa} features atmospheres with a metallicity of -0.5, while in Figure \ref{fig:COb} we have assigned a metallicity of 2.0. The models have been coloured according to their C/O ratio.

We can see from Figure \ref{fig:COa} just how much [\textit{W$_{JH}$ - H}] colour is affected by the C/O ratio, and for two of the planets here plotted we can attempt to infer whether they are consistent with oxygen- or carbon-rich model spectra.

Within its uncertainties, WASP-43b colours and magnitudes are consistent with an oxygen-rich atmosphere for all values of surface gravity, and retrievals of its metallicity have found it to be 0.3--1.7 $\times$ Solar (\cite{2014ApJ...793L..27K}, \cite{2017AJ....153...68S}). Therefore it is best matched by Figure \ref{fig:COb}, and we can see that its colour is indicative of an oxygen-rich atmosphere. This is in agreement with the upper limit set by \cite{2015arXiv150407655B}, and indeed with the recent retrieval by \cite{2019arXiv190903233I}. 

In contrast, WASP-12b's eclipse measurements coincide with the carbon-rich model atmospheres for both extremes of metallicity and all values of surface gravity. This is in agreement with \cite{2011Natur.469...64M} who found C/O $\geq$ 1 using \textit{Spitzer} secondary eclipse data. This has since been contested, with \cite{2015ApJ...814...66K} finding that the atmosphere was best fit by C/O $\approx$ 0.5 using HST transit data, but omitting \textit{Spitzer} transits due to instrument systematics. \cite{2015arXiv150407655B} also retrieved an oxygen-rich atmosphere with C/O < 0.9 for WASP-12b, even though analysis of previously unpublished \textit{Spitzer} measurements by \cite{2014ApJ...791...36S} had confirmed the original findings. We need a better understanding of the physical processes that lead to WASP-12b appearing consistent with carbon-rich model atmospheres; additionally, in order to confirm this consistency we need to ensure that the models are well calibrated to the data in this colour. See section \ref{sec:calibrations} for more detail.

We were unable to find a constraint on HAT-P-32Ab's C/O ratio in the literature, and due to the large errors on the colour we find that it is equally compatible with carbon-rich and oxygen-rich model atmospheres.

\section{Discussion}
\label{sec:discussion}

In the following sections we discuss the implications of our results, placing them in the context of unexplained low fluxes and exciting upcoming missions.

\subsection{Phosphine}
\label{sec:PH3}

The recently updated line list for phosphine published by \cite{2015MNRAS.446.2337S} puts us in a favourable position to identify planets where this gas may be present. Over the lifetime of \textit{Spitzer} there have been many unexplained low fluxes measured in Channel 2. In this section we speculate about the impact that phosphine may have for a number of eclipse measurements, and planetary environments.

The absence of PH$_3$ on the daysides of hot Jupiters due to high levels of irradiation would not preclude the possibility of its presence on the cooler and less irradiated nightsides. One such candidate is HD 189733b which has a puzzlingly low nightside flux in 4.5$\mu$m \citep{2012ApJ...754...22K, 2019ApJ...880...14S}. The lack of irradiation on the nightside may have prevented photodissociation of the molecule, and the lower temperatures would be indicative of PH$_3$ accounting for most of the atmospheric phosphorus budget \citep{2006ApJ...648.1181V}. Even if photodissociated on the dayside, Phosphorus might recombine into phosphine on the nightside, after being transported by winds.  

Alternative explanations were made for this low 4.5${\rm \mu}$m flux. For instance, Carbon monoxide (CO) also has a deep absorption feature in the 4.5$\mu$m band \citep{2007ApJS..168..140S}, and when the phase curve for HD 189733b was first observed in \textit{Spitzer}'s Channel 2, the low nightside flux was attributed to this molecule \citep{2012ApJ...754...22K}. It was initially thought that non-equilibrium chemistry would be able to explain the fact that CO was the main carbon-bearing molecule, despite the low temperature. However, recently \citep{2019ApJ...880...14S} showed that this is not the case: disequilibrium processes alone cannot account for the low fluxes in \textit{Spitzer}'s Channel 2 as excess CO is balanced by a drop in H$_2$O.

Two other hot Jupiters with similarly low 4.5$\mu$m nightside fluxes are HD 209458b \citep{2014ApJ...790...53Z} and WASP-43b \citep{2017AJ....153...68S}. 
Here too, models predict they should be significantly brighter than they are in \textit{Spitzer}'s Channel 2, as equilibrium chemistry would point to CH$_4$ being the main carbon-bearing molecule on the cooler nightside. The inclusion of PH$_3$ in these models could revise our understanding of these planets.

Considering PH$_3$ in atmospheric composition may additionally help to shed light on surprisingly shallow 4.5$\mu$m secondary eclipses measured on the daysides of far cooler planets. GJ 436b has had consecutive non-detections in \textit{Spitzer}'s Channel 2 \citep{2010Natur.464.1161S, 2014AAA...572A..73L, 2017AJ....153...86M}, pointing to a CO/CH$_4$ ratio which is considerably higher than equilibrium chemistry would predict for an object of this temperature ($\approx$700K \citep{2016MNRAS.459..789T}). `Additional absorbers' have been postulated for GJ 436b by \cite{2017AJ....153...86M} in order to resolve the apparent low flux in this wavelength, and PH$_3$ could be that absorber. 

A similar non-detection in the 4.5$\mu$m band for WASP-29b prompted claims of possible non-equilibrium abundances of CO \citep{2012DPS....4420009H}. WASP-29b is a Saturn-sized object with an equilibrium temperature of 980K; PH$_3$ could yet again provide an explanation for this excess absorption. Most recently, GJ 3470b had a minute 4.5$\mu$m flux measured by \cite{2019NatAs.tmp..361B}; this is a low metallicity, sub-Neptune sized planet with an equilibrium temperature of approximately 600K. Equilibrium chemistry once again points to Methane accounting for most of its atmospheric carbon budget, and phosphine as a convenient molecule to explain the observations.

\subsection{Upcoming Missions}
\label{sec:upcoming}

The {\it James Webb} Space Telescope (\textit{JWST}) is scheduled for launch in 2021 and is intended as a successor to the {\it Hubble} Space Telescope. A recent simulation of \textit{JWST} spectra by \cite{2017ApJ...850..199W} assessed the detectability of PH$_3$ by the telescope's NIRCam instrument and found that it \textcolor{Black}{could} be detectable in emission \textcolor{Black}{spectra} for objects of around 500K\textcolor{Black}{, and in transmission spectra for objects cooler than 1000K}. Although they concluded that for objects of 1000K or more PH$_3$ could not be resolved \textcolor{Black}{with a secondary transit}, we speculate that objects of intermediate temperatures, such as GJ 436b, might have detectable PH$_3$ due to their low 4.5$\mu$m fluxes. Additionally, the MIRI instrument will cover the wavelength range of \textit{Spitzer}'s 5.8$\mu$m band \citep{2015PASP..127..584R}; we can therefore use photometry taken in this instrument's Channel 1 to add objects highlighted in Figure \ref{fig:out2} to our PH$_3$ diagnostic colour-magnitude diagram. The extent to which these objects are offset from the brown dwarfs will indicate whether we should follow up on them using NIRCam to search for PH$_3$.  

\textcolor{Black}{While in this paper we propose that phosphine could be responsible for some of the differences in colour seen between hot Jupiters and brown dwarfs, this is not without its issues. Phosphine is hard to distinguish from CO in hotter objects, and in some of the coolest brown dwarfs, where PH$_3$ is expected to dominate the 4.5 $\mu$m band, the evidence for its presence is still scarce. {\it JWST} will therefore also allow us to explore the spectra of brown dwarfs in more detail in order to confirm the presence or absence of phosphine in these objects. The first M-band (4.5-5.1 $\mu$m) spectrum of WISE 0855, the coldest known brown dwarf, rather surprisingly did not reveal evidence of phosphine absorption; this despite being very similar to Jupiter in other respects \citep{2016ApJ...826L..17S}. This was interpreted as evidence that WISE 0855 might not share Jupiter's turbulent vertical mixing, which is responsible for the large abundances of PH$_3$ present in its observable atmosphere. \cite{2018ApJ...858...97M} later presented an L-band (3.4-4.14 $\mu$m) spectrum of WISE 0855; this too lacked the expected footprint of PH$_3$. This work was recently extended to six other brown dwarfs in the temperature range 250-750K, and phosphine absorption was still not clearly present in any of them \citep{2020arXiv200410770M}. The authors do point out that the wavelength ranges where phosphine would be most detectable (the end of the L-band and start of the M-band) coincide with the lowest signal-to-noise in their data. It is also worth noting that the centre of the phosphine absorption feature is at 4.3 $\mu$m, which falls precisely in the gap between the L and M-bands. Fortunately, {\it JWST}'s NIRCam will have both the spectral coverage and the resolution to shed new light on this mystery.}


\textit{JWST} will also be equipped with a Near Infrared Spectrograph (NIRSpec) \cite{2016AAA...592A.113D} with wavelength coverage of 0.6--5.3$\mu$m.  We will therefore be able to use our photometry tools to make diagnostics about the planets' atmospheres, using colour-magnitude diagrams by integrating \textit{JWST} spectra. While this does not replace a full atmospheric retrieval, the goal will be to identify objects that appear to be, for instance carbon-rich, or outlying the main population for one reason or other, and propose them for a more detailed follow-up as was simulated by \cite{2018AJ....156...40S}.

The C/O ratio is essential to our understanding of how and where a planet formed  \citep[e.g.][]{2011ApJ...743L..16O,  2011ApJ...743..191M, 2012ApJ...758...36M, 2017MNRAS.469.4102M}. It can also tell us whether or not a thermal inversion is likely, as carbon-rich atmospheres favour low abundances of the two molecules thought to be producing inversions \citep{2011ApJ...729...41M}: TiO and VO \citep{2008ApJ...678.1419F}. Additionally, the C/O ratio can give an indication about the habilitability of a planet \citep{2015DPS....4740403J}, as a C/O$\geq$1 causes depletion of water, even if the planet is within the habitable zone. 

The detailed spectra which \textit{JWST} will be capable of producing will also help to shed light on the atmospheres of controversial planets such as WASP-12b. This planet is predicted to have a stratosphere due to its very hot temperature \citep{2009ApJ...693.1920H}; however, so far TiO and VO have not been detected with any certainty as measurements of eclipse depths in relevant bands have been inconsistent \citep{2013MNRAS.436.2956S, 2019MNRAS.486.2397H}. Most recently, there has been close scrutiny of the orbit of WASP-12b due to the changing transit mid-points; the data now available points to orbital decay over apsidal precession \citep{2019MNRAS.482.1872B, 2020ApJ...888L...5Y}. There is also no consensus as yet on whether WASP-12b is carbon- or oxygen-rich: detections of water in transmission \citep{2015ApJ...814...66K} point to an oxygen-rich atmosphere, yet the shallow 4.5$\mu$m eclipse depth indicates CO in absorption rather than emission \citep{2014ApJ...791...36S}. 

If WASP-12b does have a thermal inversion caused by TiO, VO or another mechanism, then its position on Figures \ref{fig:COa} and \ref{fig:COb} might be misleading. \cite{2018AAA...617A.110P} explains how the dissociation of H$_2$O in the atmospheres of ultra-hot Jupiters with stratospheres leads to free hydrogen atoms capturing electrons. The resulting H$^-$ ions produce absorption features in the same spectral region as H$_2$O, which can lead to confusion when interpreting low-resolution spectra in the 1.4$\mu$m band. In the era of \textit{JWST}, we will be able to refine our interpretations of positions of ultra-hot Jupiters on our colour-magnitude diagrams in order to improve our use of them as diagnostic tools.

Looking further into the future, \textcolor{Black}{\textit{Ariel}} is planned for launch in 2028. Like \textit{JWST}, it will be equipped with near- and mid-infrared spectrographs which will allow for detailed atmospheric characterisation\footnote{\url{https://Arielspacemission.files.wordpress.com/2017/05/Ariel-ral-pl-dd-001_Ariel-payload-design-description_iss-2-01.pdf}}. In particular, the Near Infrared Spectrograph (NIRSpec) covers the wavelength ranges of the \textit{H} and \textit{W$_{JH}$} bands which will enable us to get an initial diagnostic on the C/O ratio, while the \textcolor{Black}{{\it Ariel} InfraRed Spectrometer (AIRS)} covers the 1.95--7.8$\mu$m range allowing us to choose targets to follow-up on to find PH$_3$.

\begin{figure}
	\centering
	\includegraphics[width=\columnwidth]{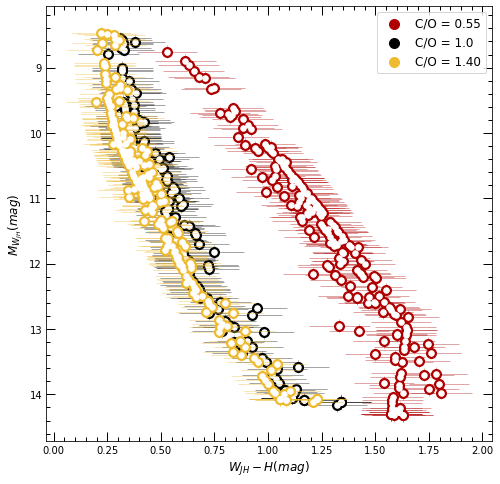}
    \caption{\textcolor{Black}{Exoplanet populations with different C/O ratios can be distinguished if they exist. Here, we show} simulations of the {\it Ariel} yield on a colour-magnitude diagram  of $M_{W_{JH}}$ vs. [$W_{JH} - H$], plotted using our function \textsc{ExoCMD\_model}. The closest matching model spectrum was selected for each planet \textcolor{Black}{on the target list,} and plotted with a C/O value of 0.55, 1.0 and 1.40 to show the spread in colour. Points are coloured according to C/O ratio.}
    \label{fig:ARIEL}
\end{figure}

\cite{2019AJ....157..242E} recently produced a list of potential targets for \textcolor{Black}{\textit{Ariel}} along with their radii and equilibrium temperatures. We searched the literature for their surface gravities, host star metallicities and host star spectral types; we were able to find all three data for 210 of the potential targets. We used this information to select the most appropriate model spectrum from \cite{2015ApJ...813...47M} and plotted them on a M$_{W_{JH}}$ vs [\textit{W$_{JH}$-H}] colour-magnitude diagram, assigning C/O ratio values of 0.55, 1 and 1.40. As the Molli\`ere models are only available in 250K increments, we interpolated the magnitudes in order to get a more realistic spread of colours. We did not have access to the \textcolor{Black}{{\it Ariel}} Radiometric Model \citep{2019EPSC...13..270M} to estimate the errors on the magnitudes; we therefore assigned an error of $\pm 0.1$ mag. This choice was not arbitrary; when computing photometry from HST/WFC3 spectra, the data with resolution comparable to what \textcolor{Black}{{\it Ariel}} will produce yielded signal-to-noise ratio of 10, equivalent to a tenth of a magnitude. We present the resulting plot in Figure \ref{fig:ARIEL}.

While the values for surface gravity, metallicity and host star spectral type had to be rounded to fit the model grid, we can see that objects with C/O ratio $\geq$1 are  distinguishable from oxygen-rich objects.

A recent paper by \cite{2020arXiv200407431M} also attempted to use colour-magnitude and colour-colour diagrams to constrain the atmospheric properties of exoplanets. In this paper, model spectra were computed using VSTAR (Versatile Software for Transfer of Atmospheric Radiation) \citep{2012MNRAS.419.1913B} for a range of values of metallicity, log g and C/O ratio. Model atmospheres were then compared with hot Jupiters in near- and mid-infrared colours using photometry in {\it $JHK_s$}-bands and {\it Spitzer's} channels 1 and 2. While planets were seen to cluster around solar values of C/O and metallicity, no definitive constraints were found as the error bars were too large. 

\subsection{Model Calibrations}
\label{sec:calibrations}

Our C/O ratio diagnostic tools rely on the availability and reliability of model spectra. We chose to use the models presented in \cite{2015ApJ...813...47M} as they covered a very wide parameter space and were publicly available. However, those models have limitations which impact the validity of our inferences. 

The first limitation we find is simply the range of temperatures available. \textit{JWST} and \textcolor{Black}{\textit{Ariel}} will both observe objects cooler than 1000K in more detail than ever before, and in order to use models to characterise these objects it is essential that we have model spectra in this temperature range.

The second limitation comes from the quality of the fit for different wavelengths. In their paper, \cite{2015ApJ...813...47M} showed a comparison between their retrieved model spectrum for HD-179833b and the many thermal emission measurements available for this planet. They showed that \textit{Spitzer} photometry at 8$\mu$m was well fit by their model, while the shorter wavelength channels were not. They also showed that while the pattern for the HST data was well fit, the measured eclipse depths were larger than those retrieved by the model.

\begin{figure*}
    \centering
    \includegraphics[width=0.8\textwidth]{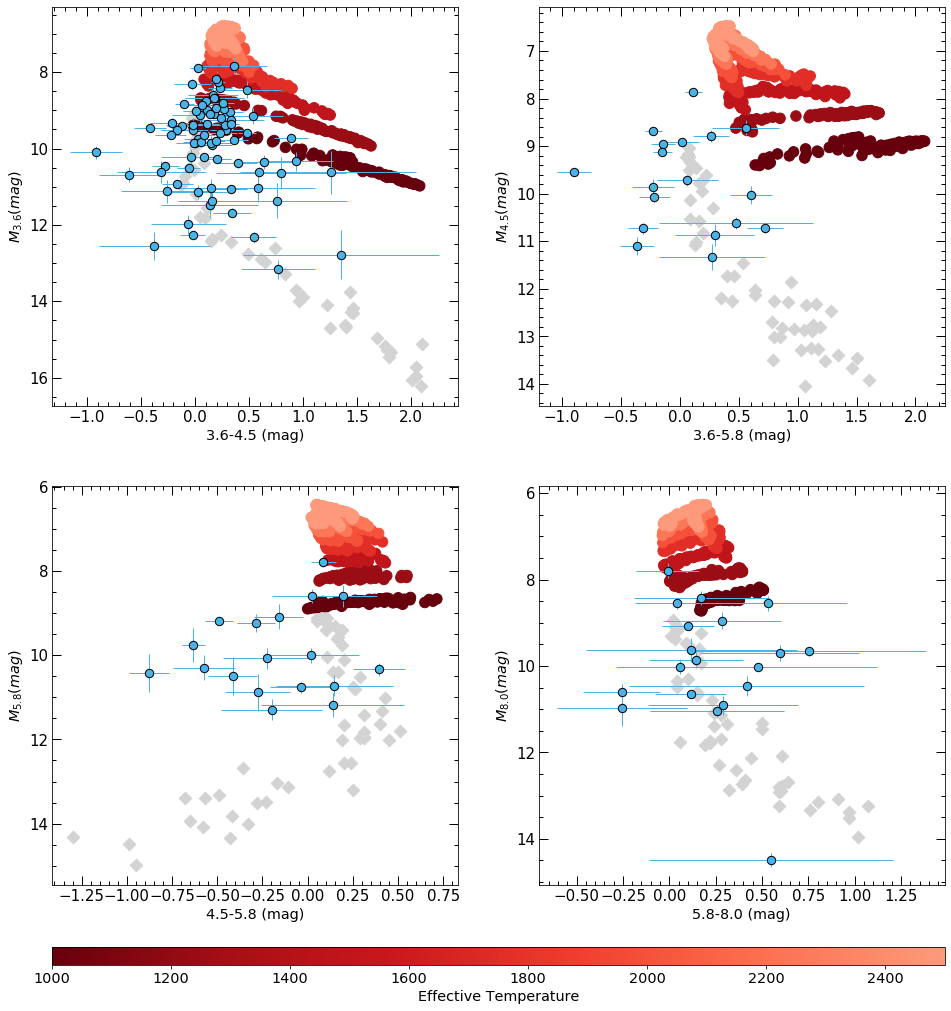}
    \caption{Colour-magnitude diagrams in {\it Spitzer} channels 1--4 \textcolor{Black}{ showing how models do not reproduce planetary properties well. Planets are plotted as blue circles in the foreground, while brown dwarfs are grey diamonds in the background.} Planetary magnitudes are not adjusted as we want to show how closely they match with model atmospheres. \textcolor{Black}{Model planetary magnitudes are plotted as edgeless circles, coloured according to effective temperature.}}
    \label{fig:offset}
\end{figure*}

We illustrate the disparity between the measured data and the model spectra in IRAC channels in Figure \ref{fig:offset}. It is clear from these plots that there is a systematic offset in all four \textit{Spitzer} bands when we compare with a large enough sample. But this also demonstrates that colour-magnitude diagrams, and indeed our tools to produce them, are especially valuable to modellers to validate their model spectra. It is also a cautionary note since similar spectra used for retrieval would be unlikely to lead to correct abundances.

The completeness of the chemistry included into the models impacts the outcome of retrievals. In particular, \cite{2017ApJ...850..150B} showed how an incomplete line list for PH$_3$, let alone its exclusion from models altogether, can lead to this crucial molecule not being detected at all. We would also need oxides of titanium and vanadium included in radiative transfer calculations in order to ensure that possible thermal inversions are explored.

\section{Conclusions}
\label{sec:conclusions}

In this paper we have presented a public Python toolkit for the plotting of near- and mid-infrared colour-magnitude diagrams. To demonstrate the functionality of the toolkit, and the usefulness of colour-magnitudes of transiting exoplanets, we have presented a selection of our newly plotted diagrams. From these we have identified some trends:

\begin{itemize}
    \item Two objects (HAT-P-2b and GJ 436b) are very blue in [3.6$\mu$m - 5.8$\mu$m] colour. Despite the fact that for each of these planets the blue colour is caused by excess emission in different bands, what they have in common is high eccentricity. Further follow-up could reveal if their colours are related to their eccentricity, and if so, whether mass and eccentricity are linked to cause excess brightness in different photometric bands. Remeasuring the mid-infrared photometry of XO-3b will also confirm why this planet does not share the blue colours of HAT-P-2b and GJ 436b.
    
    \item Objects cooler than 1000K show a wide spread in colours in [3.6 - 4.5$\mu$m] colour, which could be attributed to mass, metallicity and C/O ratio. By comparing the positions of five planets to the colours of model atmospheres we find that for C/O ratio $\geq$0.85, high metallicity causes reddening. At C/O ratios $\leq$0.75, increasing metallicity causes increasing blueness. For the planets we plotted, we also find that increasing surface gravity corresponds to redder colours.
    
    \item We attribute the comparative blueness of planets in [4.5 - 5.8$\mu$m] colour to missing PH$_3$, which absorbs prominently in the 4.5$\mu$m band. Field brown dwarfs in this temperature range could be expected to have Phospine account for most of their atmospheric phosphurus budget, yet as PH$_3$ is susceptible to photolysis by ultraviolet radiation, we believe it is feasible that hot and ultra-hot Jupiters would be missing this absorber. We propose that PH$_3$ might have been overlooked to explain several low 4.5$\mu$m fluxes, including on the dayside of GJ 436b and the nightside of HD 189733b.
    
    \item The [\textit{W$_{JH}$ - H}] colour index can be used to diagnose the C/O ratio of exoplanets. Magnitude increases in the \textit{W}-band with increased water absorption, which we attribute to lower C/O ratios. From its position on a M$_W$ vs [\textit{W$_{JH}$ - H}] diagram, we find WASP-12b to be carbon-rich.
\end{itemize}

In order to further refine constraints which derive from comparisons with model magnitudes, model spectra need to be calibrated to real data to the greatest extent possible. Our colour-magnitude diagrams are very well suited to this purpose. 

With the launch of \textit{JWST} now close, our colour-magnitude diagrams will be an invaluable tool for target selection in this new era of exoplanet atmospheric characterisation. Additionally, ESA's \textcolor{Black}{{\it Ariel}} mission will enable us to begin to study populations as a whole. Here colour-magnitude diagrams can prove essential to identify various sub-population, and to select targets from {\it Tier 1} observations to the more detailed and higher signal-to-noise {\it Tier 2 \& 3}. 

\section*{Data Availability}
\textcolor{Black}{The data used in this article are available in tables in the appendices of the article, and on Github together with the Python toolkit developed for this article at \url{https://github.com/gdransfield/ExoCMD}.}

\section*{Acknowledgements}

\textcolor{Black}{We would like the thank Joanna Barstow for her considerate and insightful review of our paper. Her feedback helped to significantly improve and clarify the manuscript.}
GD acknowledges funding from the University of Birmingham which made this research possible. 
Many thanks also go to Jean-Loup Baudino for supplying the model spectra for GJ 504b; and to Vivien Parmentier for the very helpful discussions on thermal inversions. We also thank Emil Tersiev and Brigitta Sipocz for early exploration of colour-magnitude diagrams.
This work has made use of data from the European Space Agency (ESA) mission {\it Gaia} (\url{https://www.cosmos.esa.int/gaia}), processed by the {\it Gaia} Data Processing and Analysis Consortium (DPAC,
\url{https://www.cosmos.esa.int/web/gaia/dpac/consortium}). Funding for the DPAC has been provided by national institutions, in particular the institutions participating in the {\it Gaia} Multilateral Agreement.
This publication makes use of data products from the Two Micron All Sky Survey, which is a joint project of the University of Massachusetts and the Infrared Processing and Analysis Center/California Institute of Technology, funded by the National Aeronautics and Space Administration and the National Science Foundation.
This research has made use of the NASA Exoplanet Archive, which is operated by the California Institute of Technology, under contract with the National Aeronautics and Space Administration under the Exoplanet Exploration Program. AHMJT has received funding from the European Research Council (ERC) under the European Union's Horizon 2020 research and innovation programme (grant agreement n${^\circ}$ 803193/BEBOP).




\bibliographystyle{mnras}
\bibliography{CMDs.bib} 



\appendix

\section{recovering magnitudes}

\subsection{Host star magnitudes in new bands}
\label{sec:pickles-ap}

Where a host star apparent magnitude was not available, we made use of standard spectra from the Pickles Atlas\footnote{\url{http://www.stsci.edu/hst/instrumentation/reference-data-for-calibration-and-tools/astronomical-catalogs/pickles-atlas}}. In order to calculate the magnitudes of the parent stars, we began by searching \url{exoplanet.eu} \citep{2011AAA...532A..79S} for the spectral type of the host star. We acquired the Pickles spectrum corresponding to the parent star's spectral type; we then integrated the flux in 2MASS \textit{J}, \textit{H} and \textit{K} bands. For all magnitudes computed in this paper, we use the Vega-Magnitude system, setting its apparent magnitude to zero in all bands. 

\begin{figure*}
	\centering
	\includegraphics[width=\textwidth]{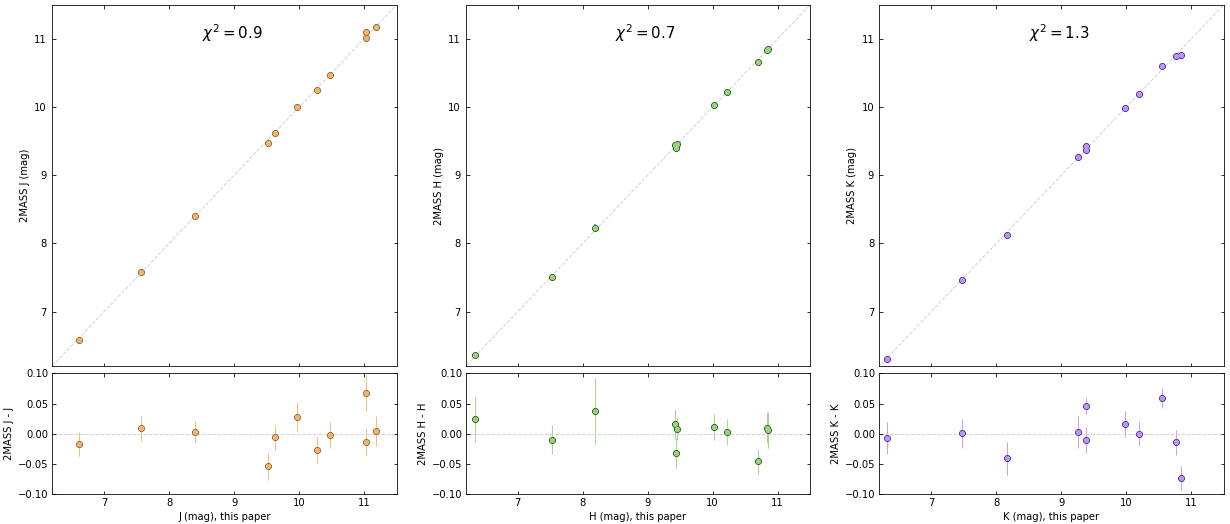}
    \caption{2MASS magnitudes of parent stars from the literature compared with our recovered mags using standard spectra from the Pickles Atlas.}
    \label{fig:recover}
\end{figure*}

In order to compute Vega's flux in \textit{J}, \textit{H} and \textit{K} we obtained Kurucz's high resolution spectrum\footnote{\url{http://kurucz.harvard.edu/stars/vega/}} which we then integrated in all three bands. Finally, we recovered \textit{H} and \textit{K} magnitudes from \textit{J}, \textit{J} and \textit{H} from \textit{K}, and \textit{J} and \textit{K} from \textit{H}. Where the spectral type of the parent star was not certain in the literature, we chose the spectrum which was a best fit in terms of temperature and recovered magnitudes. We present one-to-one plots of these recovered magnitudes in Figure \ref{fig:recover}. 


\subsection{Brown dwarf magnitudes from SpeX spectra}
\label{sec:spex-ap}

\begin{figure*}
	\centering
	\includegraphics[width=\textwidth]{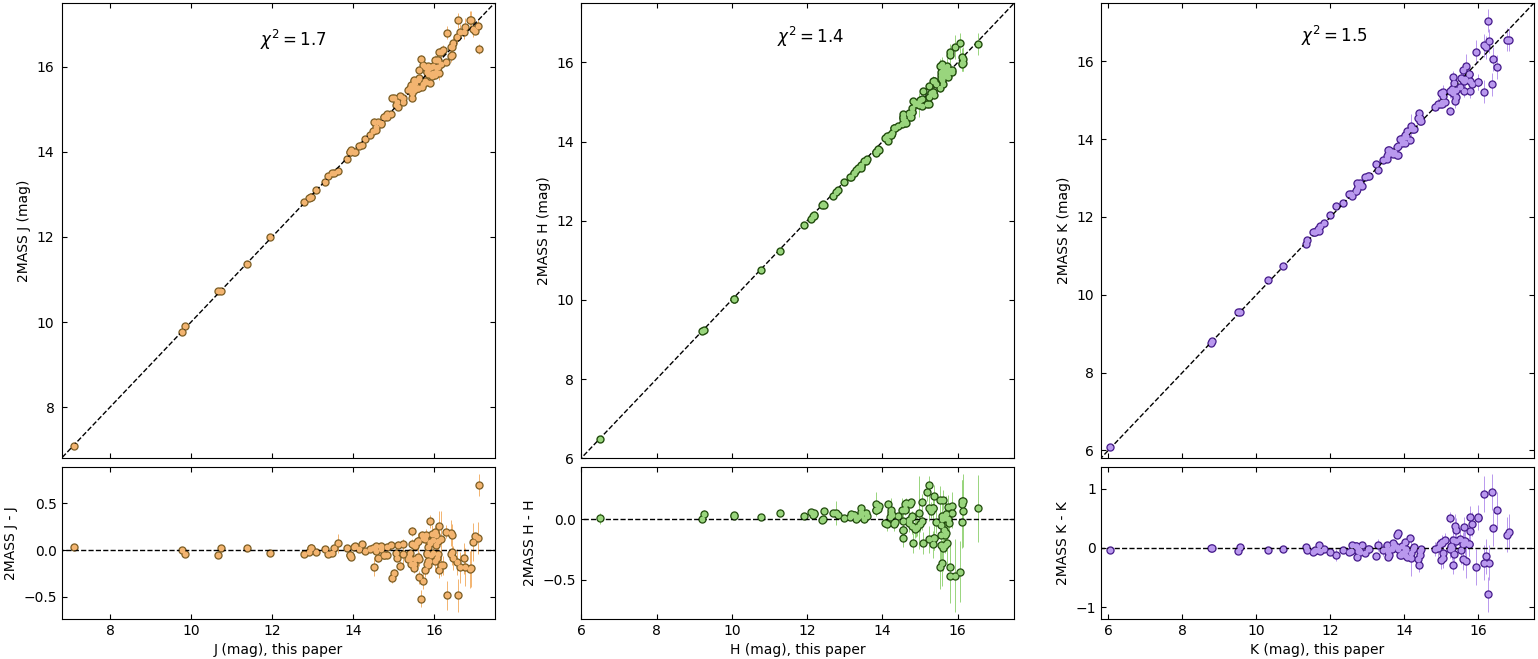}
    \caption{One-to-one plots of recovered \textit{J}, \textit{H} and \textit{K} magnitudes for brown dwarfs using SpeX spectra.}
    \label{fig:HKsynth}
\end{figure*}

We initially downloaded all 597 files available containing spectra for objects spanning the M to T spectral classes. 2MASS \textit{J}, \textit{H} and \textit{K} magnitudes were provided for the majority of them; we discarded any data where they were missing. We calculated apparent magnitudes using Vega as a reference once again, integrating the Kurucz spectrum in each band. In order to test the validity of our method, we first combined the integrated Vega flux with the provided 2MASS magnitudes of each star to recover synthetic \textit{J}, \textit{H} and \textit{K} magnitudes for the brown dwarfs. As each of the spectra are provided normalised, an important step in the determination of the magnitudes was to calculate a scale factor from each. These scale factors were determined by computing the band-integrated flux for each star in the same units as Vega's flux, working backwards with each of the 2MASS magnitudes:
\begin{equation}
\centering
    F_{bd} = F_{vega} \times 10^{\frac{2MASS\, (mag)}{-2.5}},
\end{equation}
where F$_{bd}$ is the band-integrated brown dwarf flux in Vega's units, F$_{vega}$ is Vega's band-integrated flux, and 2MASS (mag) is the 2MASS \textit{J, H} or \textit{K} magnitude of the brown dwarf. Dividing these fluxes by those obtained by integrating the spectra of the brown dwarfs yielded the required scale factors in all bands, which we then averaged. We used these to scale up the fluxes obtained by integrating the \textit{J}, \textit{H} and \textit{K} fluxes. All fluxes were also scaled using the spectral response function of each of the photometric bands. Errors are propagated throughout. We present one-to-one plots of recovered \textit{J}, \textit{H} and \textit{K} magnitudes in Figure \ref{fig:HKsynth}, along with reduced $\chi^2$. In these plots we can see that 2MASS \textit{K} magnitudes are recovered with the greatest uncertainties; plots of a random sample of SpeX spectra reveal that generally the noise is very large at longer wavelengths which leads to very low signal-to-noise ratio on these magnitudes.

\section{Magnitude scaling}
\label{sec:scaling}

\begin{figure}
	\centering
	\includegraphics[width=\columnwidth]{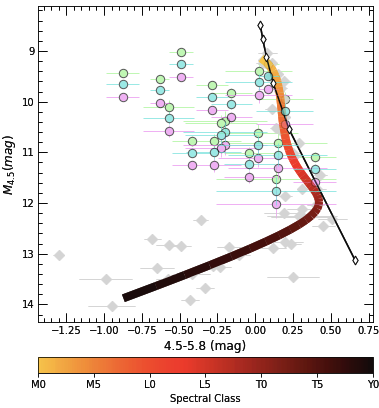}
    \caption{\textcolor{Black}{Colour-magnitude diagram with planetary magnitudes scaled by different factors: blue circles indicate scaling to a 0.9R$_J$ object, pink circles indicate scaling to 0.8R$_J$, and green circles are planets scaled to a Jupiter sized object (R = R$_J$). }}
    \label{fig:Scaling}
\end{figure}

\textcolor{Black}{In figure \ref{fig:Scaling} we present a colour-magnitude diagram where each planet has been scaled to a different size; this is to show the impact of choosing a different scaling factor. The size of the planets only affects their vertical positions, and small changes in the scale factors do not have a huge impact. In fact, provided all planets are scaled to the same size as each other, the main result is to entirely remove the effect of the differing planetary sizes. This in itself allows better comparison with brown dwarfs as they are considerably more homogeneous in size than planets. }

\section{Tooklit Walkthrough}
\label{sec:toolkit}

Users can interact with the ExoCMD\_toolkit via our Jupyter applet \footnote{\url{https://github.com/gdransfield/ExoCMD}}. This makes the experience far easier as plot options can be selected via drop-down boxes and tick-boxes rather than having to write lines of code. In this section we will show how the applet works and some of the available options. 

\begin{figure*}
	\centering
	\includegraphics[width=\textwidth]{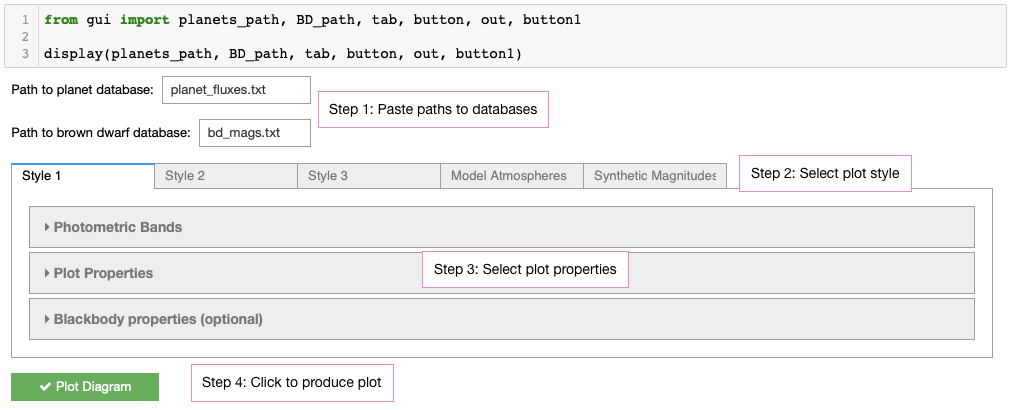}
    \caption{Applet view on first opening the Jupyter Notebook. }
    \label{fig:wholeapp}
\end{figure*}

Figure \ref{fig:wholeapp} shows the applet as it appears on first opening the notebook. The code can be toggled on and off in order to see how the applet is interacting with the .py modules we have written. There are two boxes where users can paste the paths to the planet and brown dwarf databases after cloning the repository them from Github. These are prefilled, but can be edited if the names of the databases are changed. Once all selections for a particular style of plot have been made, users should click the `Plot diagram' button. Diagrams will appear in the notebook and can be saved using the `Save figure' button. 

\begin{figure*}
	\centering
	\includegraphics[width=\textwidth]{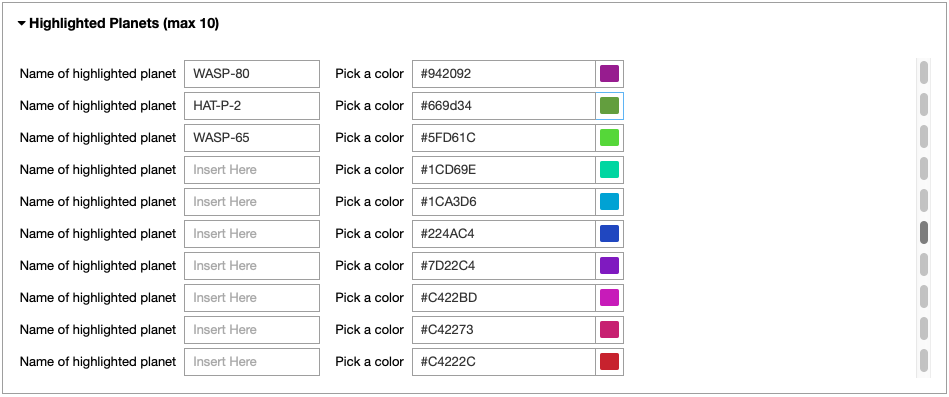}
    \caption{Selecting planets to highlight on a colour-magnitude diagram. The names are not case-sensitive but there must be hyphens between letters and numbers. Planet letters are omitted in this version. }
    \label{fig:highlight}
\end{figure*}

The `Style 1' tab corresponds to the function ExoCMD\_1, which produces colour-magnitude diagrams in the style of those presented in \cite{2014MNRAS.444..711T}. `Style 2' calls the function ExoCMD\_2 and outputs plots in the same style as Figure \ref{fig:PH3}. For the function ExoCMD\_3, which allows users to highlight specific planets in a different colour as in Figures \ref{fig:out1} and \ref{fig:out2}, users should select the `Style 3' tab. Here it is necessary to enter the name of the host star placing a hyphen between letters and numbers. See Figure \ref{fig:highlight} for an example. 

\begin{figure*}
	\centering
	\includegraphics[width=\textwidth]{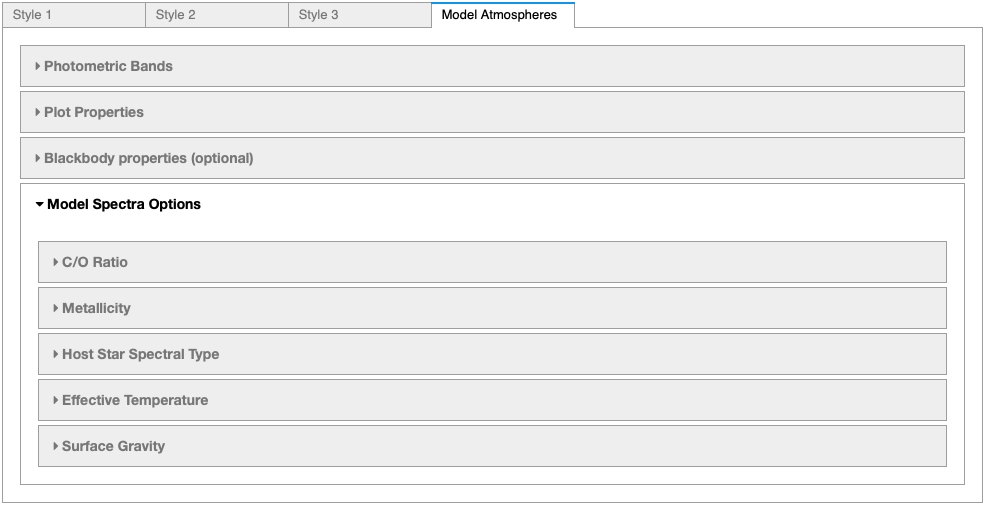}
    \caption{`Style 3' and `Model Atmospheres' both have an extra element compared with `Style 1' and `Style 2'. Here we show the view for `Model Atmospheres' with a closed accordion.}
    \label{fig:style4}
\end{figure*}

The tab labelled `Model Atmospheres' allows users to create colour-magnitude diagrams using the function ExoCMD\_model in order to compare synthetic photometry with real planets. There are many options to choose from when setting the model spectra. Some defaults are always selected as constraining nothing will lead to photometry being computed for all 10,640 available spectra. While this can be done, it will be time consuming so users should only select `all' for each parameter if that is what they wish to do. Figure \ref{fig:style4} shows the layout of the accordion for this style of diagram.

The tab on the far right, labelled `Synthetic Magnitudes', accesses a database of synthetic photometry derived from SpeX spectra. Plots produced will call the function \textsc{ExoCMD\_synth}; available bands are {2MASS JHK}, along with  {\it Sloan-z'}, {\it NB1190, NB2090} and {\it W$_{JH}$}.

With the exception of `Synthetic Magnitudes', all four styles of plot have the same photometric bands available. They can also have a 0.9R$_J$ blackbody added to the diagram, have a colourbar included or not, and can have planetary magnitudes adjusted to the size of a 0.9R$_J$ object.

\section{Planet database}
\label{sec:database}

In these tables we present the planetary databases. Table \ref{tab:composite} contains eccentricities, radii, equilibrium temperatures, host star spectral types and astrometric distances to planets included in our database. Table \ref{tab:fluxes} contains secondary eclipse measurements in near- and mid-infrared bands for all planets we have included on our colour-magnitude diagrams. Both tables can be downloaded in .txt or .xlsx form from Githib. 
\onecolumn
\begin{landscape}
\begin{longtable}[c]{@{}ccccccc@{}}
\toprule
\textbf{Planet} &
  \textbf{Eccentricity} &
  \textbf{Radius (R$_J$)} &
  \textbf{Equilibrium Temperature (K)} &
  \textbf{Host Star Spectral Type} &
  \textbf{Distance (pc)} &
  \textbf{References} \\* \midrule
\endfirsthead
\endhead
\bottomrule
\endfoot
\endlastfoot
CoRoT-1 b &
  0 &
  1.49$\pm$0.08 &
  1898$\pm$50 &
  G0 V &
  787.91$\pm$24.18 &
  \citenum{2008AAA...482L..17B}    \citenum{2018AJ....156...58B} \\
CoRoT-2 b &
  0.0143$\pm$0.0077 &
  1.466$^{+0.042}_{-0.044}$ &
  1521$\pm$18 &
  G7V &
  213.28$\pm$2.49 &
  \citenum{2010AAA...511A...3G}  \citenum{2012MNRAS.426.1291S}  \citenum{2018AJ....156...58B} \citenum{2014AAA...565L...1P}\\
GJ 3470 b &
  0.017$\pm$0.016 &
  0.408$\pm$0.016 &
  594$\pm$97 &
  M1.5 &
  29.42$\pm$0.05 &
  \citenum{2014MNRAS.443.1810B} \citenum{2016MNRAS.463.2574A}  \citenum{2012AAA...546A..27B} \citenum{2018AJ....156...58B} \\
GJ 436 b &
  0.13827$\pm$0.00018 &
  0.372$\pm$0.015 &
  686$\pm$10 &
  M2.5 V &
  9.75$\pm$0.01 &
  \citenum{2014AcA....64..323M}  \citenum{2016MNRAS.459..789T} \citenum{2004ApJ...617..580B} \citenum{2018AJ....156...58B} \\
HAT-P-1 b &
  0 &
  1.319$\pm$0.019 &
  1322$^{+14}_{-15}$ &
  G0 V &
  158.98$\pm$0.98 &
  \citenum{2018AJ....156..213M} \citenum{2014MNRAS.437...46N}  \citenum{2008ApJ...686..649J} \citenum{2018AJ....156...58B} \\
HAT-P-13 b &
  0.0133$\pm$0.0041 &
  1.272$\pm$0.065 &
  1740$\pm$27 &
  G4 &
  246.81$\pm$2.23 &
  \citenum{2010ApJ...718..575W}  \citenum{2016MNRAS.459..789T} \citenum{2009ApJ...707..446B} \citenum{2018AJ....156...58B} \\
HAT-P-18 b &
  0.084$\pm$0.048 &
  0.995$\pm$0.052 &
  852$\pm$28 &
  K2 &
  161.4$\pm$0.61 &
  \citenum{2011ApJ...726...52H}    \citenum{2018AJ....156...58B} \\
HAT-P-19 b &
  0.067$\pm$0.042 &
  1.132$\pm$0.072 &
  1010$\pm$42 &
  K1 &
  202.08$\pm$1.47 &
  \citenum{2011ApJ...726...52H}    \citenum{2018AJ....156...58B} \\
HAT-P-2 b &
  0.5172$\pm$0.0019 &
  0.951$^{+0.039}_{-0.053}$ &
  1540$\pm$30 &
  F8 &
  127.77$\pm$0.43 &
  \citenum{2018AJ....156..213M} \citenum{2008AAA...481..529L} \citenum{2010MNRAS.401.2665P} \citenum{2018AJ....156...58B} \\
HAT-P-20 b &
  0.015$\pm$0.005 &
  0.867$\pm$0.033 &
  970$\pm$23 &
  K3 &
  71.04$\pm$0.2 &
  \citenum{ 2011ApJ...742..116B}    \citenum{2018AJ....156...58B} \\
HAT-P-23 b &
  0.11$\pm$0.04 &
  1.09$\pm$0.23 &
  1951$\pm$30 &
  G0 &
  364.81$\pm$4.77 &
  \citenum{2017AJ....153..136S}  \citenum{2015AAA...577A..54C} \citenum{ 2011ApJ...742..116B} \citenum{2018AJ....156...58B} \\
HAT-P-3 b &
  0 &
  0.94$\pm$0.07 &
  1189$\pm$16 &
  K &
  134.55$\pm$0.48 &
  \citenum{2017AJ....153..136S}  \citenum{2012MNRAS.426.1291S} \citenum{2007ApJ...666L.121T} \citenum{2018AJ....156...58B} \\
HAT-P-30 b &
  0.04$\pm$0.02 &
  1.44$\pm$0.15 &
  1630$\pm$42 &
  F7 &
  213.99$\pm$2.22 &
  \citenum{2017AJ....153..136S}  \citenum{2011ApJ...735...24J} \citenum{ 2018yCat.5153....0L} \citenum{2018AJ....156...58B} \\
HAT-P-32 b &
  0.159$^{+0.051}_{-0.028}$ &
  1.98$\pm$0.045 &
  1836$\pm$7 &
  - &
  289.21$\pm$5.35 &
  \citenum{2019AJ....157...82W}    \citenum{2018AJ....156...58B} \\
HAT-P-33 b &
  0 &
  1.85$\pm$0.49 &
  1782$\pm$28 &
  F4 &
  396.11$\pm$7.65 &
  \citenum{2017AJ....153..136S}  \citenum{2011ApJ...742...59H} \citenum{ 2018yCat.5153....0L} \citenum{2018AJ....156...58B} \\
HAT-P-4 b &
  0 &
  1.274$^{+0.049}_{-0.06}$ &
  1686$^{+30}_{-26}$ &
  G1 V &
  320.45$\pm$2.85 &
  \citenum{2008ApJ...677.1324T}   \citenum{2007ApJ...670L..41K} \citenum{2018AJ....156...58B} \\
HAT-P-40 b &
  0 &
  1.52$\pm$0.17 &
  1770$\pm$33 &
  F &
  464.47$\pm$6.54 &
  \citenum{2017AJ....153..136S}  \citenum{2012AJ....144..139H}  \citenum{2018AJ....156...58B} \\
HAT-P-41 b &
  0 &
  2.05$\pm$0.5 &
  1941$\pm$38 &
  F &
  348.18$\pm$4.53 &
  \citenum{2017AJ....153..136S}  \citenum{2012AJ....144..139H}  \citenum{2018AJ....156...58B} \\
HAT-P-6 b &
  0 &
  1.48$\pm$0.15 &
  1675$^{+32}_{-31}$ &
  F8 V &
  275.36$\pm$3.64 &
  \citenum{2017AJ....153..136S}  \citenum{2008ApJ...677.1324T}  \citenum{2018AJ....156...58B} \citenum{2013MNRAS.433.2097F} \\
HAT-P-7 b &
  0 &
  1.51$\pm$0.21 &
  2733$\pm$21 &
  F6 V &
  341.08$\pm$2.43 &
  \citenum{2017AJ....153..136S}  \citenum{2013ApJ...764L..22M}  \citenum{2018AJ....156...58B} \citenum{2013MNRAS.433.2097F} \\
HAT-P-8 b &
  0 &
  1.4$\pm$0.13 &
  1700$\pm$35 &
  F6 &
  211.55$\pm$1.71 &
  \citenum{2017AJ....153..136S}  \citenum{2009ApJ...704.1107L} \citenum{ 2018yCat.5153....0L} \citenum{2018AJ....156...58B} \\
HD 149026 b &
  0 &
  0.74$\pm$0.02 &
  1626$^{+69}_{-37}$ &
  G0 &
  75.86$\pm$0.17 &
  \citenum{2017AJ....153..136S}  \citenum{2010MNRAS.408.1689S} \citenum{1918AnHar..91....1C} \citenum{2018AJ....156...58B} \\
HD 189733 b &
  0 &
  1.13$\pm$0.01 &
  1209$\pm$11 &
  K2 V &
  19.76$\pm$0.01 &
  \citenum{2017AJ....153..136S}  \citenum{2019PASP..131k5003A} \citenum{2003AJ....126.2048G} \citenum{2018AJ....156...58B} \\
HD 209458 b &
  0 &
  1.39$\pm$0.02 &
  1459$\pm$12 &
  F8 &
  48.3$\pm$0.12 &
  \citenum{2017AJ....153..136S}  \citenum{2010MNRAS.408.1689S} \citenum{1918AnHar..91....1C} \citenum{2018AJ....156...58B} \\
KELT-1 b &
  0.0099$^{+0.01}_{-0.0069}$ &
  1.11$^{+0.032}_{-0.022}$ &
  - &
  F5 &
  268.43$\pm$3.06 &
  \citenum{2012ApJ...761..123S}    \citenum{2018AJ....156...58B} \\
KELT-2 A b &
  0 &
  1.35$\pm$0.08 &
  - &
  F8 &
  134.06$\pm$0.8 &
  \citenum{2017AJ....153..136S}   \citenum{1918AnHar..91....1C} \citenum{2018AJ....156...58B} \\
KELT-3 b &
  0 &
  1.56$\pm$0.11 &
  1816$^{+37}_{-39}$ &
  F &
  210.25$\pm$5.54 &
  \citenum{2017AJ....153..136S}  \citenum{2013ApJ...773...64P}  \citenum{2018AJ....156...58B} \\
KELT-7 b &
  0 &
  1.6$\pm$0.06 &
  2048$\pm$27 &
  F &
  136.68$\pm$0.94 &
  \citenum{2017AJ....153..136S}  \citenum{2015AJ....150...12B}  \citenum{2018AJ....156...58B} \\
Kepler-13A b &
  $0.00064\substack{+0.00012\\-0.00016}$ &
  1.512$\pm$0.035 &
  2550$\pm$80 &
  A &
  519.1$\pm$30.7 &
  \citenum{2015ApJ...804..150E}    \citenum{2018AJ....156...58B} \citenum{2014ApJ...788...92S}\\
Kepler-5 b &
  0.043 &
  1.426$^{+0.036}_{-0.051}$ &
  1750$\pm$20 &
  F5V &
  899.78$\pm$16.77 &
  \citenum{2017AAA...602A.107B} \citenum{2015ApJ...804..150E}   \citenum{2018AJ....156...58B} \citenum{2016AAA...594A..39F}\\
Kepler-6 b &
  0.06 &
  1.304$^{+0.018}_{-0.033}$ &
  1460$\pm$10 &
  - &
  587.04$\pm$5.07 &
  \citenum{2017AAA...602A.107B} \citenum{2015ApJ...804..150E}   \citenum{2018AJ....156...58B} \\
Qatar-1 b &
  0 &
  1.143$^{+0.026}_{-0.025}$ &
  1418$^{+28}_{-27}$ &
  K &
  185.62$\pm$0.8 &
  \citenum{2017AJ....153...78C}    \citenum{2018AJ....156...58B} \citenum{2011MNRAS.417..709A}\\
TrES-1 b &
  0 &
  1.13$\pm$0.06 &
  1140$^{+13}_{-12}$ &
  K0 V &
  159.66$\pm$0.74 &
  \citenum{2017AJ....153..136S} \citenum{2008ApJ...677.1324T} \citenum{2004ApJ...613L.153A} \citenum{2018AJ....156...58B} \\
TrES-2 b &
  0 &
  1.36$\pm$0.08 &
  1466$\pm$9 &
  G0 V &
  215.32$\pm$1.05 &
  \citenum{2017AJ....153..136S}  \citenum{2019MNRAS.486.2290O} \citenum{2006ApJ...651L..61O} \citenum{2018AJ....156...58B} \\
TrES-3 b &
  0 &
  1.336$^{+0.031}_{-0.037}$ &
  1638$\pm$22 &
  G4 V &
  231.34$\pm$1.3 &
  \citenum{2009ApJ...691.1145S}  \citenum{2011MNRAS.417.2166S} \citenum{2007ApJ...663L..37O} \citenum{2018AJ....156...58B} \\
TrES-4 b &
  0 &
  1.61$\pm$0.18 &
  1778$\pm$22 &
  F8 V &
  515.98$\pm$7.03 &
  \citenum{2017AJ....153..136S}  \citenum{2016MNRAS.459..789T} \citenum{2007ApJ...667L.195M} \citenum{2018AJ....156...58B} \\
WASP-1 b &
  0 &
  1.483$^{+0.024}_{-0.034}$ &
  1812$\pm$14 &
  F7 V &
  393.07$\pm$10.8 &
  \citenum{2014AcA....64...27M}  \citenum{2016MNRAS.459..789T} \citenum{2007MNRAS.375..951C} \citenum{2018AJ....156...58B} \\
WASP-10 b &
  0.0473$^{+0.0034}_{-0.0029}$ &
  1.08$\pm$0.02 &
  1370$\pm$50 &
  K5 V &
  141$\pm$0.75 &
  \citenum{2014ApJ...785..126K} \citenum{2009ApJ...692L.100J}  \citenum{2015AAA...576A..42S} \citenum{2018AJ....156...58B} \\
WASP-100 b &
  0 &
  1.33$\pm$0.14 &
  2190$\pm$140 &
  F2 &
  364.41$\pm$2.75 &
  \citenum{2017AJ....153..136S}  \citenum{2014MNRAS.440.1982H}  \citenum{2018AJ....156...58B} \\
WASP-101 b &
  0 &
  1.43$\pm$0.09 &
  1560$\pm$35 &
  F6 &
  201.22$\pm$1.15 &
  \citenum{2017AJ....153..136S}  \citenum{2014MNRAS.440.1982H}  \citenum{2018AJ....156...58B} \\
WASP-103 b &
  0.15 &
  1.528$^{+0.073}_{-0.047}$ &
  2508$^{+75}_{-70}$ &
  F8 V &
  883.3$\pm$120.5 &
  \citenum{2017AAA...602A.107B} \citenum{2014AAA...562L...3G}   \citenum{2018AJ....156...58B} \\
WASP-104 b &
  0 &
  1.137$\pm$0.037 &
  1516$\pm$39 &
  G8 &
  185.93$\pm$1.48 &
  \citenum{2014AAA...570A..64S}    \citenum{2018AJ....156...58B} \\
WASP-12 b &
  0.0447 &
  1.937$\pm$0.056 &
  2593$\pm$57 &
  G0 V &
  427.25$\pm$6.07 &
  \citenum{2016MNRAS.459..789T} \citenum{2019AJ....158...39C}   \citenum{2018AJ....156...58B} \\
WASP-121 b &
  0 &
  1.865$\pm$0.044 &
  2358$\pm$52 &
  F6 V &
  269.9$\pm$1.58 &
  \citenum{2016MNRAS.458.4025D}    \citenum{2018AJ....156...58B} \\
WASP-131 b &
  0 &
  1.22$\pm$0.05 &
  1460$\pm$30 &
  G0 &
  200.08$\pm$2.66 &
  \citenum{2017MNRAS.465.3693H}    \citenum{2018AJ....156...58B} \\
WASP-14 b &
  0.09 &
  1.38$\pm$0.08 &
  1872$\pm$29 &
  F5V &
  162$\pm$0.81 &
  \citenum{2017AJ....153..136S}  \citenum{2015MNRAS.451.4139R}  \citenum{2018AJ....156...58B} \citenum{2011AAA...529A.136E}\\
WASP-17 b &
  0 &
  1.87$\pm$0.24 &
  1755$\pm$28 &
  F4 &
  405.91$\pm$8.78 &
  \citenum{2017AJ....153..136S}  \citenum{2012MNRAS.426.1338S} \citenum{2010AAA...524A..25T} \citenum{2018AJ....156...58B} \\
WASP-18 b &
  0 &
  1.191$\pm$0.038 &
  2413$\pm$44 &
  F6 IV &
  123.48$\pm$0.37 &
  \citenum{2017AJ....153..136S} \citenum{2019AJ....157..178S}  \citenum{1978mcts.book.....H} \citenum{2018AJ....156...58B} \\
WASP-19 b &
  0.002$^{+0.014}_{-0.002}$ &
  1.392$\pm$0.04 &
  2520 &
  G8 V &
  268.33$\pm$1.72 &
  \citenum{2016ApJ...823..122W}   \citenum{2010ApJ...708..224H} \citenum{2018AJ....156...58B} \\
WASP-2 b &
  0.0054$^{+0.009}_{-0.0044}$ &
  1.081$^{+0.041}_{-0.04}$ &
  1311$^{+52}_{-50}$ &
  K1 V &
  153.24$\pm$1.64 &
  \citenum{2014ApJ...785..126K} \citenum{2019PASP..131k5003A}  \citenum{2007MNRAS.375..951C} \citenum{2018AJ....156...58B} \\
WASP-24 b &
  0 &
  1.38$\pm$0.16 &
  1772$\pm$29 &
  F8/9 &
  322.11$\pm$4.5 &
  \citenum{2017AJ....153..136S}  \citenum{2014MNRAS.444..776S}  \citenum{2018AJ....156...58B} \citenum{2011AAA...529A.136E}\\
WASP-26 b &
  0 &
  1.21$\pm$0.15 &
  1650$\pm$24 &
  G0 &
  252.76$\pm$4.76 &
  \citenum{2017AJ....153..136S}  \citenum{2014MNRAS.444..776S}  \citenum{2018AJ....156...58B} \citenum{2011AAA...529A.136E}\\
WASP-3 b &
  0 &
  1.42$\pm$0.17 &
  2020$\pm$35 &
  F7 V &
  231.16$\pm$1.65 &
  \citenum{2017AJ....153..136S}  \citenum{2011MNRAS.417.2166S}  \citenum{2018AJ....156...58B} \citenum{2008MNRAS.385.1576P}\\
WASP-33 b &
  0 &
  1.593$\pm$0.074 &
  2782$\pm$41 &
  A5 &
  121.94$\pm$1 &
  \citenum{2017AJ....153..136S} \citenum{2019AJ....158...39C}  \citenum{1918AnHar..91....1C} \citenum{2018AJ....156...58B} \\
WASP-36 b &
  0.019 &
  1.327$\pm$0.021 &
  1733$\pm$19 &
  G2 &
  386.35$\pm$5.26 &
  \citenum{2017AAA...602A.107B} \citenum{2016MNRAS.459.1393M}   \citenum{2018AJ....156...58B} \citenum{2012AJ....143...81S}\\
WASP-39 b &
  0 &
  1.27$\pm$0.04 &
  1166$\pm$14 &
  G &
  213.98$\pm$1.76 &
  \citenum{2011AAA...531A..40F}  \citenum{2018AAA...613A..41M}  \citenum{2018AJ....156...58B} \citenum{2018AJ....155...29W}\\
WASP-4 b &
  0 &
  1.321$\pm$0.039 &
  1673$\pm$17 &
  G7 V &
  267.21$\pm$3.77 &
  \citenum{2019AJ....157..217B}  \citenum{2012MNRAS.426.1291S} \citenum{2008ApJ...675L.113W} \citenum{2018AJ....156...58B} \\
WASP-43 b &
  0 &
  0.93$^{+0.07}_{-0.09}$ &
  1427$\pm$9 &
  K7 V &
  86.75$\pm$0.33 &
  \citenum{2011AAA...535L...7H}  \citenum{2017AAA...601A..53E} \citenum{2015AAA...576A..42S} \citenum{2018AJ....156...58B} \\
WASP-46 b &
  0.022 &
  1.174$\pm$0.037 &
  1636$\pm$44 &
  G6 V &
  375.31$\pm$4.46 &
  \citenum{2017AAA...602A.107B} \citenum{2016MNRAS.456..990C} \citenum{2016MNRAS.456..990C}  \citenum{2018AJ....156...58B} \\
WASP-48 b &
  0 &
  1.5$\pm$0.2 &
  2035$\pm$52 &
  G &
  454.14$\pm$4.47 &
  \citenum{2017AJ....153..136S}  \citenum{2016MNRAS.459..789T}  \citenum{2018AJ....156...58B} \\
WASP-5 b &
  0.038$^{+0.026}_{-0.018}$ &
  1.087$^{+0.068}_{-0.071}$ &
  1706$^{+52}_{-48}$ &
  G4 V &
  309.14$\pm$3.41 &
  \citenum{2009AAA...496..259G}   \citenum{2008MNRAS.387L...4A} \citenum{2018AJ....156...58B} \\
WASP-6 b &
  0.05 &
  1.03$\pm$0.1 &
  1184$\pm$16 &
  G8 V &
  197.12$\pm$1.63 &
  \citenum{2017AJ....153..136S}  \citenum{2015MNRAS.450.1760T}  \citenum{2018AJ....156...58B} \\
WASP-62 b &
  0 &
  1.32$\pm$0.08 &
  1475$^{+25}_{-20}$ &
  F7 &
  175.63$\pm$0.59 &
  \citenum{2017AJ....153..136S}  \citenum{2017MNRAS.464..810B}  \citenum{2018AJ....156...58B} \citenum{2012MNRAS.426..739H}\\
WASP-63 b &
  0 &
  1.41$\pm$0.14 &
  - &
  G8 &
  290.68$\pm$2.03 &
  \citenum{2017AJ....153..136S}    \citenum{2018AJ....156...58B} \citenum{2012MNRAS.426..739H}\\
WASP-64 b &
  0 &
  1.271$\pm$0.039 &
  1689$\pm$49 &
  G7 &
  369.93$\pm$3.03 &
  \citenum{2013AAA...552A..82G}    \citenum{2018AJ....156...58B} \\
WASP-65 b &
  0 &
  1.112$\pm$0.059 &
  1480$\pm$10 &
  G6 &
  273.7$\pm$2.73 &
  \citenum{2013AAA...559A..36G}    \citenum{2018AJ....156...58B} \\
WASP-67 b &
  0 &
  1.15$\pm$0.11 &
  1003$\pm$20 &
  K0 V &
  189.47$\pm$1.56 &
  \citenum{2017AJ....153..136S}  \citenum{2014AAA...568A.127M}  \citenum{2018AJ....156...58B} \citenum{2012MNRAS.426..739H}\\
WASP-69 b &
  0 &
  1.11$\pm$0.04 &
  963$\pm$18 &
  K5 &
  49.96$\pm$0.13 &
  \citenum{2017AJ....153..136S}  \citenum{2017AAA...608A.135C} \citenum{2014MNRAS.445.1114A} \citenum{2018AJ....156...58B} \\
WASP-74 b &
  0 &
  1.36$\pm$0.1 &
  1910$\pm$40 &
  F9 &
  149.22$\pm$1.15 &
  \citenum{2017AJ....153..136S}  \citenum{2015AJ....150...18H}  \citenum{2018AJ....156...58B} \\
WASP-76 b &
  0 &
  1.83$^{+0.06}_{-0.04}$ &
  2160$\pm$40 &
  F7 &
  194.46$\pm$6.21 &
  \citenum{2016AAA...585A.126W}    \citenum{2018AJ....156...58B} \\
WASP-77 A b &
  0 &
  1.38$\pm$0.09 &
  1674$\pm$24 &
  G8 V &
  105.17$\pm$1.21 &
  \citenum{2017AJ....153..136S}  \citenum{2016MNRAS.459..789T} \citenum{2013PASP..125...48M} \citenum{2018AJ....156...58B} \\
WASP-78 b &
  0 &
  1.93$\pm$0.45 &
  2470$^{+54}_{-56}$ &
  F8 &
  754.26$\pm$17.1 &
  \citenum{2017AJ....153..136S}  \citenum{2017MNRAS.464..810B}  \citenum{2018AJ....156...58B} \citenum{2012AAA...547A..61S}\\
WASP-79 b &
  0 &
  1.67$\pm$0.15 &
  1716$^{+26}_{-24}$ &
  F5 &
  246.69$\pm$1.82 &
  \citenum{2017AJ....153..136S}  \citenum{2017MNRAS.464..810B}  \citenum{2018AJ....156...58B} \citenum{2012AAA...547A..61S}\\
WASP-8 b &
  0.31 &
  1.13$\pm$0.05 &
  - &
  G8 V &
  89.96$\pm$0.36 &
  \citenum{2017AJ....153..136S}  \citenum{2015AAA...576A..42S}  \citenum{2018AJ....156...58B} \\
WASP-80 b &
  0.002$^{+0.01}_{-0.002}$ &
  0.999$^{+0.03}_{-0.031}$ &
  825$\pm$19 &
  K7 V &
  49.79$\pm$0.12 &
  \citenum{2015MNRAS.450.2279T}   \citenum{2013AAA...551A..80T} \citenum{2018AJ....156...58B} \\
WASP-87 b &
  0 &
  1.385$\pm$0.06 &
  - &
  F5 &
  298.39$\pm$3.62 &
  \citenum{2016ApJ...823...29A}    \citenum{2018AJ....156...58B} \citenum{2014arXiv1410.3449A}\\
WASP-94 A b &
  0 &
  1.58$\pm$0.13 &
  1604$^{+25}_{-22}$ &
  F8 &
  211.21$\pm$2.51 &
  \citenum{2017AJ....153..136S}  \citenum{2014AAA...572A..49N}  \citenum{2018AJ....156...58B} \\
WASP-97 b &
  0 &
  1.14$\pm$0.06 &
  1555$\pm$40 &
  G5 &
  151.07$\pm$0.51 &
  \citenum{2017AJ....153..136S}  \citenum{2014MNRAS.440.1982H}  \citenum{2018AJ....156...58B} \\
XO-1 b &
  0 &
  1.14$\pm$0.07 &
  1210$\pm$16 &
  G1 V &
  163.55$\pm$0.62 &
  \citenum{2017AJ....153..136S}  \citenum{2010MNRAS.408.1689S} \citenum{2006ApJ...648.1228M} \citenum{2018AJ....156...58B} \\
XO-2 N b &
  0.028$^{+0.038}_{-0.022}$ &
  0.993$\pm$0.012 &
  1328$^{+17}_{-28}$ &
  K0 V &
  154.27$\pm$1.46 &
  \citenum{2014ApJ...785..126K} \citenum{2012ApJ...761....7C} \citenum{2012MNRAS.426.1291S} \citenum{2007ApJ...671.2115B} \citenum{2018AJ....156...58B} \\
XO-3 b &
  0.29 &
  1.41$\pm$0.12 &
  1729$\pm$34 &
  F5 V &
  213.05$\pm$2.72 &
  \citenum{2017AJ....153..136S}  \citenum{2010MNRAS.408.1689S} \citenum{2007AAS...210.9605J} \citenum{2018AJ....156...58B} \\
XO-4 b &
  0 &
  1.25$\pm$0.08 &
  1630$^{+169}_{-36}$ &
  F5 V &
  272.65$\pm$2.91 &
  \citenum{2017AJ....153..136S}  \citenum{2010MNRAS.408.1689S} \citenum{2008arXiv0805.2921M} \citenum{2018AJ....156...58B} \\* \bottomrule
\caption{Composite planet data which we have in this paper. Compiled with the help of the NASA Exoplanet Archive.}
\label{tab:composite}\\
\end{longtable}
\end{landscape}
\twocolumn

\onecolumn
\begin{landscape}
\begin{longtable}[c]{@{}ccccccccccccc@{}}
\toprule
\textbf{Planet} &
  \textbf{{\it J}} &
  \textbf{{\it H}} &
  \textbf{{\it K}} &
  \textbf{3.6$\mu$m} &
  \textbf{4.5$\mu$m} &
  \textbf{5.8$\mu$m} &
  \textbf{8.0$\mu$m} &
  \textbf{{\it W$_{JH}$}} &
  \textbf{{\it H$_s$}} &
  \textbf{References} &
  \textbf{} &
  \textbf{} \\* \midrule
\endfirsthead
\multicolumn{12}{c}%
{{\bfseries Table \thetable\ continued from previous page}} \\
\endhead
\bottomrule
\endfoot
\endlastfoot
CoRoT-1 b &
  - &
  0.145$\pm$0.049 &
  0.336$\pm$0.042 &
  0.415$\pm$0.042 &
  0.482$\pm$0.042 &
  - &
  - &
  - &
  - &
  \citenum{2012ApJ...744..122Z} \citenum{2009ApJ...707.1707R} \citenum{2011ApJ...726...95D} &
   &
   \\
CoRoT-2 b &
  - &
  - &
  0.16$\pm$0.09 &
  0.355$\pm$0.02 &
  0.5$\pm$0.02 &
  - &
  0.51$\pm$0.059 &
  - &
  - &
  \citenum{2010AJ....139.1481A} \citenum{2011ApJ...726...95D} &
   &
   \\
GJ 3470 b &
  - &
  - &
  - &
  0.0115$^{+0.0027}_{0.0026}$ &
  0.0003$\pm$0.0022 &
  - &
  - &
  - &
  - &
  \citenum{2019NatAs.tmp..361B} &
   &
   \\
GJ 436 b &
  - &
  - &
  - &
  0.041$\pm$0.003 &
  - &
  0.033$\pm$0.014 &
  0.054$\pm$0.008 &
  - &
  - &
  \citenum{2010Natur.464.1161S} &
   &
   \\
HAT-P-1 b &
  - &
  - &
  0.109$\pm$0.025 &
  0.08$\pm$0.008 &
  0.135$\pm$0.022 &
  0.203$\pm$0.031 &
  0.238$\pm$0.04 &
  - &
  - &
  \citenum{2011AAA...528A..49D} \citenum{2010ApJ...708..498T} &
   &
   \\
HAT-P-13 b &
  - &
  - &
  - &
  0.0801$\pm$0.0081 &
  0.19$\pm$0.0124 &
  - &
  - &
  - &
  - &
  \citenum{2017ApJ...836..143H} \citenum{2019arXiv190107040G} &
   &
   \\
HAT-P-18 b &
  - &
  - &
  - &
  0.0437$^{+0.0146}_{-0.0144}$ &
  0.0326$^{+0.0144}_{-0.0147}$ &
  - &
  - &
  - &
  - &
  \citenum{2019arXiv190800014W} &
   &
   \\
HAT-P-19 b &
  - &
  - &
  - &
  0.062$\pm$0.014 &
  0.062$\pm$0.016 &
  - &
  - &
  - &
  - &
  \citenum{2015ApJ...810..118K} &
   &
   \\
HAT-P-2 b &
  - &
  - &
  - &
  0.0996$\pm$0.0072 &
  0.1031$\pm$0.0061 &
  0.071$^{+0.029}_{-0.013}$ &
  0.1392$^{+0.0095}_{-0.0095}$ &
  - &
  - &
  \citenum{2013ApJ...766...95L} &
   &
   \\
HAT-P-20 b &
  - &
  - &
  - &
  0.0615$\pm$0.0082 &
  0.1096$\pm$0.0077 &
  - &
  - &
  - &
  - &
  \citenum{2015ApJ...805..132D} &
   &
   \\
HAT-P-23 b &
  - &
  - &
  0.234$\pm$0.046 &
  0.248$\pm$0.019 &
  0.309$\pm$0.026 &
  - &
  - &
  - &
  - &
  \citenum{2014ApJ...781..109O} \citenum{2014ApJ...781..109O} &
   &
   \\
HAT-P-3 b &
  - &
  - &
  - &
  0.112$^{+0.015}_{-0.03}$ &
  0.094$^{+0.016}_{-0.009}$ &
  - &
  - &
  - &
  - &
  \citenum{2013ApJ...770..102T} &
   &
   \\
HAT-P-30 b &
  - &
  - &
  - &
  0.1584$\pm$0.0107 &
  0.1825$\pm$0.0147 &
  - &
  - &
  - &
  - &
  \citenum{2019arXiv190107040G} &
   &
   \\
HAT-P-32 b &
  0.04$\pm$0.014 &
  0.09$\pm$0.033 &
  0.178$\pm$0.057 &
  0.364$\pm$0.016 &
  0.438$\pm$0.02 &
  - &
  - &
  0.051$\pm$0.013 &
  0.057$\pm$0.013 &
  {\it this work} \citenum{2014ApJ...796..115Z} &
   &
   \\
HAT-P-33 b &
  - &
  - &
  - &
  0.1603$\pm$0.0127 &
  0.1835$\pm$0.0199 &
  - &
  - &
  - &
  - &
  \citenum{2019arXiv190107040G} &
   &
   \\
HAT-P-4 b &
  - &
  - &
  - &
  0.142$^{+0.014}_{-0.016}$ &
  0.122$^{+0.012}_{-0.014}$ &
  - &
  - &
  - &
  - &
  \citenum{2013ApJ...770..102T} &
   &
   \\
HAT-P-40 b &
  - &
  - &
  - &
  0.0988$\pm$0.0168 &
  0.1057$\pm$0.0145 &
  - &
  - &
  - &
  - &
  \citenum{2019arXiv190107040G} &
   &
   \\
HAT-P-41 b &
  - &
  - &
  - &
  0.1829$\pm$0.0319 &
  0.2278$\pm$0.0177 &
  - &
  - &
  - &
  - &
  \citenum{2019arXiv190107040G} &
   &
   \\
HAT-P-6 b &
  - &
  - &
  - &
  0.117$\pm$0.008 &
  0.106$\pm$0.006 &
  - &
  - &
  - &
  - &
  \citenum{2012ApJ...746..111T} &
   &
   \\
HAT-P-7 b &
  - &
  - &
  - &
  0.098$\pm$0.017 &
  0.159$\pm$0.022 &
  0.245$\pm$0.031 &
  0.262$\pm$0.027 &
  - &
  - &
  \citenum{2010ApJ...710...97C} &
   &
   \\
HAT-P-8 b &
  - &
  - &
  - &
  0.131$^{+0.007}_{-0.01}$ &
  0.111$^{+0.008}_{-0.007}$ &
  - &
  - &
  - &
  - &
  \citenum{2012ApJ...746..111T} &
   &
   \\
HD 149026 b &
  - &
  - &
  - &
  0.04$\pm$0.003 &
  0.034$\pm$0.006 &
  0.044$\pm$0.01 &
  0.052$\pm$0.006 &
  - &
  - &
  \citenum{2012ApJ...754..136S} &
   &
   \\
HD 189733 b &
  - &
  - &
  - &
  0.256$\pm$0.014 &
  0.214$\pm$0.02 &
  0.31$\pm$0.034 &
  0.344$^{+0.0036}_{-0.0036}$ &
  - &
  - &
  \citenum{2008ApJ...686.1341C} &
   &
   \\
HD 209458 b &
  0.091$\pm$0.018 &
  - &
  - &
  0.106$^{+0.007}_{-0.008}$ &
  0.133$\pm$0.011 &
  0.142$^{+0.059}_{-0.058}$ &
  0.215$\pm$0.012 &
  0.073$\pm$0.018 &
  0.147$\pm$0.018 &
  {\it this work}   \citenum{2015MNRAS.451..680E} &
   &
   \\
KELT-1 b &
  - &
  - &
  0.16$^{+0.018}_{0.02}$ &
  0.195$\pm$0.01 &
  0.2$\pm$0.012 &
  - &
  - &
  - &
  - &
  \citenum{2015ApJ...802...28C} \citenum{2014ApJ...783..112B} &
   &
   \\
KELT-2 A b &
  - &
  - &
  - &
  0.0572$^{+0.0045}_{-0.0046}$ &
  0.0616$^{+0.0044}_{-0.0045}$ &
  - &
  - &
  - &
  - &
  \citenum{2018AJ....156..133P} &
   &
   \\
KELT-3 b &
  - &
  - &
  - &
  0.1766$\pm$0.0097 &
  0.1656$\pm$0.0104 &
  - &
  - &
  - &
  - &
  \citenum{2019arXiv190107040G} &
   &
   \\
KELT-7 b &
  - &
  - &
  - &
  0.1688$\pm$0.0046 &
  0.1896$\pm$0.0057 &
  - &
  - &
  - &
  - &
  \citenum{2019arXiv190107040G} &
   &
   \\
Kepler-13 Ab &
  0.0724$\pm$0.0071 &
  - &
  0.122$\pm$0.051 &
  0.156$\pm$0.031 &
  0.222$\pm$0.023 &
  - &
  - &
  0.071$\pm$0.08 &
  0.086$\pm$0.008 &
  {\it this work} \citenum{2014ApJ...788...92S} &
   &
   \\
Kepler-5 b &
  - &
  - &
  - &
  0.103$\pm$0.017 &
  0.107$\pm$0.015 &
  - &
  - &
  - &
  - &
  \citenum{2011ApJS..197...11D} &
   &
   \\
Kepler-6 b &
  - &
  - &
  - &
  0.069$\pm$0.027 &
  0.151$\pm$0.019 &
  - &
  - &
  - &
  - &
  \citenum{2011ApJS..197...11D} &
   &
   \\
Qatar-1 b &
  - &
  - &
  0.136$\pm$0.034 &
  0.149$\pm$0.051 &
  0.273$\pm$0.049 &
  - &
  - &
  - &
  - &
  \citenum{2015ApJ...802...28C} \citenum{2018AAA...610A..55G} &
   &
   \\
TrES-1 b &
  - &
  - &
  - &
  0.083$\pm$0.024 &
  0.094$\pm$0.024 &
  0.162$\pm$0.042 &
  0.213$\pm$0.042 &
  - &
  - &
  \citenum{2014ApJ...797...42C} &
   &
   \\
TrES-2 b &
  - &
  - &
  0.062$^{+0.013}_{-0.011}$ &
  0.127$\pm$0.021 &
  0.23$\pm$0.024 &
  0.199$\pm$0.054 &
  0.359$\pm$0.06 &
  - &
  - &
  \citenum{2010ApJ...717.1084C} \citenum{2010ApJ...710.1551O} &
   &
   \\
TrES-3 b &
  - &
  - &
  0.133$^{+0.018}_{-0.016}$ &
  0.356$\pm$0.035 &
  0.372$\pm$0.054 &
  0.449$\pm$0.097 &
  0.475$\pm$0.046 &
  0.047$\pm$0.028 &
  - &
  \citenum{2010ApJ...718..920C} \citenum{2010ApJ...711..374F} &
   &
   \\
TrES-4 b &
  - &
  - &
  - &
  0.137$\pm$0.011 &
  0.148$\pm$0.016 &
  0.261$\pm$0.059 &
  0.318$\pm$0.044 &
  - &
  - &
  \citenum{2009ApJ...691..866K} &
   &
   \\
WASP-1 b &
  - &
  - &
  - &
  0.184$\pm$0.016 &
  0.217$\pm$0.017 &
  0.274$\pm$0.058 &
  0.474$\pm$0.046 &
  - &
  - &
  \citenum{2010arXiv1004.0836W} &
   &
   \\
WASP-10 b &
  - &
  - &
  0.137$^{+0.013}_{-0.019}$ &
  0.1$\pm$0.011 &
  0.146$\pm$0.016 &
  - &
  - &
  - &
  - &
  \citenum{2015AAA...574A.103C} \citenum{2015ApJ...810..118K} &
   &
   \\
WASP-100 &
  - &
  - &
  - &
  0.1267$\pm$0.0098 &
  0.172$\pm$0.0119 &
  - &
  - &
  - &
  - &
  \citenum{2019arXiv190107040G} &
   &
   \\
WASP-101 &
  - &
  - &
  - &
  0.1161$\pm$0.0111 &
  0.1194$\pm$0.0113 &
  - &
  - &
  - &
  - &
  \citenum{2019arXiv190107040G} &
   &
   \\
WASP-103 b &
  0.131$\pm$0.02 &
  - &
  0.3567$^{+0.04}_{-0.035}$ &
  0.3702$\pm$0.0256 &
  0.4711$\pm$0.0339 &
  - &
  - &
  0.146$\pm$0.027 &
  0.16$\pm$0.03 &
  {\it this work}  \citenum{2018MNRAS.474.2334D} \citenum{2019arXiv190107040G} &
   &
   \\
WASP-104 b &
  - &
  - &
  - &
  0.1709$\pm$0.0195 &
  0.2643$\pm$0.0303 &
  - &
  - &
  - &
  - &
  \citenum{2019arXiv190107040G} &
   &
   \\
WASP-12 b &
  0.139$\pm$0.03 &
  0.176$^{+0.016}_{-0.021}$ &
  0.296$\pm$0.014 &
  0.421$\pm$0.011 &
  0.428$\pm$0.012 &
  0.696$\pm$0.06 &
  0.696$\pm$0.096 &
  0.198$\pm$0.022 &
  0.205$\pm$0.022 &
  {\it this work} \citenum{2011AJ....141...30C} \citenum{2015ApJ...802...28C} \citenum{2014ApJ...791...36S} &
   &
   \\
WASP-121 b &
  0.107$\pm$0.011 &
  - &
  - &
  0.367$\pm$0.013 &
  0.4684$\pm$0.0121 &
  - &
  - &
  0.125$\pm$0.011 &
  0.13$\pm$0.11 &
  {\it this work}   \citenum{2017Natur.548...58E} \citenum{2019arXiv190107040G} &
   &
   \\
WASP-131 b &
  - &
  - &
  - &
  0.0364$\pm$0.0097 &
  0.0282$\pm$0.0078 &
  - &
  - &
  - &
  - &
  \citenum{2019arXiv190107040G} &
   &
   \\
WASP-14 b &
  - &
  - &
  - &
  0.1816$\pm$0.0067 &
  0.2284$\pm$0.009 &
  - &
  0.181$\pm$0.022 &
  - &
  - &
  \citenum{2019arXiv190107040G} &
   &
   \\
WASP-17 b &
  - &
  - &
  - &
  - &
  0.229$\pm$0.013 &
  - &
  0.237$\pm$0.039 &
  - &
  - &
  \citenum{2011MNRAS.416.2108A} &
   &
   \\
WASP-18 b &
  0.094$\pm$0.004 &
  - &
  0.13$\pm$0.03 &
  0.2973$\pm$0.007 &
  0.3858$\pm$0.0113 &
  0.37$\pm$0.03 &
  0.41$\pm$0.02 &
  0.108$\pm$0.004 &
  0.113$\pm$0.004 &
  {\it this work}  \citenum{2015MNRAS.454.3002Z} \citenum{2017ApJ...850L..32S} &
   &
   \\
WASP-19 b &
  - &
  0.276$\pm$0.044 &
  0.287$\pm$0.02 &
  0.483$\pm$0.025 &
  0.572$\pm$0.03 &
  0.65$\pm$0.11 &
  0.73$\pm$0.12 &
  - &
  - &
  \citenum{2013MNRAS.430.3422A} \citenum{2014MNRAS.445.2746Z} \citenum{2013MNRAS.430.3422A} &
   &
   \\
WASP-2 b &
  - &
  - &
  - &
  0.083$\pm$0.035 &
  0.169$\pm$0.017 &
  0.192$\pm$0.077 &
  0.285$\pm$0.059 &
  - &
  - &
  \citenum{2010arXiv1004.0836W} &
   &
   \\
WASP-24 b &
  - &
  - &
  - &
  0.159$\pm$0.013 &
  0.202$\pm$0.018 &
  - &
  - &
  - &
  - &
  \citenum{2012AAA...545A..93S} &
   &
   \\
WASP-26 b &
  - &
  - &
  - &
  0.126$\pm$0.013 &
  0.149$\pm$0.016 &
  - &
  - &
  - &
  - &
  \citenum{2013MNRAS.432..693M} &
   &
   \\
WASP-3 b &
  - &
  - &
  0.193$\pm$0.014 &
  0.209$^{+0.04}_{-0.028}$ &
  0.282$\pm$0.012 &
  - &
  0.328$^{+0.086}_{-0.055}$ &
  - &
  - &
  \citenum{2015ApJ...802...28C} \citenum{2014MNRAS.441.3666R} &
   &
   \\
WASP-33 b &
  0.11$\pm$0.03 &
  - &
  0.244$^{+0.027}_{-0.02}$ &
  0.26$\pm$0.05 &
  0.41$\pm$0.02 &
  - &
  - &
  0.123$\pm$0.025 &
  0.13$\pm$0.03 &
  {\it this work}  \citenum{2013AAA...550A..54D} \citenum{2012ApJ...754..106D} &
   &
   \\
WASP-36 b &
  - &
  - &
  0.13$\pm$0.04 &
  - &
  - &
  - &
  - &
  - &
  - &
  \citenum{2015MNRAS.454.3002Z} &
   &
   \\
WASP-39 b &
  - &
  - &
  - &
  0.088$\pm$0.015 &
  0.096$\pm$0.018 &
  - &
  - &
  - &
  - &
  \citenum{2015ApJ...810..118K} &
   &
   \\
WASP-4 b &
  - &
  - &
  0.185$^{+0.014}_{-0.013}$ &
  0.319$\pm$0.031 &
  0.343$\pm$0.027 &
  - &
  - &
  - &
  - &
  \citenum{2011AAA...530A...5C} \citenum{2011ApJ...727...23B} &
   &
   \\
WASP-43 b &
  0.042$\pm$0.0045 &
  0.103$\pm$0.017 &
  0.194$\pm$0.029 &
  0.3773$\pm$0.0138 &
  0.3866$\pm$0.0195 &
  - &
  - &
  0.036$\pm$0.004 &
  0.056$\pm$0.004 &
  {\it this work} \citenum{2013ApJ...770...70W}  \citenum{2019arXiv190107040G} &
   &
   \\
WASP-46 b &
  0.129$\pm$0.055 &
  0.194$\pm$0.078 &
  0.26$^{+0.05}_{-0.03}$ &
  0.136$\pm$0.0701 &
  0.4446$\pm$0.0589 &
  - &
  - &
  - &
  - &
  \citenum{2014AAA...567A...8C} \citenum{2015MNRAS.454.3002Z} \citenum{2019arXiv190107040G} &
   &
   \\
WASP-48 b &
  - &
  0.047$\pm$0.016 &
  0.109$\pm$0.027 &
  0.176$\pm$0.013 &
  0.214$\pm$0.02 &
  - &
  - &
  - &
  - &
  \citenum{2014ApJ...781..109O}  \citenum{2014ApJ...781..109O} &
   &
   \\
WASP-5 b &
  0.168$^{+0.05}_{-0.052}$ &
  - &
  0.2$\pm$0.02 &
  0.197$\pm$0.028 &
  0.237$\pm$0.024 &
  - &
  - &
  - &
  - &
  \citenum{2014AAA...563A..40C}  \citenum{2015MNRAS.454.3002Z} \citenum{2013ApJ...773..124B} &
   &
   \\
WASP-6 b &
  - &
  - &
  - &
  0.094$\pm$0.019 &
  0.115$\pm$0.022 &
  - &
  - &
  - &
  - &
  \citenum{2015ApJ...810..118K} &
   &
   \\
WASP-62 b &
  - &
  - &
  - &
  0.1616$\pm$0.0146 &
  0.1359$\pm$0.013 &
  - &
  - &
  - &
  - &
  \citenum{2019arXiv190107040G} &
   &
   \\
WASP-63 b &
  - &
  - &
  - &
  0.0522$\pm$0.0095 &
  0.0533$\pm$0.0128 &
  - &
  - &
  - &
  - &
  \citenum{2019arXiv190107040G} &
   &
   \\
WASP-64 b &
  - &
  - &
  - &
  0.2859$\pm$0.027 &
  0.2071$\pm$0.0471 &
  - &
  - &
  - &
  - &
  \citenum{2019arXiv190107040G} &
   &
   \\
WASP-65 b &
  - &
  - &
  - &
  0.1587$\pm$0.0245 &
  0.0724$\pm$0.0318 &
  - &
  - &
  - &
  - &
  \citenum{2019arXiv190107040G} &
   &
   \\
WASP-67 b &
  - &
  - &
  - &
  0.022$\pm$0.013 &
  0.08$\pm$0.018 &
  - &
  - &
  - &
  - &
  \citenum{2015ApJ...810..118K} &
   &
   \\
WASP-69 b &
  - &
  - &
  - &
  0.0421$\pm$0.0029 &
  0.0463$\pm$0.0039 &
  - &
  - &
  - &
  - &
  \citenum{2019arXiv190800014W} &
   &
   \\
WASP-74 b &
  - &
  - &
  - &
  0.1446$\pm$0.0066 &
  0.2075$\pm$0.01 &
  - &
  - &
  - &
  - &
  \citenum{2019arXiv190107040G} &
   &
   \\
WASP-76 b &
  - &
  - &
  - &
  0.2645$\pm$0.0063 &
  0.3345$\pm$0.0082 &
  - &
  - &
  - &
  - &
  \citenum{2019arXiv190107040G} &
   &
   \\
WASP-77 b &
  - &
  - &
  - &
  0.1845$\pm$0.0094 &
  0.2362$\pm$0.0127 &
  - &
  - &
  - &
  - &
  \citenum{2019arXiv190107040G} &
   &
   \\
WASP-78 b &
  - &
  - &
  - &
  0.2001$\pm$0.0218 &
  0.2013$\pm$0.0351 &
  - &
  - &
  - &
  - &
  \citenum{2019arXiv190107040G} &
   &
   \\
WASP-79 b &
  - &
  - &
  - &
  0.1394$\pm$0.0088 &
  0.1783$\pm$0.0106 &
  - &
  - &
  - &
  - &
  \citenum{2019arXiv190107040G} &
   &
   \\
WASP-8 b &
  - &
  - &
  - &
  0.113$\pm$0.018 &
  0.0692$\pm$0.0068 &
  - &
  0.093$\pm$0.023 &
  - &
  - &
  \citenum{2013ApJ...768...42C} &
   &
   \\
WASP-80 b &
  - &
  - &
  - &
  0.455$\pm$0.1 &
  0.944$\pm$0.064 &
  - &
  - &
  - &
  - &
  \citenum{2015MNRAS.450.2279T} &
   &
   \\
WASP-87 b &
  - &
  - &
  - &
  0.2077$\pm$0.0127 &
  0.2705$\pm$0.0137 &
  - &
  - &
  - &
  - &
  \citenum{2019arXiv190107040G} &
   &
   \\
WASP-94A b &
  - &
  - &
  - &
  0.0867$\pm$0.0059 &
  0.0995$\pm$0.0093 &
  - &
  - &
  - &
  - &
  \citenum{2019arXiv190107040G} &
   &
   \\
WASP-97 b &
  - &
  - &
  - &
  0.1359$\pm$0.0084 &
  0.1534$\pm$0.0101 &
  - &
  - &
  - &
  - &
  \citenum{2019arXiv190107040G} &
   &
   \\
XO-1 b &
  - &
  - &
  - &
  0.086$\pm$0.007 &
  0.122$\pm$0.009 &
  0.261$\pm$0.031 &
  0.21$\pm$0.029 &
  - &
  - &
  \citenum{2008ApJ...684.1427M} &
   &
   \\
XO-2 N b &
  - &
  - &
  - &
  0.081$\pm$0.017 &
  0.098$\pm$0.02 &
  0.167$\pm$0.036 &
  0.133$\pm$0.049 &
  - &
  - &
  \citenum{2009ApJ...701..514M} &
   &
   \\
XO-3 b &
  - &
  - &
  - &
  0.101$\pm$0.004 &
  0.158$\pm$0.0036 &
  0.134$\pm$0.049 &
  0.15$\pm$0.036 &
  - &
  - &
  \citenum{2010ApJ...711..111M} \citenum{2014ApJ...794..134W} &
   &
   \\
XO-4 b &
  - &
  - &
  - &
  0.056$^{+0.012}_{-0.006}$ &
  0.135$\pm$0.01 &
  - &
  - &
  - &
  - &
  \citenum{2012ApJ...746..111T}  \\* \bottomrule
\caption{Planet secondary eclipse measurements in near- and mid-infrared bands, including our {\it W$_{JH}$} band and {\it H$_s$}.}
\label{tab:fluxes}\\
\end{longtable}
\end{landscape}
\twocolumn

\section{Updated Plots}
\label{sec:updates}

In this section we present updated versions of all near- and mid-infrared plots presented in \cite{2014MNRAS.444..711T}

\begin{figure*}
	\centering
	\includegraphics[width=\textwidth]{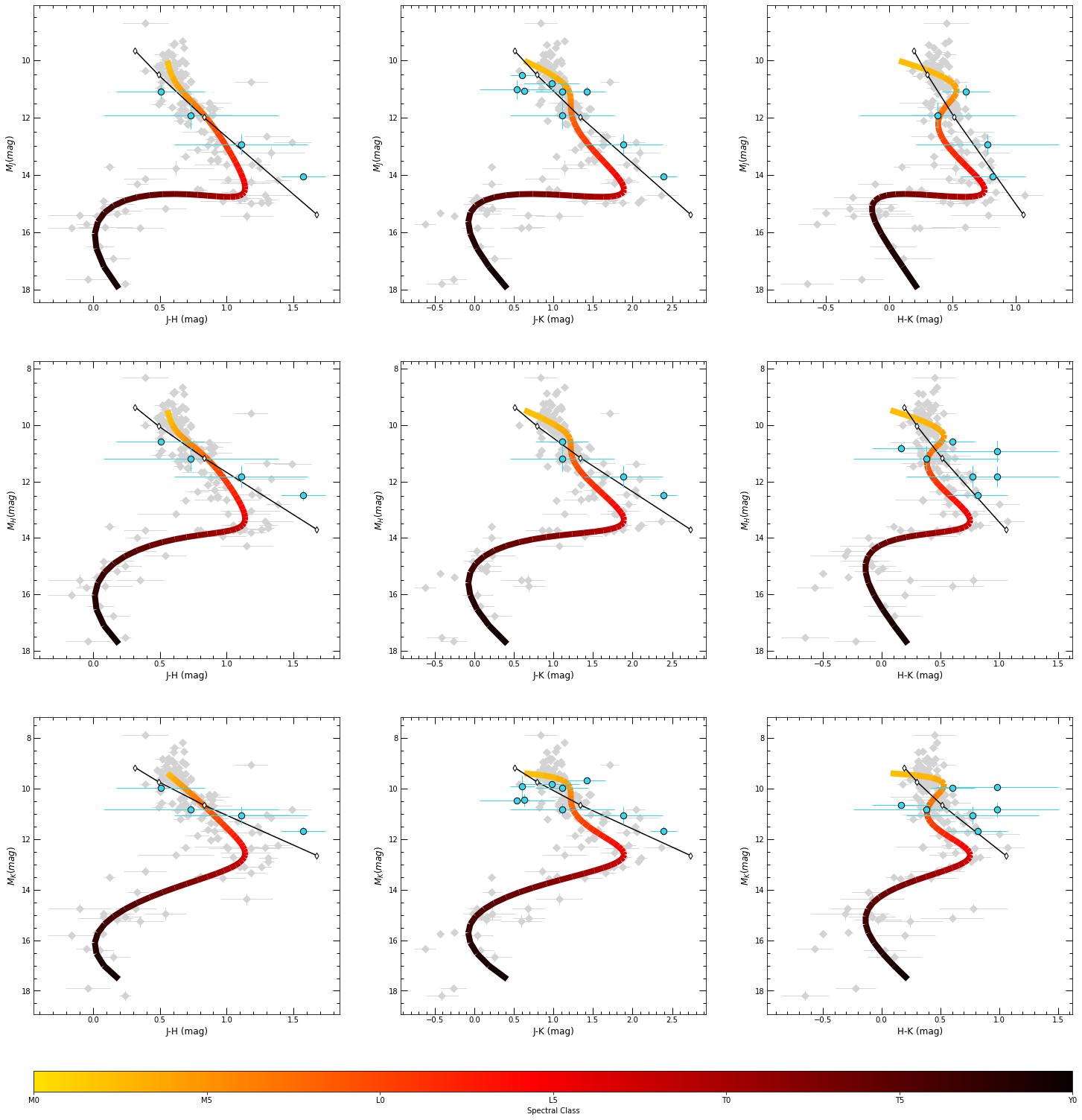}
    \caption{Updated near-infrared colour-magnitude diagrams in near-infrared photometric bands, similar to those first published in \protect\cite{2014MNRAS.444..711T}, plotted using our \textsc{ExoCMD\_2} plotting function. Magnitudes have been adjusted to coincide with an object of radius $\rm0.9 R_J$ for easier comparison with brown dwarfs. We have also plotted in black the location of a $\rm0.9 R_J$ blackbody as a black line; the white-filled diamonds show the position of the blackbody at temperatures of 1500K, 2500K, 3500K and 4500K. The polynomial representing the mean sequence of the brown dwarfs has been coloured according to spectral type.}
    \label{fig:nir}
\end{figure*}

In Figure \ref{fig:nir} we present updated colour-magnitude diagrams in the {\it 2MASS} photometric bands, {\it J}, {\it H} and {\it K}. The absolute magnitudes of planets have been scaled to a $\rm0.9R_J$ object coincide with the typical size of a brown dwarf. Additionally, we have plotted the location of a $\rm0.9R_J$ blackbody at temperatures of 1500K, 2500K, 3500K and 4500K. The mean position of the brown dwarf sequence is shown by a polynomial coloured by spectral type, computed using coefficients provided by \cite{2012ApJS..201...19D}. Due to the low number of {\it J}-band fluxes measured for exoplanets, we have supplemented the data with our photometry computed from HST/WFC3 low resolution spectra.

The small number of planets presented in \cite{2014MNRAS.444..711T} appeared to be equally compatible with the ultra-cool dwarfs and the blackbody sequence. This remains the case, which is in agreements also with the near-infrared colour-magnitude diagrams presented recently in \cite{2019AJ....157..101M}. \cite{2018AAA...617A.110P} writes that objects belonging to the subclass of ultra-hot Jupiters should lack any emission and absorption features, instead resembling a blackbody. At this time is is still impossible to disentangle which family of objects the planets resemble most; all objects which appear excessively blue or red have errors which make interpretations ambiguous. However, in Section \ref{sec:COratio} we see that in our newly created mid-infrared, the {\it $W_{JH}$}-band, we begin to see a departure from both the blackbody sequence and the narrow spread in colour of the M and L brown dwarfs.

\begin{figure*}
	\centering
	\includegraphics[width=\textwidth]{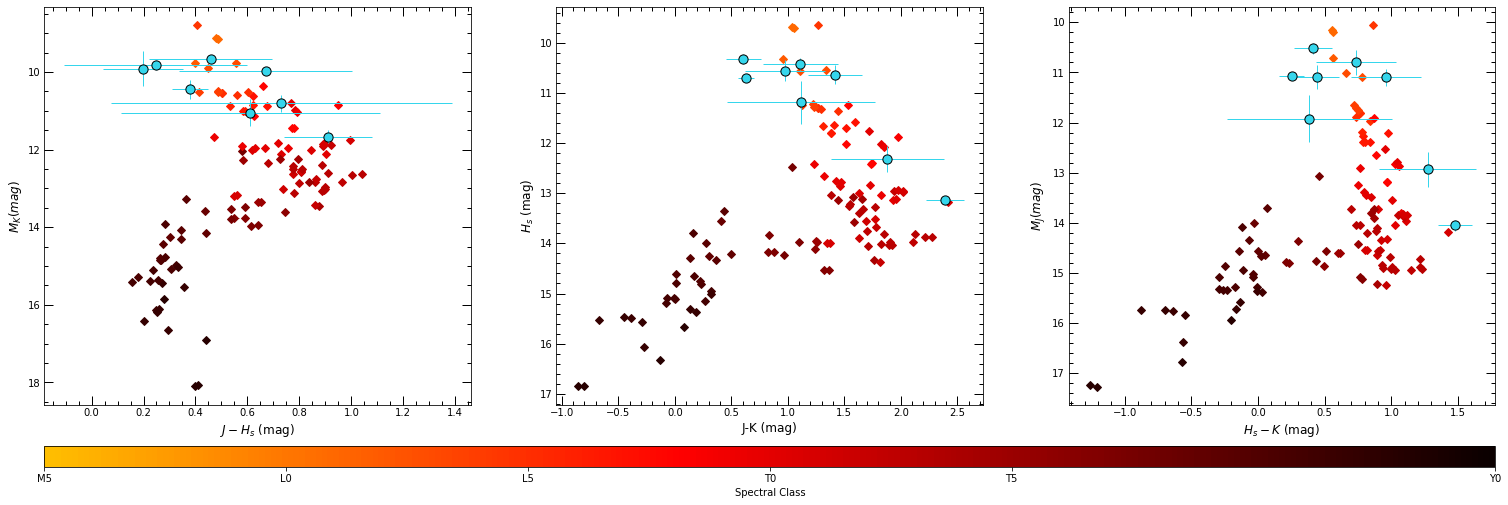}
    \caption{\textcolor{Black}{Near-infrared colour-magnitude diagrams plotted using our \textsc{ExoCMD\_synth}. Planets are shown as blue circles in the foreground, whereas ultra-cool dwarfs are plotted as diamonds in the background coloured according to spectral type. This is similar to Fig.~\ref{fig:nir}, but here using the {\it H$_s$} band instead.}}
    \label{fig:rising}
\end{figure*}

\textcolor{Black}{In Figure \ref{fig:rising} we present three near-infrared colour-magnitude diagrams produced using \textsc{ExoCMD\_synth} using {\it H$_s$)} photometry instead of {\it 2MASS H}. These plots represent the rising diagonal in Figure \ref{fig:nir} and we are able to showcase the positions of several more planets which have low resolution spectra measured with {\it HST}'s WFC3 instrument.}

\begin{figure*}
	\centering
	\includegraphics[width=\textwidth]{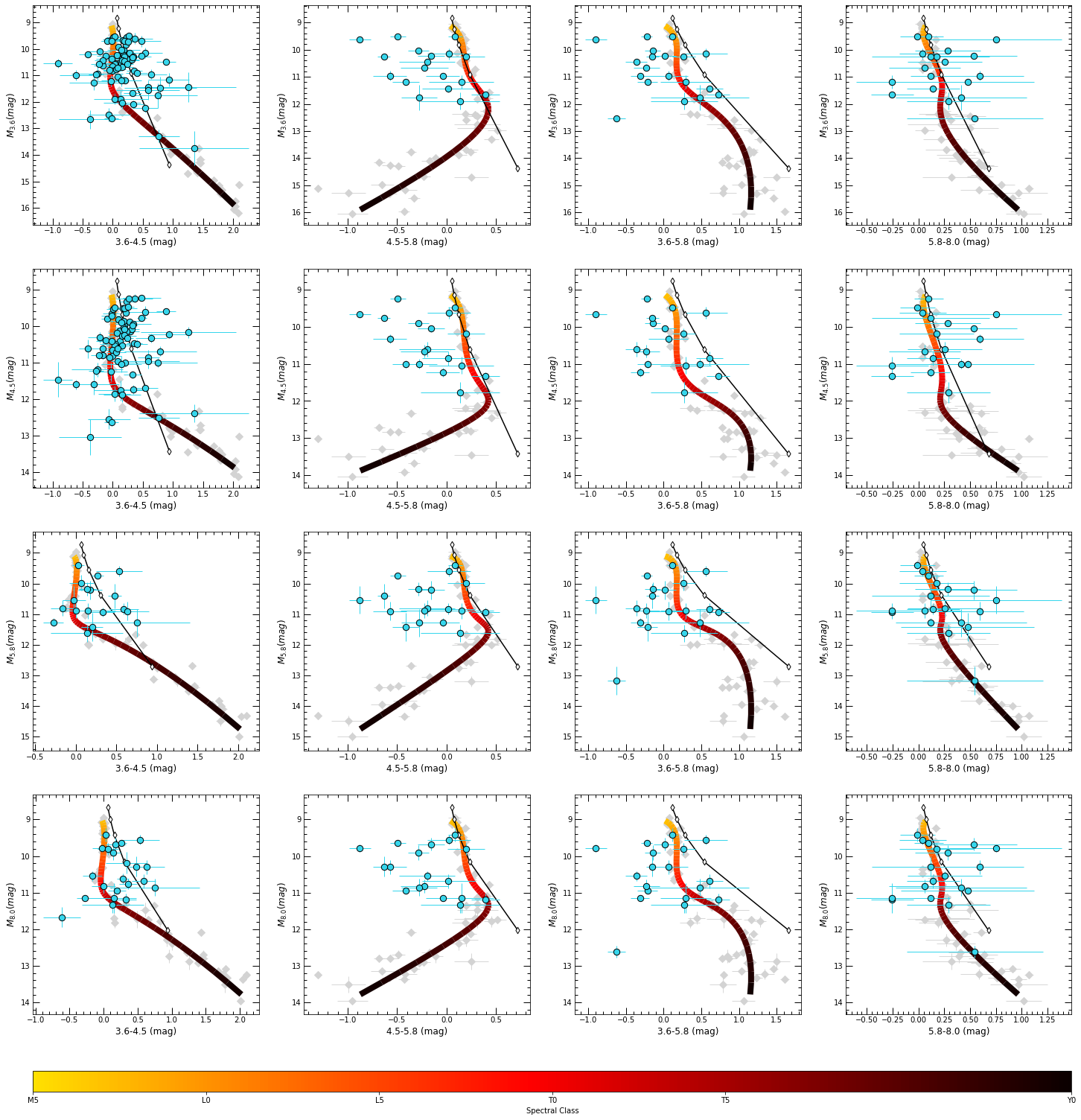}
    \caption{Updated near-infrared colour-magnitude diagrams in mid-infrared photometric bands, similar to those first published in \protect\cite{2014MNRAS.444..711T}, plotted using our \textsc{ExoCMD\_2} plotting function. Magnitudes have been adjusted to coincide with an object of radius $ \rm 0.9 R_J$ for more straight-forward comparison with the brown dwarf sequence. As before, we have also plotted a $0.9 \rm R_J$ blackbody with a black line, highlighting temperatures of 750K, 1750K, 2750K, 3750K and 4750K. The polynomial representing the mean sequence of the brown dwarfs has been coloured according to spectral type. }
    \label{fig:mir}
\end{figure*}

In Figure \ref{fig:mir} we present updated colour-magnitudes in {\it Spitzer's} mid-infrared channels 1-4. Absolute magnitudes have once again been scaled to facilitate comparison with brown dwarfs. In black we show the blackbody sequence, with the unfilled diamonds indicating the position of a $\rm0.9R_J$ blackbody of temperature 750K, 1750K, 2750K, 3750K and 4750K. We continue to plot the mean sequence of brown dwarfs using polynomial coefficients as above.

The biggest increase in measurements can be seen in the top two plots on the far left, principally due to the contribution from \cite{2019arXiv190107040G}. There is some increased scatter evident in these plots, and in $M_{4.5}$ vs [3.6$\mu$m - 4.5$\mu$m] in particular we are beginning to see an emerging sub-population of cooler objects which depart from both the brown dwarfs and the blackbodies. We explore this further in Section \ref{sec:outliers}.

\section{Brown Dwarf Photometry}
\label{sec:bd_photometry}

Here we present a summary of the photometry we computed from SpeX spectra for this paper, along with astrometric distances derived from parallaxes found in \cite{2012ApJS..201...19D}.

\onecolumn
\begin{landscape}
\begin{longtable}[c]{@{}lllllllllll@{}}
\toprule
\multicolumn{1}{c}{Name} &
  \multicolumn{1}{c}{Distance (pc)} &
  \multicolumn{8}{c}{Apparent magnitudes (mag)} &
  Refs \\* \midrule
\endfirsthead
\multicolumn{11}{c}%
{{\bfseries Table \thetable\ continued from previous page}} \\
\endhead
\bottomrule
\endfoot
\endlastfoot
 &
   &
  m$_J$ &
  m$_H$ &
  m$_K$ &
  m$_{W_{JH}}$ &
  m$_{H_s}$ &
  m$_{NB1190}$ &
  m$_{NB2090}$ &
  m$_{z'}$ &
   \\
2MASP J0345432+254023 &
  26.95$\pm$0.36 &
  13.96$\pm$0 &
  13.23$\pm$0 &
  12.69$\pm$0 &
  13.89$\pm$0 &
  13.46$\pm$0 &
  14.14$\pm$0 &
  12.84$\pm$0 &
  16.02$\pm$0 &
  \citenum{2006AJ....131.1007B}, \citenum{2002AJ....124.1170D} \\
2MASS J00345157+0523050 &
  38.3$\pm$2.76 &
  15.12$\pm$0 &
  14.23$\pm$0 &
  13.52$\pm$0 &
  15.16$\pm$0 &
  14.5$\pm$0 &
  15.4$\pm$0 &
  13.7$\pm$0 &
  17.58$\pm$0 &
  \citenum{2004AJ....127.2856B}, \citenum{2016ApJ...833...96L} \\
2MASS J00501994-3322402 &
  10.68$\pm$0.41 &
  15.42$\pm$0 &
  15.2$\pm$0 &
  15.1$\pm$0 &
  17.4$\pm$0.01 &
  15.1$\pm$0 &
  16.45$\pm$0 &
  14.51$\pm$0 &
  18.6$\pm$0 &
  \citenum{2006ApJ...639.1095B}, \citenum{2004AJ....127.2948V} \\
2MASS J02572581-3105523 &
  30.18$\pm$3.83 &
  16.61$\pm$0 &
  15.56$\pm$0 &
  14.98$\pm$0 &
  16.93$\pm$0.01 &
  15.8$\pm$0 &
  17.01$\pm$0.01 &
  14.97$\pm$0 &
  18.97$\pm$0.01 &
  \citenum{2007AJ....133.2320S}, \citenum{2004AJ....127.2948V} \\
2MASS J04070752+1546457 &
  5.71$\pm$0.06 &
  15.55$\pm$0 &
  15.38$\pm$0 &
  15.68$\pm$0 &
  18.34$\pm$0 &
  15.11$\pm$0 &
  17.2$\pm$0 &
  14.92$\pm$0 &
  18.88$\pm$0 &
  \citenum{2008ApJ...681..579B}, \citenum{2012ApJS..201...19D} \\
2MASS J05160945-0445499 &
  9.06$\pm$0.33 &
  14.42$\pm$0 &
  13.43$\pm$0 &
  12.79$\pm$0 &
  14.62$\pm$0 &
  13.68$\pm$0 &
  14.79$\pm$0.01 &
  12.85$\pm$0 &
  16.66$\pm$0 &
  \citenum{2008ApJ...681..579B}, \citenum{2012ApJ...752...56F} \\
2MASS J06244595-4521548 &
  8.7$\pm$0.3 &
  13.49$\pm$0 &
  12.43$\pm$0 &
  11.67$\pm$0 &
  13.59$\pm$0 &
  12.74$\pm$0 &
  13.74$\pm$0 &
  11.79$\pm$0 &
  15.97$\pm$0 &
  \citenum{2007AJ....133.2320S}, \citenum{2016ApJS..225...10F} \\
2MASS J09393548-2448279 &
  8.89$\pm$0.07 &
  15.68$\pm$0 &
  15.61$\pm$0 &
  15.61$\pm$0 &
  18.54$\pm$0.02 &
  15.4$\pm$0 &
  17.02$\pm$0.01 &
  14.9$\pm$0 &
  19.04$\pm$0 &
  \citenum{2006ApJ...637.1067B}, \citenum{2012ApJS..201...19D} \\
2MASS J09490860-1545485 &
  16.96$\pm$0.28 &
  13.53$\pm$0 &
  12.66$\pm$0 &
  12.01$\pm$0 &
  13.38$\pm$0 &
  14.11$\pm$0 &
  13.77$\pm$0.01 &
  12.15$\pm$0 &
  15.74$\pm$0 &
  \citenum{2006ApJ...637.1067B}, \citenum{2014AJ....147...94D} \\
2MASS J10073369-4555147 &
  16.72$\pm$2.27 &
  15.44$\pm$0 &
  14.4$\pm$0 &
  13.67$\pm$0 &
  15.42$\pm$0 &
  12.85$\pm$0 &
  15.84$\pm$0 &
  13.77$\pm$0 &
  17.78$\pm$0 &
  \citenum{2007AJ....134.1162L}, \citenum{2012ApJ...752...56F} \\
2MASS J11061197+2754225 &
  10.56$\pm$0.42 &
  15.94$\pm$0 &
  15.82$\pm$0 &
  16.23$\pm$0 &
  18.33$\pm$0.01 &
  14.47$\pm$0 &
  16.98$\pm$0.01 &
  15.65$\pm$0.01 &
  19.11$\pm$0 &
  \citenum{2007AJ....134.1162L}, \citenum{2004AJ....127.2948V} \\
2MASS J11263991-5003550 &
  11.01$\pm$0.27 &
  15.93$\pm$0 &
  15.81$\pm$0.01 &
  15.74$\pm$0.01 &
  18.11$\pm$0.02 &
  14.63$\pm$0 &
  17.36$\pm$0.03 &
  15.02$\pm$0.01 &
  19.58$\pm$0 &
  \citenum{2008ApJ...674..451B}, \citenum{2003AJ....126..975T} \\
2MASS J12373919+6526148 &
  30.2$\pm$4.39 &
  16.73$\pm$0 &
  15.74$\pm$0.01 &
  15.04$\pm$0 &
  16.97$\pm$0.01 &
  15.68$\pm$0 &
  17.26$\pm$0.02 &
  15.12$\pm$0.01 &
  19.08$\pm$0.01 &
  \citenum{2007ApJ...655..522L}, \citenum{2004AJ....127.2948V} \\
2MASS J13313310+3407583 &
  12.94$\pm$0.25 &
  12.95$\pm$0 &
  12.15$\pm$0 &
  11.56$\pm$0 &
  12.96$\pm$0 &
  15.57$\pm$0.01 &
  13.31$\pm$0 &
  11.7$\pm$0 &
  15.09$\pm$0 &
  \citenum{2010ApJS..190..100K}, \citenum{2014PASP..126...15W} \\
2MASS J15200224-4422419A &
  12.94$\pm$0.25 &
  12.95$\pm$0 &
  12.15$\pm$0 &
  11.57$\pm$0 &
  12.97$\pm$0 &
  15.96$\pm$0.01 &
  13.15$\pm$0 &
  11.71$\pm$0 &
  15.1$\pm$0 &
  \citenum{2007ApJ...658..557B}, \citenum{2014PASP..126...15W} \\
2MASS J15200224-4422419B &
  17.49$\pm$0.28 &
  13.86$\pm$0 &
  12.98$\pm$0 &
  12.35$\pm$0 &
  13.84$\pm$0 &
  12.36$\pm$0 &
  14.08$\pm$0.01 &
  12.49$\pm$0.01 &
  16.06$\pm$0 &
  \citenum{2007ApJ...658..557B}, \citenum{2016AJ....152...24W} \\
2MASS J15462718-3325111 &
  14.08$\pm$0.6 &
  15.39$\pm$0 &
  14.88$\pm$0 &
  14.9$\pm$0 &
  16.72$\pm$0 &
  13.23$\pm$0 &
  16.08$\pm$0 &
  14.52$\pm$0 &
  18.46$\pm$0 &
  \citenum{2008ApJ...681..579B}, \citenum{2013AAA...560A..52M} \\
2MASS J16150413+1340079 &
  14.5$\pm$0.72 &
  15.88$\pm$0 &
  15.69$\pm$0 &
  15.65$\pm$0 &
  17.72$\pm$0.01 &
  14.95$\pm$0 &
  16.79$\pm$0.01 &
  15.06$\pm$0 &
  18.95$\pm$0 &
  \citenum{2007AJ....134.1162L}, \citenum{2004AJ....127.2948V} \\
2MASS J17502484-0016151 &
  9.49$\pm$0.68 &
  15.63$\pm$0 &
  15.54$\pm$0 &
  16.02$\pm$0 &
  18.02$\pm$0.01 &
  15.61$\pm$0 &
  17.02$\pm$0 &
  15.42$\pm$0 &
  18.85$\pm$0 &
  \citenum{2010ApJ...710.1142B}, \citenum{2012ApJ...752...56F} \\
2MASS J18212815+1414010 &
  10.57$\pm$0.27 &
  15.7$\pm$0 &
  15.65$\pm$0 &
  15.57$\pm$0.01 &
  18.01$\pm$0.02 &
  15.38$\pm$0 &
  17.16$\pm$0.02 &
  14.81$\pm$0.01 &
  19.07$\pm$0 &
  \citenum{2008ApJ...686..528L}, \citenum{2012ApJS..201...19D} \\
2MASS J18283572-4849046 &
  10.03$\pm$0.67 &
  14.68$\pm$0 &
  13.55$\pm$0 &
  12.83$\pm$0 &
  14.7$\pm$0 &
  15.43$\pm$0.01 &
  15.11$\pm$0 &
  12.87$\pm$0 &
  16.96$\pm$0 &
  \citenum{2004AJ....127.2856B}, \citenum{2013AJ....146..161M} \\
2MASS J21392676+0220226 &
  34.58$\pm$1.49 &
  15.52$\pm$0.01 &
  14.33$\pm$0.01 &
  13.55$\pm$0.01 &
  15.36$\pm$0.01 &
  13.82$\pm$0 &
  15.96$\pm$0.02 &
  13.69$\pm$0.03 &
  18.18$\pm$0.01 &
  \citenum{2006ApJ...637.1067B}, \citenum{2016ApJ...833...96L} \\
2MASS J21403907+3655563 &
  12.31$\pm$0.07 &
  14.49$\pm$0 &
  13.35$\pm$0 &
  12.57$\pm$0 &
  14.46$\pm$0 &
  14.57$\pm$0.01 &
  14.76$\pm$0 &
  12.66$\pm$0 &
  16.98$\pm$0 &
  \citenum{2010ApJS..190..100K}, \citenum{2016AJ....152...24W} \\
2MASS J21481633+4003594 &
  5.34$\pm$0.13 &
  15.87$\pm$0.01 &
  15.7$\pm$0.01 &
  16.73$\pm$0.01 &
  18.59$\pm$0.03 &
  15.55$\pm$0 &
  17.36$\pm$0.02 &
  16.07$\pm$0.04 &
  19.11$\pm$0.01 &
  \citenum{2010ApJS..190..100K}, \citenum{2008ApJ...689L..53B} \\
2MASS J21513839-4853542 &
  5.34$\pm$0.13 &
  15.9$\pm$0 &
  15.72$\pm$0 &
  16.7$\pm$0.01 &
  18.5$\pm$0.02 &
  13.59$\pm$0 &
  17.31$\pm$0.02 &
  16.17$\pm$0.02 &
  18.89$\pm$0 &
  \citenum{2006ApJ...637.1067B}, \citenum{2008ApJ...689L..53B} \\
2MASS J21542494-1023022 &
  18.08$\pm$2.16 &
  16.07$\pm$0 &
  15.32$\pm$0 &
  15.24$\pm$0 &
  16.87$\pm$0.01 &
  15.49$\pm$0 &
  16.45$\pm$0.01 &
  14.99$\pm$0.01 &
  18.99$\pm$0 &
  \citenum{2007AJ....134.1162L}, \citenum{2012ApJ...752...56F} \\
2MASS J22282889-4310262 &
  18.08$\pm$2.16 &
  16.09$\pm$0 &
  15.32$\pm$0 &
  15.22$\pm$0 &
  16.78$\pm$0 &
  15.45$\pm$0 &
  16.52$\pm$0 &
  14.94$\pm$0 &
  18.99$\pm$0 &
  \citenum{2004AJ....127.2856B}, \citenum{2012ApJ...752...56F} \\
2MASS J22425317+2542573 &
  14.08$\pm$1.03 &
  15.76$\pm$0 &
  15.6$\pm$0.01 &
  15.53$\pm$0.01 &
  17.58$\pm$0.02 &
  15.49$\pm$0.01 &
  16.76$\pm$0.02 &
  14.91$\pm$0.02 &
  18.81$\pm$0.01 &
  \citenum{2010ApJ...710.1142B}, \citenum{2012ApJ...752...56F} \\
2MASS J23512200+3010540 &
  16.84$\pm$0.47 &
  14.03$\pm$0 &
  13.29$\pm$0 &
  12.79$\pm$0 &
  14.43$\pm$0 &
  14.26$\pm$0 &
  14.29$\pm$0 &
  12.9$\pm$0.01 &
  16.36$\pm$0 &
  \citenum{2010ApJS..190..100K}, \citenum{2014AJ....147...94D} \\
2MASSI J0117474-340325 &
  10.41$\pm$0.52 &
  15.82$\pm$0 &
  15.73$\pm$0 &
  16.26$\pm$0.01 &
  18.18$\pm$0.01 &
  13.56$\pm$0 &
  17.07$\pm$0.01 &
  15.73$\pm$0.01 &
  19.03$\pm$0 &
  \citenum{2008ApJ...681..579B}, \citenum{2004AJ....127.2948V} \\
2MASSI J0243137-245329 &
  28.11$\pm$1.23 &
  14.21$\pm$0 &
  13.39$\pm$0 &
  12.76$\pm$0 &
  14.01$\pm$0 &
  15.57$\pm$0 &
  14.39$\pm$0 &
  12.92$\pm$0 &
  16.35$\pm$0 &
  \citenum{2004AJ....127.2856B}, \citenum{2016ApJ...833...96L} \\
2MASSI J0328426+230205 &
  18.52$\pm$0.04 &
  13.6$\pm$0 &
  12.76$\pm$0 &
  12.19$\pm$0 &
  13.59$\pm$0 &
  13.6$\pm$0 &
  13.87$\pm$0 &
  12.33$\pm$0 &
  16.01$\pm$0 &
  \citenum{2008ApJ...681..579B}, \citenum{2014AAA...565A..20S} \\
2MASSI J0415195-093506 &
  18.52$\pm$0.04 &
  14.58$\pm$0 &
  13.78$\pm$0 &
  13.25$\pm$0 &
  14.75$\pm$0 &
  12.97$\pm$0 &
  14.83$\pm$0 &
  13.35$\pm$0 &
  16.54$\pm$0 &
  \citenum{2004AJ....127.2856B}, \citenum{2014AAA...565A..20S} \\
2MASSI J0439010-235308 &
  11.36$\pm$0.25 &
  15.57$\pm$0 &
  15.43$\pm$0 &
  15.56$\pm$0.01 &
  17.64$\pm$0.01 &
  14$\pm$0 &
  16.51$\pm$0.01 &
  15.01$\pm$0.01 &
  18.68$\pm$0 &
  \citenum{2007ApJ...659..655B}, \citenum{2003AJ....126..975T} \\
2MASSI J0652307+471034 &
  14.58$\pm$1.36 &
  16.16$\pm$0 &
  16.16$\pm$0.01 &
  16.24$\pm$0.01 &
  18.24$\pm$0.02 &
  15.39$\pm$0 &
  17.15$\pm$0.02 &
  15.64$\pm$0.01 &
  19.2$\pm$0.01 &
  \citenum{2010ApJ...710.1142B}, \citenum{2012ApJ...752...56F} \\
2MASSI J0727182+171001 &
  9.22$\pm$0.22 &
  13.3$\pm$0 &
  12.41$\pm$0 &
  11.84$\pm$0 &
  13.63$\pm$0 &
  16.01$\pm$0.01 &
  13.69$\pm$0 &
  11.93$\pm$0 &
  15.52$\pm$0 &
  \citenum{2006ApJ...639.1095B}, \citenum{2011AJ....141...54A} \\
2MASSI J0825196+211552 &
  9.38$\pm$0.02 &
  13.4$\pm$0 &
  12.39$\pm$0 &
  11.68$\pm$0 &
  13.57$\pm$0 &
  12.69$\pm$0 &
  13.8$\pm$0 &
  11.83$\pm$0 &
  15.78$\pm$0 &
  \citenum{2010ApJ...710.1142B}, \citenum{2016MNRAS.455..357S} \\
2MASSI J0847287-153237 &
  11.38$\pm$0.26 &
  15.21$\pm$0 &
  15$\pm$0 &
  15.04$\pm$0 &
  17.17$\pm$0 &
  12.69$\pm$0 &
  16.17$\pm$0.01 &
  14.53$\pm$0.01 &
  18.27$\pm$0 &
  \citenum{2006AJ....132.2074M}, \citenum{2013MNRAS.433.2054S} \\
2MASSI J0937347+293142 &
  9.85$\pm$0.19 &
  15.05$\pm$0.01 &
  14.19$\pm$0.01 &
  13.74$\pm$0.01 &
  15.9$\pm$0.01 &
  14.93$\pm$0 &
  15.66$\pm$0.04 &
  13.66$\pm$0.03 &
  17.96$\pm$0 &
  \citenum{2006ApJ...639.1095B}, \citenum{2013MNRAS.433.2054S} \\
2MASSI J1010148-040649 &
  9.85$\pm$0.19 &
  15.09$\pm$0 &
  14.17$\pm$0 &
  13.72$\pm$0 &
  15.93$\pm$0 &
  14.5$\pm$0.02 &
  15.52$\pm$0 &
  13.63$\pm$0 &
  17.95$\pm$0 &
  \citenum{2006ApJ...639.1114R}, \citenum{2013MNRAS.433.2054S} \\
2MASSI J1047538+212423 &
  87.49$\pm$7.65 &
  15.8$\pm$0 &
  15.05$\pm$0 &
  14.47$\pm$0 &
  15.7$\pm$0.01 &
  15.24$\pm$0.01 &
  15.97$\pm$0.01 &
  14.62$\pm$0.01 &
  17.38$\pm$0.01 &
  \citenum{2008ApJ...681..579B}, \citenum{2016ApJ...833...96L} \\
2MASSI J1217110-031113 &
  9.9$\pm$0.17 &
  14.15$\pm$0 &
  12.81$\pm$0 &
  11.73$\pm$0 &
  14.07$\pm$0 &
  13.16$\pm$0 &
  14.64$\pm$0 &
  11.9$\pm$0 &
  16.88$\pm$0 &
  \citenum{2006ApJ...639.1095B}, \citenum{2016ApJ...833...96L} \\
2MASSI J1711457+223204 &
  16.67$\pm$1.06 &
  15.66$\pm$0 &
  15.29$\pm$0 &
  15.36$\pm$0 &
  17.07$\pm$0.01 &
  15.36$\pm$0 &
  16.31$\pm$0 &
  14.94$\pm$0 &
  18.9$\pm$0 &
  \citenum{2010ApJ...710.1142B}, \citenum{2013MNRAS.433.2054S} \\
2MASSI J1807159+501531 &
  10.86$\pm$0.31 &
  15.56$\pm$0 &
  15.47$\pm$0 &
  15.29$\pm$0 &
  17.44$\pm$0 &
  16.47$\pm$0 &
  16.68$\pm$0.02 &
  14.59$\pm$0.02 &
  18.76$\pm$0 &
  \citenum{2008ApJ...681..579B}, \citenum{2013MNRAS.433.2054S} \\
2MASSI J2104149-103736 &
  20.86$\pm$1.19 &
  14.77$\pm$0 &
  13.79$\pm$0 &
  13.04$\pm$0.01 &
  14.59$\pm$0.01 &
  15.32$\pm$0.01 &
  15.19$\pm$0.01 &
  13.17$\pm$0.04 &
  17.06$\pm$0.01 &
  \citenum{2010ApJ...710.1142B}, \citenum{2014PASP..126...15W} \\
2MASSI J2254188+312349 &
  20.86$\pm$1.19 &
  14.78$\pm$0 &
  13.8$\pm$0 &
  13.03$\pm$0 &
  14.65$\pm$0 &
  13.99$\pm$0 &
  15.08$\pm$0 &
  13.18$\pm$0 &
  17.05$\pm$0 &
  \citenum{2004AJ....127.2856B}, \citenum{2014PASP..126...15W} \\
2MASSI J2356547-155310 &
  24.26$\pm$0.83 &
  15.89$\pm$0 &
  14.8$\pm$0 &
  13.94$\pm$0 &
  15.85$\pm$0 &
  15.05$\pm$0 &
  16.23$\pm$0 &
  14.06$\pm$0 &
  18.29$\pm$0 &
  \citenum{2006ApJ...637.1067B}, \citenum{2016ApJ...833...96L} \\
2MASSs J0850359+105716 &
  7.47$\pm$0.03 &
  12.8$\pm$0 &
  11.91$\pm$0 &
  11.33$\pm$0 &
  12.9$\pm$0 &
  13.87$\pm$0 &
  13.13$\pm$0.01 &
  11.42$\pm$0.01 &
  15.06$\pm$0 &
  \citenum{2011AJ....141...70B}, \citenum{2016AJ....152...24W} \\
2MASSW J1239272+551537 &
  15.24$\pm$0.49 &
  15.84$\pm$0 &
  14.69$\pm$0 &
  13.95$\pm$0 &
  15.95$\pm$0.01 &
  12.13$\pm$0.01 &
  16.22$\pm$0.01 &
  13.98$\pm$0.01 &
  18.03$\pm$0 &
  \citenum{2010ApJ...710.1142B}, \citenum{2002AJ....124.1170D} \\
2MASSW J1507476-162738 &
  15.24$\pm$0.49 &
  15.85$\pm$0 &
  14.7$\pm$0 &
  13.92$\pm$0 &
  15.82$\pm$0 &
  14.95$\pm$0 &
  16.26$\pm$0 &
  13.96$\pm$0.01 &
  18.15$\pm$0 &
  \citenum{2007ApJ...659..655B}, \citenum{2002AJ....124.1170D} \\
2MASSW J1632291+190441 &
  25.84$\pm$0.47 &
  15.95$\pm$0 &
  14.77$\pm$0 &
  13.93$\pm$0 &
  15.93$\pm$0 &
  15.02$\pm$0 &
  16.35$\pm$0.01 &
  14.01$\pm$0 &
  18.43$\pm$0.01 &
  \citenum{2007ApJ...659..655B}, \citenum{2012ApJS..201...19D} \\
2MASSW J1728114+394859 &
  33.22$\pm$0.88 &
  16.45$\pm$0 &
  15.27$\pm$0 &
  14.44$\pm$0 &
  16.51$\pm$0.01 &
  15.55$\pm$0.01 &
  16.77$\pm$0.01 &
  14.55$\pm$0.01 &
  18.78$\pm$0.01 &
  \citenum{2011AJ....141...70B}, \citenum{2012ApJS..201...19D} \\
DENIS J124514.1-442907 &
  78.99$\pm$12.92 &
  14.55$\pm$0 &
  13.85$\pm$0 &
  13.28$\pm$0 &
  14.59$\pm$0 &
  14.14$\pm$0 &
  14.73$\pm$0 &
  13.46$\pm$0.01 &
  16.63$\pm$0 &
  \citenum{2007ApJ...669L..97L}, \citenum{2013ApJ...762..118W} \\
GJ 1001B &
  12.98$\pm$0.35 &
  13.09$\pm$0 &
  12.09$\pm$0 &
  11.37$\pm$0 &
  13.12$\pm$0 &
  12.33$\pm$0 &
  13.41$\pm$0 &
  11.5$\pm$0 &
  15.54$\pm$0 &
  \citenum{2007ApJ...659..655B}, \citenum{2014AJ....147...94D} \\
Gl 337CD &
  20.36$\pm$0.22 &
  15.58$\pm$0 &
  14.56$\pm$0 &
  14.03$\pm$0 &
  15.82$\pm$0 &
  14.77$\pm$0.01 &
  15.98$\pm$0 &
  14.01$\pm$0 &
  17.9$\pm$0 &
  \citenum{2010ApJ...710.1142B}, \citenum{2007AAA...474..653V} \\
Gl 584C &
  17.86$\pm$0.25 &
  16.13$\pm$0 &
  14.96$\pm$0 &
  14.23$\pm$0 &
  16.2$\pm$0.01 &
  15.23$\pm$0 &
  16.38$\pm$0.01 &
  14.26$\pm$0.01 &
  18.49$\pm$0.01 &
  \citenum{2010ApJ...710.1142B}, \citenum{2007AAA...474..653V} \\
Gliese 417BC &
  21.93$\pm$0.21 &
  14.58$\pm$0 &
  13.5$\pm$0 &
  12.73$\pm$0 &
  14.62$\pm$0 &
  13.78$\pm$0 &
  14.99$\pm$0 &
  12.89$\pm$0 &
  17.08$\pm$0 &
  \citenum{2010ApJ...710.1142B}, \citenum{2007AAA...474..653V} \\
Gliese 570D &
  5.84$\pm$0.03 &
  15.2$\pm$0 &
  15.12$\pm$0.01 &
  15.47$\pm$0 &
  18.13$\pm$0 &
  14.91$\pm$0.01 &
  16.79$\pm$0.02 &
  14.73$\pm$0.01 &
  18.39$\pm$0 &
  \citenum{2004AJ....127.2856B}, \citenum{2007AAA...474..653V} \\
HD 89744B &
  39.43$\pm$0.48 &
  14.86$\pm$0 &
  14.1$\pm$0 &
  13.56$\pm$0 &
  14.8$\pm$0 &
  14.3$\pm$0 &
  15.19$\pm$0 &
  13.73$\pm$0 &
  16.92$\pm$0 &
  \citenum{2008ApJ...681..579B}, \citenum{2007AAA...474..653V} \\
HN Peg &
  17.89$\pm$0.14 &
  16.02$\pm$0 &
  15.28$\pm$0 &
  15.05$\pm$0 &
  17.21$\pm$0 &
  15.49$\pm$0 &
  16.54$\pm$0 &
  14.84$\pm$0 &
  18.85$\pm$0 &
  \citenum{2007ApJ...654..570L}, \citenum{2007AAA...474..653V} \\
LHS 2924 &
  11.01$\pm$0.16 &
  11.97$\pm$0 &
  11.26$\pm$0 &
  10.73$\pm$0 &
  11.87$\pm$0 &
  11.48$\pm$0 &
  12.1$\pm$0 &
  10.88$\pm$0 &
  13.91$\pm$0 &
  \citenum{2006AJ....131.1007B}, \citenum{1992AJ....103..638M} \\
LHS 3566 &
  17.46$\pm$0.82 &
  11.37$\pm$0 &
  10.76$\pm$0 &
  10.34$\pm$0 &
  11.2$\pm$0 &
  10.89$\pm$0 &
  11.59$\pm$0 &
  10.46$\pm$0 &
  12.76$\pm$0 &
  \citenum{2004AJ....127.2856B}, \citenum{2005AJ....130..337C} \\
LP 944-20 &
  6.41$\pm$0.04 &
  10.74$\pm$0 &
  10.04$\pm$0 &
  9.52$\pm$0 &
  10.72$\pm$0 &
  10.28$\pm$0 &
  11.05$\pm$0 &
  9.68$\pm$0 &
  12.72$\pm$0 &
  \citenum{2008ApJ...681..579B}, \citenum{2014AJ....147...94D} \\
SDSS J000013.54+255418.6 &
  14.12$\pm$0.38 &
  15.1$\pm$0 &
  14.71$\pm$0 &
  14.82$\pm$0 &
  16.81$\pm$0 &
  14.75$\pm$0 &
  15.77$\pm$0 &
  14.36$\pm$0 &
  18.04$\pm$0 &
  \citenum{2006ApJ...637.1067B}, \citenum{2012ApJS..201...19D} \\
SDSS J015141.69+124429.6 &
  21.4$\pm$1.54 &
  16.51$\pm$0 &
  15.57$\pm$0 &
  15.27$\pm$0 &
  16.94$\pm$0.01 &
  15.77$\pm$0 &
  16.75$\pm$0 &
  15.17$\pm$0 &
  18.93$\pm$0 &
  \citenum{2004AJ....127.2856B}, \citenum{2004AJ....127.2948V} \\
SDSS J020742.48+000056.2 &
  34.13$\pm$4.66 &
  16.38$\pm$0.01 &
  15.99$\pm$0.01 &
  15.95$\pm$0.01 &
  17.92$\pm$0.02 &
  16.01$\pm$0.01 &
  17$\pm$0.02 &
  15.57$\pm$0.03 &
  19.56$\pm$0.01 &
  \citenum{2006ApJ...637.1067B}, \citenum{2010AAA...524A..38M} \\
SDSS J032553.17+042540.1 &
  17.99$\pm$3.53 &
  16.36$\pm$0 &
  16.12$\pm$0 &
  16.35$\pm$0 &
  18.24$\pm$0 &
  16.06$\pm$0 &
  17.3$\pm$0.02 &
  15.85$\pm$0 &
  19.51$\pm$0 &
  \citenum{2006AJ....131.2722C}, \citenum{2012ApJ...752...56F} \\
SDSS J074201.41+205520.5 &
  15.04$\pm$1.94 &
  15.75$\pm$0 &
  15.56$\pm$0 &
  15.74$\pm$0 &
  17.77$\pm$0.01 &
  15.49$\pm$0 &
  16.56$\pm$0.01 &
  15.22$\pm$0.01 &
  18.7$\pm$0 &
  \citenum{2010ApJ...710.1142B}, \citenum{2012ApJ...752...56F} \\
SDSS J083048.80+012831.1 &
  23.2$\pm$3.28 &
  16.39$\pm$0 &
  16.12$\pm$0 &
  16.26$\pm$0 &
  18.36$\pm$0.01 &
  16.12$\pm$0 &
  17.14$\pm$0.01 &
  15.79$\pm$0.01 &
  19.46$\pm$0 &
  \citenum{2010ApJ...710.1142B}, \citenum{2012ApJ...752...56F} \\
SDSS J103026.78+021306.4 &
  40.32$\pm$17.23 &
  16.93$\pm$0 &
  15.89$\pm$0 &
  15.27$\pm$0 &
  17.11$\pm$0.01 &
  16.13$\pm$0.01 &
  17.33$\pm$0.01 &
  15.31$\pm$0.01 &
  19.3$\pm$0.01 &
  \citenum{2010ApJ...710.1142B}, \citenum{2012ApJ...752...56F} \\
SDSS J104335.08+121314.1 &
  14.6$\pm$2.26 &
  16.06$\pm$0 &
  14.92$\pm$0 &
  14.24$\pm$0 &
  16.24$\pm$0 &
  15.16$\pm$0 &
  16.3$\pm$0 &
  14.25$\pm$0 &
  18.43$\pm$0 &
  \citenum{2010ApJS..190..100K}, \citenum{2012ApJ...752...56F} \\
SDSS J104409.43+042937.6 &
  19.49$\pm$3.84 &
  15.98$\pm$0 &
  14.89$\pm$0 &
  14.2$\pm$0 &
  16.09$\pm$0 &
  15.12$\pm$0 &
  16.36$\pm$0 &
  14.25$\pm$0 &
  18.29$\pm$0 &
  \citenum{2010ApJ...710.1142B}, \citenum{2012ApJ...752...56F} \\
SDSS J104842.84+011158.5 &
  13.91$\pm$1.43 &
  12.91$\pm$0 &
  12.17$\pm$0 &
  11.6$\pm$0 &
  12.92$\pm$0 &
  12.38$\pm$0 &
  13.24$\pm$0 &
  11.74$\pm$0 &
  14.99$\pm$0 &
  \citenum{2008ApJ...681..579B}, \citenum{2016ApJS..225...10F} \\
SDSS J115553.86+055957.5 &
  17.27$\pm$3.04 &
  15.74$\pm$0 &
  14.7$\pm$0 &
  14.04$\pm$0 &
  15.91$\pm$0 &
  14.94$\pm$0 &
  15.97$\pm$0 &
  14.09$\pm$0 &
  18.03$\pm$0 &
  \citenum{2010ApJ...710.1142B}, \citenum{2012ApJ...752...56F} \\
SDSS J120747.17+024424.8 &
  22.47$\pm$6.16 &
  15.47$\pm$0 &
  14.55$\pm$0 &
  14.09$\pm$0 &
  15.93$\pm$0 &
  14.79$\pm$0 &
  15.73$\pm$0 &
  14.07$\pm$0 &
  17.83$\pm$0 &
  \citenum{2007AJ....134.1162L}, \citenum{2012ApJ...752...56F} \\
SDSS J133148.92-011651.4 &
  14.86$\pm$2.78 &
  15.41$\pm$0 &
  14.56$\pm$0 &
  14.04$\pm$0 &
  15.89$\pm$0.01 &
  14.84$\pm$0 &
  15.8$\pm$0.01 &
  14.12$\pm$0.01 &
  17.66$\pm$0.01 &
  \citenum{2010ApJ...710.1142B}, \citenum{2013AJ....146..161M} \\
SDSS J141659.78+500626.4 &
  45.66$\pm$1.29 &
  17.03$\pm$0 &
  16.05$\pm$0 &
  15.4$\pm$0.01 &
  17.19$\pm$0.01 &
  16.3$\pm$0 &
  17.21$\pm$0.02 &
  15.48$\pm$0.04 &
  19.59$\pm$0 &
  \citenum{2006AJ....131.2722C}, \citenum{2007AAA...474..653V} \\
SDSS J151114.66+060742.9 &
  27.25$\pm$4.75 &
  15.9$\pm$0 &
  15.13$\pm$0 &
  14.45$\pm$0 &
  16.28$\pm$0 &
  15.26$\pm$0.01 &
  16.24$\pm$0 &
  14.51$\pm$0.01 &
  18.37$\pm$0 &
  \citenum{2010ApJ...710.1142B}, \citenum{2012ApJ...752...56F} \\
SDSS J152103.24+013142.7 &
  24.21$\pm$4.22 &
  16.29$\pm$0 &
  15.58$\pm$0 &
  15.45$\pm$0 &
  17.26$\pm$0.01 &
  15.75$\pm$0.01 &
  16.81$\pm$0.01 &
  15.19$\pm$0.01 &
  19.35$\pm$0.01 &
  \citenum{2010ApJ...710.1142B}, \citenum{2012ApJ...752...56F} \\
SDSS J175805.46+463311.9 &
  14.09$\pm$0.37 &
  16.09$\pm$0 &
  15.91$\pm$0 &
  15.77$\pm$0 &
  17.99$\pm$0.02 &
  15.76$\pm$0 &
  17.47$\pm$0 &
  15.11$\pm$0.01 &
  19.35$\pm$0 &
  \citenum{2006ApJ...639.1095B}, \citenum{2007AAA...474..653V} \\
SDSS J202820.32+005226.5 &
  30.08$\pm$1.19 &
  14.29$\pm$0 &
  13.41$\pm$0 &
  12.76$\pm$0 &
  14.32$\pm$0 &
  13.63$\pm$0 &
  14.47$\pm$0 &
  12.9$\pm$0 &
  16.49$\pm$0 &
  \citenum{2004AJ....127.2856B}, \citenum{2016AJ....152...24W} \\
SDSS J204749.61-071818.3 &
  30.12$\pm$4.99 &
  16.81$\pm$0 &
  15.76$\pm$0 &
  15.23$\pm$0 &
  17.07$\pm$0.01 &
  15.97$\pm$0 &
  17.09$\pm$0.01 &
  15.22$\pm$0.01 &
  19.12$\pm$0.01 &
  \citenum{2010ApJ...710.1142B}, \citenum{2013MNRAS.433.2054S} \\
SDSSp J003259.36+141036.6 &
  33.18$\pm$5.68 &
  16.77$\pm$0 &
  15.65$\pm$0 &
  15$\pm$0 &
  16.92$\pm$0.01 &
  15.93$\pm$0 &
  17.12$\pm$0.01 &
  15$\pm$0.01 &
  19.23$\pm$0.01 &
  \citenum{2010ApJ...710.1142B}, \citenum{2004AJ....127.2948V} \\
SDSSp J010752.33+004156.1 &
  15.59$\pm$1.1 &
  15.87$\pm$0 &
  14.56$\pm$0 &
  13.6$\pm$0 &
  15.73$\pm$0 &
  14.83$\pm$0 &
  16.25$\pm$0.01 &
  13.69$\pm$0 &
  18.26$\pm$0.01 &
  \citenum{2010ApJ...710.1142B}, \citenum{2004AJ....127.2948V} \\
SDSSp J010752.33+004156.1 &
  15.59$\pm$1.1 &
  15.84$\pm$0 &
  14.57$\pm$0 &
  13.63$\pm$0 &
  15.7$\pm$0 &
  14.84$\pm$0 &
  16.27$\pm$0 &
  13.72$\pm$0 &
  18.36$\pm$0 &
  \citenum{2010ApJ...710.1142B}, \citenum{2004AJ....127.2948V} \\
SDSSp J053951.99-005902.0 &
  13.14$\pm$0.37 &
  13.98$\pm$0 &
  13.12$\pm$0 &
  12.56$\pm$0 &
  14.14$\pm$0 &
  13.35$\pm$0 &
  14.33$\pm$0 &
  12.66$\pm$0 &
  16.25$\pm$0 &
  Unpublished, \citenum{2004AJ....127.2948V} \\
SDSSp J083008.12+482847.4 &
  13.09$\pm$0.59 &
  15.42$\pm$0 &
  14.33$\pm$0 &
  13.71$\pm$0 &
  15.91$\pm$0 &
  14.64$\pm$0 &
  15.86$\pm$0 &
  13.72$\pm$0 &
  17.69$\pm$0 &
  \citenum{2008ApJ...681..579B}, \citenum{2004AJ....127.2948V} \\
SDSSp J083717.22-000018.3 &
  29.67$\pm$11.84 &
  16.96$\pm$0.01 &
  16.08$\pm$0.01 &
  15.71$\pm$0.01 &
  17.54$\pm$0.02 &
  16.33$\pm$0.01 &
  17.32$\pm$0.02 &
  15.71$\pm$0.02 &
  19.57$\pm$0.01 &
  \citenum{2006ApJ...637.1067B}, \citenum{2004AJ....127.2948V} \\
SDSSp J125453.90-012247.4 &
  11.78$\pm$0.26 &
  14.92$\pm$0 &
  14.07$\pm$0 &
  13.82$\pm$0 &
  15.86$\pm$0 &
  14.33$\pm$0 &
  15.51$\pm$0.01 &
  13.66$\pm$0 &
  17.68$\pm$0 &
  \citenum{2004AJ....127.2856B}, \citenum{2002AJ....124.1170D} \\
SDSSp J132629.82-003831.5 &
  20.01$\pm$2.53 &
  16.23$\pm$0 &
  15.02$\pm$0 &
  14.1$\pm$0 &
  16.1$\pm$0 &
  15.32$\pm$0 &
  16.49$\pm$0 &
  14.21$\pm$0 &
  18.59$\pm$0 &
  \citenum{2010ApJ...710.1142B}, \citenum{2004AJ....127.2948V} \\
SDSSp J162414.37+002915.6 &
  11$\pm$0.15 &
  15.52$\pm$0 &
  15.42$\pm$0 &
  15.59$\pm$0 &
  17.68$\pm$0.01 &
  15.3$\pm$0.01 &
  16.47$\pm$0 &
  15$\pm$0 &
  18.63$\pm$0 &
  \citenum{2006ApJ...639.1095B}, \citenum{2003AJ....126..975T} \\
SDSSp J175032.96+175903.9 &
  27.59$\pm$3.45 &
  16.2$\pm$0 &
  15.68$\pm$0 &
  15.8$\pm$0 &
  17.47$\pm$0.01 &
  15.77$\pm$0 &
  16.7$\pm$0.01 &
  15.46$\pm$0.01 &
  18.92$\pm$0 &
  \citenum{2004AJ....127.2856B}, \citenum{2004AJ....127.2948V} \\
VB 10 &
  5.95$\pm$0.02 &
  9.88$\pm$0 &
  9.26$\pm$0 &
  8.77$\pm$0 &
  9.86$\pm$0 &
  9.43$\pm$0 &
  10.14$\pm$0 &
  8.9$\pm$0 &
  11.54$\pm$0 &
  \citenum{2004AJ....127.2856B}, \citenum{2016AJ....152...24W} \\
VB 8 &
  6.48$\pm$0.02 &
  9.78$\pm$0 &
  9.2$\pm$0 &
  8.81$\pm$0 &
  9.79$\pm$0 &
  9.38$\pm$0 &
  10.03$\pm$0 &
  8.94$\pm$0 &
  11.2$\pm$0 &
  \citenum{2008ApJ...681..579B}, \citenum{2016AJ....152...24W} \\
Wolf 359 &
  2.42$\pm$0.01 &
  7.11$\pm$0 &
  6.49$\pm$0 &
  6.06$\pm$0 &
  6.9$\pm$0 &
  6.62$\pm$0 &
  7.35$\pm$0 &
  6.17$\pm$0 &
  8.55$\pm$0 &
  \citenum{2008ApJ...681..579B}, \citenum{2016AJ....152...24W} \\* \bottomrule
\caption{Photometry computed from SpeX spectra using our \textsc{Synth.ph} module.}
\label{tab:my-table}\\
\end{longtable}
\end{landscape}
\twocolumn

\bsp	
\label{lastpage}
\end{document}